\documentclass[a4paper, twoside]{thesis} 

\usepackage{a4}
\usepackage{amsmath}
\usepackage{amsfonts}
\usepackage{amssymb}
\usepackage{theorem}
\usepackage[dvips]{color}
\usepackage{bm}
\usepackage{graphicx}

\begin{document}

\frontmatter

\thispagestyle{empty}

.

  \vspace{7 cm}

\begin{center}

  {\bf \huge An Introduction to}\\
  \vspace{1 cm}
  {\bf \huge the Formalism of Quantum Information}

\end{center}

  \vspace{4 cm}

\begin{flushright}

  {\bf \large Notes by}\\
  \vspace{1 cm}

  {\bf \Large Carlos Navarrete-Benlloch}

\end{flushright}


\newpage
\thispagestyle{empty}
\vspace*{18cm}

\tableofcontents

\mainmatter

\pagebreak

\chapter{Introduction to the notes}
Quantum information is an emerging field which has attracted a lot of attention in the last fifteen years or so. It is a broad subject which covers from the most applied questions (e.g., how to build quantum computers or secure cryptographic systems), to the most theoretical problems concerning the formalism of quantum mechanics, its complexity, and its potential to go beyond classical physics.

I have written these notes as an introduction to the formalism of quantum information, trying to show how it is a natural progression from an undergraduate course on quantum mechanics (just as I felt it was for me). Hence, these are mainly notes about the language of quantum information and the meaning of its terms, not about its applications.

Not showing many examples of the application of the tools that are introduced in the notes, the reader might feel that some definitions are just `highly complicated ways of writing simple quantum mechanical expressions', even finding hard to understand the motivation behind introducing concepts like 'entanglement' or `quantum operations'; questions along the line of ``so what?'' might cross the reader's mind, and under such circumstances I can only ask for his/her patience: this is a work in progress and I hope to keep improving the notes with practical examples, applications, and implementations... but for the time being, just trust me when I say that there is a proper reason for introducing every concept appearing in the notes, as well as for writing each expression the way it is written.

Also, let me remark that these notes do not intend to be a comprehensive and exhaustive review about the field, but just a logical sequence of ideas, concepts and definitions, building up towards the topics I've worked on within the continuous variables formalism, which actually takes over most of the notes. In this way, most of the results are stated without proof, and I haven't been careful with references; as I said, this is a work in progress, and I intend to add such things in the future. In any case, most of the topics that I introduce have been thoroughly covered in previous books and reviews \cite{Nielsen00book,HorodeckiReview,EisertUN,Braunstein05,Weedbrook12}, and with this notes I only intend to give my point of view, as well as to introduce the different topics and objects in exactly the way I work with and think about them, hoping it will be useful, especially for the people who are starting in the field.

This notes originated in 2010 as a way of studying for my close collaboration with Ra\'ul Garc\'ia-Patr\'on and Nicolas J. Cerf, to whom I am deeply grateful for guidance and incredibly clarifying discussions. They benefited greatly as well from the weeks I spent in Doha with Hyunchul Nha and his group in 2012, especially from discussions with him and Ho-Joon Kim.

Munich, April 21, 2015.

\begin{flushright}
Carlos Navarrete-Benlloch
\end{flushright}

\chapter{Quantum mechanics in an isolated system}
\label{Isolated}
The intention of this initial chapter is to introduce the axioms of quantum
mechanics as we will use them during these notes. These are introduced from a
very pragmatic viewpoint, not trying to justify them or motivate them from a
physical perspective, for a reasoned presentation of them I recommend reading
\cite{NavarreteThesis} (Appendix A, as well as the references therein), as the
tone and notation of these notes are fully correlated with that reference. The
mathematics behind quantum mechanics (Hilbert spaces) are assumed to be known,
although I provide a summary of them in Appendix \ref{MathematicsQuantum} as a
reminder, and in order to fix the notation.

\section{The state of the system}

In the mathematical framework of quantum mechanics, a \textit{Hilbert space}
$\mathcal{H}$ is associated to any physical system. The state of the system is
completely specified by a \textit{density operator} $\hat{\rho}$, that is, a
self-adjoint ($\hat{\rho}=\hat{\rho}^{\dagger}$), positive ($\langle\psi
|\hat{\rho}|\psi\rangle\geq0$ $\forall|\psi\rangle\in\mathcal{H}$), and
unit-trace ($\mathrm{tr}\{\hat{\rho}\}=1$) operator.

Density operators can always be written as a convex mixture of projectors,
that is,
\begin{equation}
\hat{\rho}=\sum_{k=1}^{N}w_{k}|\varphi_{k}\rangle\langle\varphi_{k}|
\end{equation}
where $\{w_{k}\}_{k=1,2,...,N}$ is a probability distribution ($w_{k}\geq0$
$\forall k$ and $\sum_{k=1}^{N}w_{k}=1$) and the states $\{|\varphi_{k}%
\rangle\}_{k=1,2,...,N}$ are normalized, but may not be orthogonal. The set
$\{w_{k},|\varphi_{k}\rangle\}_{k=1,2,...,N}$ is known as an \textit{ensemble
decomposition} of the state $\hat{\rho}$, and it is not unique (in the sense
that different ensembles can lead to the same $\hat{\rho}$). Not that $N$ does
not have to coincide with the dimension of the Hilbert space $\mathcal{H}$.

It can be proved that two ensembles $\{w_{k},|\varphi_{k}\rangle
\}_{k=1,2,...,N}$ and $\{v_{l},|\psi_{l}\rangle\}_{l=1,2,...,M}$ give rise to
the same density operator $\hat{\rho}$ if and only if there exists a
left-unitary matrix\footnote{$\mathcal{U}$ is left-unitary if $\mathcal{U}%
^{\dagger}\mathcal{U}=\mathcal{I}$ but $\mathcal{UU}^{\dagger}\neq\mathcal{I}%
$, where $\mathcal{I}$ is the identity matrix of the corresponding dimension.
It is easy to prove that finite-dimensional left-unitary matrices are
unitary.} $\mathcal{U}$ such that \cite{Nielsen00book}%
\begin{equation}
\sqrt{w_{k}}|\varphi_{k}\rangle=\sum_{l}\mathcal{U}_{kl}\sqrt{v_{l}}|\psi
_{l}\rangle,
\end{equation}
where if $k\neq l$, then $|k-l|$ zeros must be included in the ensemble with
less states, so that $\mathcal{U}$ is a square matrix.

When only one state $|\varphi\rangle$ contributes to the mixture, then
$\hat{\rho}=|\varphi\rangle\langle\varphi|$ and we say that the state is
\textit{pure}; otherwise, the state is \textit{mixed}. A necessary and
sufficient condition for $\hat{\rho}$ to be pure is $\hat{\rho}^{2}=\hat{\rho
}$.

In the next chapters we will learn that the mixedness of a state always comes
from the fact that some of the information of the system has been lost to some
other inaccessible system with which it has interacted for a while before
becoming isolated. In other words, the state of a system is pure only when it
has no correlations at all with other systems.

\section{Evolution of the system}

Quantum mechanics contemplates two different types of evolution of a closed
system which is initially in some state $\hat{\rho}$:

\begin{itemize}
\item \textbf{Unitary. }If the system is allowed to evolve during a time $t$
without `observing it', its state changes to%
\begin{equation}
\hat{\rho}^{\prime}=\hat{U}\hat{\rho}\hat{U}^{\dagger},
\end{equation}
where%
\begin{equation}
\hat{U}=\exp(\hat{H}t/\mathrm{i}\hbar),
\end{equation}
is a unitary operator, being $\hat{H}$ a self-adjoint operator corresponding
to the Hamiltonian of the system.

\item \textbf{Projective measurements. }When one measures an observable with
associated self-adjoint operator $\hat{Y}=\sum_{j}y_{j}\hat{P}_{j}$, where
$\hat{P}_{j}=|y_{j}\rangle\langle y_{j}|$ are the projectors onto its
eigenvectors\footnote{We will only consider the case of observables having
non-degenerate eigenvalues.} and $y_{j}\in%
\mathbb{R}
$, the outcome $y_{j}$ appears with probability
\begin{equation}
p_{j}=\mathrm{tr}\{\hat{\rho}\hat{P}_{j}\}=\langle y_{j}|\hat{\rho}%
|y_{j}\rangle.
\end{equation}
If the outcome is recorded by the observer, the state of the system changes
(\textit{collapses}) to%
\begin{equation}
\hat{\rho}_{j}^{\prime}=p_{j}^{-1}\hat{P}_{j}\hat{\rho}\hat{P}_{j}%
=|y_{j}\rangle\langle y_{j}|,\label{CollapsedState1}%
\end{equation}
and the measurement is called \textit{selective}. If, on the other hand, the
outcome of the measurement is unknown to the observer, the system is left in
the mixture%
\begin{equation}
\hat{\rho}^{\prime}=\sum_{j}p_{j}|y_{j}\rangle\langle y_{j}|=\sum_{j}\hat
{P}_{j}\hat{\rho}\hat{P}_{j},\label{CollapsedState2}%
\end{equation}
and the measurement is called \textit{non-selective}.\newline It might seem
strange to see that the state of the system is different after the measurement
depending on whether the observer learns the outcome or not; however, note
that the state (\ref{CollapsedState2}) is statistically indistinguishable from
keeping track of the outcomes at any idividual realization of the measurement,
and assume that the state collapses after each measurement to the
corresponding state (\ref{CollapsedState1}). That is, on what concerns to the
statistics of any experiment, selective and non-selective measurements are equivalent.
\end{itemize}

Note that these two types of evolution are fundamentally different: while
unitary evolution is reversible, projective measurements are not.

\section{The von Neumann entropy}

The \textit{von Neumann entropy} of a state $\hat{\rho}$ is defined as%
\begin{equation}
S[\hat{\rho}]=-\mathrm{tr}\{\hat{\rho}\log\hat{\rho}\}.
\end{equation}
Given the diagonal representation of the state%
\begin{equation}
\hat{\rho}=\sum_{j=1}^{d}\lambda_{j}|r_{j}\rangle\langle r_{j}|,
\end{equation}
where the set $\{|r_{j}\rangle\}_{j=1,2,...,d}$ forms an orthonormal basis of
the Hilbert space $\mathcal{H}$ (which therefore has dimension $d$), the von
Neumann entropy is just the Shannon entropy of the distribution\footnote{In
the following we use the notations $\mathbf{p}=\operatorname{col}(p_{1}%
,p_{2},...,p_{d})$ and $\{p_{j}\}_{j=1,2,...,d}$ for a probability
distribution interchangeably.} $\boldsymbol{\lambda}=\operatorname{col}%
(\lambda_{1},\lambda_{2},...,\lambda_{d})$, that is,%
\begin{equation}
S[\hat{\rho}]=-\sum_{j=1}^{d}\lambda_{j}\log\lambda_{j}.
\end{equation}

For a pure state $|\varphi\rangle\langle\varphi|$ the entropy is zero, while
it has a miximum $\log d$ for the \textit{maximally mixed state}%
\begin{equation}
\hat{\rho}_{\mathrm{MM}}=\frac{1}{d}\hat{I}.\label{RhoMM}%
\end{equation}
Hence, the entropy can be understood as a measure of the mixedness of the
state\footnote{The case $d\rightarrow\infty$ will be discussed later when
studying in detail infinite-dimensional spaces.}. Note that when the state of
the system is $\hat{\rho}_{\mathrm{MM}}$ and a given observable $\hat{Y}$
acting on $\mathcal{H}$ is measured, all its eigenvalues are equally likely to
appear as an outcome of the measurement, that is,%
\begin{equation}
p_{j}=\langle y_{j}|\hat{\rho}|y_{j}\rangle=1/d\text{ \ }\forall j.\text{
\ \ \ [flat distribution]}%
\end{equation}
This reinforces the interpretation of mixedness as due to some kind of
information loss.

The entropy is a concave functional of density operators, that is, given the
convex mixture of states
\begin{equation}
\hat{\rho}=\sum_{k}w_{k}\hat{\rho}_{k},
\end{equation}
where $\{w_{k}\}_{k=1,2,...,N}$ is a probability distribution and $\{\hat
{\rho}_{k}\}_{k=1,2,...,N}$ is a set of density operators, we have%
\begin{equation}
S[\hat{\rho}]\geq\sum_{k=1}^{N}w_{k}S[\hat{\rho}_{k}].
\end{equation}
This is easily proved by writing each density matrix in its diagonal form, and
using the concavity of the $-x\log x$ function.

Finally, it is interesting to note that the entropy does not change with
unitary evolution, can only increase with non-selective projective
measurements, and can only decrease with selective projective measurements
(indeed, when no degeneracies are present, it collapses to zero, as the state
becomes pure).

\chapter{Bipartite systems and entanglement}
\label{Entanglement}
\section{Entangled states}

Consider two systems $A$ and $B$ (named after Alice and Bob, two observers
which are able to interact locally with their respective system) with
associated Hilbert spaces $\mathcal{H}_{A}$ and $\mathcal{H}_{B}$. Imagine
that the systems \textit{A} and \textit{B} interact during some time in such a
way that they cannot be described anymore by independent states $\hat{\rho
}_{A}$ and $\hat{\rho}_{B}$ acting on $\mathcal{H}_{A}$ and $\mathcal{H}_{B}$,
respectively, but by a state $\hat{\rho}_{AB}$ acting on the joint space
$\mathcal{H}_{A}\otimes\mathcal{H}_{B}$. The question is, is it possible to
reproduce the statistics of the measurements performed by Alice on system
\textit{A} via some state $\hat{\rho}_{A}$ defined in $\mathcal{H}_{A}$ only?
This question has a positive and \textit{unique} answer: this state is given
by the \textit{reduced density operator}%
\begin{equation}
\hat{\rho}_{A}=\mathrm{tr}_{B}\{\hat{\rho}_{AB}\},
\end{equation}
that is, by performing the partial trace\footnote{Given an orthonormal basis
$\{|b_{j}\rangle\}_{j}$ of $\mathcal{H}_{B}$, this is defined by%
\begin{equation}
\mathrm{tr}\{\hat{\rho}_{AB}\}=\sum_{j}\langle b_{j}|\hat{\rho}_{AB}%
|b_{j}\rangle,
\end{equation}
which is indeed an operator acting on $\mathcal{H}_{A}$.} over system's
\textit{B} subspace onto the joint state. Of course, the same applies to Bob,
whose system is described locally by the reduced density operator $\hat{\rho
}_{B}=\mathrm{tr}_{A}\{\hat{\rho}_{AB}\}$.

If the state of the joint system is of the type $\hat{\rho}_{AB}%
^{(\mathrm{t})}=\hat{\rho}_{A}\otimes\hat{\rho}_{B}$, that is, a tensor
product of two arbitrary density operators, the actions performed by Alice on
system $A$ won't affect Bob's system, the statistics of which are given by
$\hat{\rho}_{B}$, no matter the actual state $\hat{\rho}_{A}$. In this case
$A$ and $B$ are \textit{uncorrelated}. For any other type of joint state, $A$
and $B$ will share some kind of correlation.

Correlations are not strange in classical mechanics; hence, a problem of
paramount relevance in quantum mechanics is to understand which type of
correlations can appear at a classical level, and which are purely quantum.
This is because only if the latter are present, one can expect to find quantum
mechanical effects such as violation of local realism, computational
algorithms exponentially faster than the fastest ones known in classical
computer science, or unconditionally secure cryptography.

Intuitively, the state of the system will induce only classical correlations
between $A$ and $B$ if and only if it can be written as%
\begin{equation}
\hat{\rho}_{AB}^{(\mathrm{s})}=\sum_{k}w_{k}\hat{\rho}_{A}^{(k)}\otimes
\hat{\rho}_{B}^{(k)}, \label{RhoSep}%
\end{equation}
where the $\hat{\rho}^{(k)}$'s are density operators and $\{w_{k}%
\}_{k=1,2,...}$ is a probability distribution, because it means that Alice and
Bob can share a classical machine which randomly picks a value of $k$
according to the distribution $\{w_{k}\}_{k}$ to trigger the preparation of
the states $\hat{\rho}_{A}^{(k)}$ and $\hat{\rho}_{B}^{(k)}$, what can be done
locally, and hence cannot induce further correlations. If the process is
automatized so that Alice and Bob do not learn the outcome $k$ of the random
number generator, the mixture $\hat{\rho}_{AB}^{(\mathrm{s})}$ is prepared. In
other words, the state does not contain quantum correlations if it can be
prepared using only local operations and classical communication\footnote{We
will learn the exact meaning of this class of operations in the next chapter.}.

There is yet another way of justifying that states which cannot be written in
the separable form (\ref{RhoSep}) will make $A$ and $B$ share quantum
correlations. The idea is that one of the fundamental differences between
classical and quantum mechanics is the \textit{superposition principle}, that
is, \textit{interferences}. Hence, it is intuitive that correlations will have
a quantum nature when the joint state of the systems exploits the concept of
\textquotedblleft superposition of joint states\textquotedblright, that is,
when it cannot be written as a tensor product of two independent states of the
systems, or as a purely classical statistical mixture of these, which
corresponds exactly to (\ref{RhoSep}).

States of the type $\hat{\rho}_{AB}^{(\mathrm{s})}$ are called
\textit{separable}. Any non-separable state will induce quantum correlations
between $A$ and $B$; these correlations which cannot be generated by classical
means are known as \textit{entanglement}, and states which are not separable
are called \textit{entangled states}.

\section{Characterizing and quantifying entanglement}

In general, given a mixed state $\hat{\rho}_{AB}$ acting on $\mathcal{H}%
_{A}\otimes\mathcal{H}_{B}$, it is hard to find out whether it is separable or
not, the difficulty coming from distinguishing between quantum and classical
correlations. Indeed, the best known criterion for separability, the
\textit{Peres-Horodecki criterion}, yields only necessary and sufficient
conditions when $d_{A}\times d_{B}\leq6$ (note that this includes the case of
two qubits), and for a reduced class of states in infinite-dimensional Hilbert
spaces (Gaussian states, see Chapter \ref{ContinuousVariables}). This criterion states
that a necessary condition for the separability of a density operator is that
it remains being positive after the operation of partial transposition, that
is, given%
\begin{equation}
\hat{\rho}_{AB}=\sum_{jklm}\rho_{jl,km}|a_{j},b_{l}\rangle\langle a_{k}%
,b_{m}|,
\end{equation}
where $\{|a_{j},b_{l}\rangle\doteqdot|a_{j}\rangle\otimes|b_{l}\rangle
\}_{j,l}$ is an orthonormal basis of $\mathcal{H}_{A}\otimes\mathcal{H}_{B}$,
\begin{equation}
\hat{\rho}_{AB}^{T_{A}}=\sum_{jklm}\rho_{kl,jm}|a_{j},b_{l}\rangle\langle
a_{k},b_{m}|,
\end{equation}
is a positive operator. We will learn more about this criterion and some more
when studying infinite dimensional Hilbert spaces in Chapter \ref{ContinuousVariables}.

A very different problem is that of quantifying the level of correlations present in
the state, and more importantly, how much of these correspond to entanglement.
Even though we pretty much understand the conditions that a proper
\textit{entanglement measure} $E[\hat{\rho}_{AB}]$ must satisfy, we have not
found a completely satisfactory one for general states \cite{EisertUN} (either
they do not satisfy all the conditions, or/and can only be efficiently
computed for restricted classes of states). It is not the intention of these
introductory notes to explain all these measures and up to what point they are
satisfactory, but let me spend a few words on this topic for completeness (see
\cite{EisertUN,HorodeckiReview} for more details).

The basic conditions that a good entanglement measure $E[\hat{\rho}_{AB}]$
should satisfy are quite intuitive:

\begin{enumerate}
\item $E[\hat{\rho}_{AB}]$ is positive definite and equal to zero for
separable states.

\item Given the mixture $\hat{\rho}_{AB}=\sum_{j}p_{j}\hat{\rho}_{j}$, where
$\{p_{j}\}_{j}$ is a probability distribution and $\{\hat{\rho}_{j}\}_{j}$ are
density operators acting on $\mathcal{H}_{A}\otimes\mathcal{H}_{B}$,
$E[\hat{\rho}_{AB}]\leq\sum_{j}p_{j}E[\hat{\rho}_{j}]$, which is as to say
that the entanglement of a collection of states cannot be increased by not
knowing which of them has been prepared.

\item At least on average, the entanglement can only decrease when Alice and
Bob apply protocols involving only local operations and classical communication.
\end{enumerate}

These three conditions define what is known as an \textit{entanglement
monotone}. By themselves, they are not enough to define a unique entanglement
measure even for pure states. However, by adding two more conditions known as
\textit{weak additivity} and \textit{weak continuity} \cite{EisertUN}, which
find an intuitive justification in the asymptotic limit of having infinitely
many copies of the state, it is possible to prove that the
\textit{entanglement entropy}, which measures how mixed is left the reduced
density operator of one of the parties after tracing out the other, is the
unique entanglement measure of pure states $\hat{\rho}_{AB}=|\psi\rangle
_{AB}\langle\psi|$; given the reduced density operator $\hat{\rho}%
_{A}=\mathrm{tr}_{B}\{|\psi\rangle_{AB}\langle\psi|\}$ or $\hat{\rho}%
_{B}=\mathrm{tr}_{A}\{|\psi\rangle_{AB}\langle\psi|\}$, this entanglement
measure can be evaluated as%
\begin{equation}
E[|\psi\rangle_{AB}]=S[\hat{\rho}_{A}]=S[\hat{\rho}_{B}],\label{EntEntr}%
\end{equation}
where the equality of the von Neumann entropies of the reduced states will be
clear after the following section. Hence, the problem of quantifying the
entanglement is basically solved for pure bipartite states. The pure states
whose corresponding reduced states are maximally mixed are known as
\textit{maximally entangled states}.

In the case of mixed states, the entanglement entropy is not even an
entanglement monotone, as the entanglement entropy of the separable state
$\hat{\rho}_{A}\otimes\hat{\rho}_{B}$ is just the entropy of the mixed states
$\hat{\rho}_{A}$ or $\hat{\rho}_{B}$, which is not zero except for pure states
(note that, furthermore, it depends on which mode is traced out). However,
there are many quantifiers which are entanglement monotones:

\begin{itemize}
\item Possibly the most natural entanglement measure for mixed states is the
\textit{distillable entanglement}. Suppose that we give $N$ copies of the
mixed state $\hat{\rho}_{AB}$ to Alice and Bob; the process of
\textit{distillation }refers to the conversion of these copies to copies of
maximally entangled states via protocols involving only local operations and
classical communication. The distillable entanglement is defined as the
maximum number of maximally entangled states that can be distilled from
infinitely many copies of $\hat{\rho}_{AB}$. Apart from an entanglement
monotone, it can also be shown to satisfy the weak additivity and weak
continuity conditions, and to be equal to the entanglement entropy for pure
states. Its drawback is that it requires a maximization over all the possible
distillation protocols, and it is therefore very hard to evaluate
(paraphrasing \cite{EisertUN}: it is a problem ranging from \textquotedblleft
difficult to hopeless\textquotedblright.)

\item The \textit{entanglement of formation} can be seen as the dual of the
distillable entanglement: it measures the number of maximally entangled states
that are needed to prepare infinitely many copies of the mixed state. It can
be evaluated as%
\begin{equation}
E_{F}[\hat{\rho}_{AB}]=\underset{\{w_{k},|\psi_{k}\rangle\}_{k}}{\min}\sum
_{k}w_{k}E[|\psi_{k}\rangle],
\end{equation}
where the minimization is performed over all the possible ensemble
decompositions of $\hat{\rho}_{AB}$, what makes the measure very hard to
evaluate again. Nevertheless, closed formulas have been obtained for the
entanglement of formation of the general state of two qubits ($d_{A}\times
d_{B}=2\times2$), as well as for reduced classes of higher-dimensional
bipartite states with strong symmetries.

\item There is an entanglement monotone which can be evaluated fairly
efficiently, as it does not require any optimization procedure: the
\textit{logarithmic negativity}. In loose terms, it quantifies how much the
state $\hat{\rho}_{AB}$ violates the Peres-Horodecki criterion via%
\begin{equation}
E_{N}[\hat{\rho}_{AB}]=\log||\hat{\rho}_{AB}^{T_{A}}||_{1}=\log\left[
1+\sum_{j}(|\tilde{\lambda}_{j}|-\tilde{\lambda}_{j})\right]  ,
\end{equation}
where $\{\tilde{\lambda}_{j}\}_{j}$ are the eigenvalues of $\hat{\rho}%
_{AB}^{T_{A}}$, and $||\hat{A}||_{1}=\mathrm{tr}\sqrt{\hat{A}\hat{A}^{\dagger
}}$ denotes the so-called \textit{trace norm}. The problem with this measure
is that it does not collapse to the entanglement entropy (\ref{EntEntr}) for
pure states, and is not weakly additive in general.
\end{itemize}

\section{Schmidt decomposition and purifications}

It is possible to show \cite{Nielsen00book} that any pure bipartite state can
be written in the form%
\begin{equation}
|\psi\rangle_{AB}=\sum_{j=1}^{d}\sqrt{\lambda_{j}}|u_{j}\rangle\otimes
|v_{j}\rangle,
\end{equation}
where $\{\lambda_{j}\}_{j=1,2,...,d}$ is a probability distribution, and
$\{|u_{j}\rangle\}_{j=1,2,...,d}$ and $\{|v_{j}\rangle\}_{j=1,2,...,d}$ are
orthonormal bases of $\mathcal{H}_{A}$ and $\mathcal{H}_{B}$, respectively,
which we assume to have the same dimension without loss of generality. With
the state written in this form (known as \textit{Schmidt decomposition}), it
is completely trivial to evaluate the entanglement entropy: as the reduced
density operators are diagonal, that is,%
\begin{equation}
\hat{\rho}_{A}=\sum_{j=1}^{d}\lambda_{j}|u_{j}\rangle\langle u_{j}|\text{,
\ \ \ \ and \ \ \ \ }\hat{\rho}_{B}=\sum_{j=1}^{d}\lambda_{j}|v_{j}%
\rangle\langle v_{j}|,
\end{equation}
their von Neumann entropies give%
\begin{equation}
E[|\psi\rangle_{AB}]=-\sum_{j=1}^{d}\lambda_{j}\log\lambda_{j}.
\end{equation}

The Schmidt decomposition allows us to introduce the concept of
\textit{purification}: given a system $A$ with associated Hilbert space
$\mathcal{H}_{A}$ in a mixed state with diagonal representation $\hat{\rho
}_{A}=\sum_{j}\lambda_{j}|r_{j}\rangle\langle r_{j}|$, we can always introduce
another system $B$ laying in a Hilbert space $\mathcal{H}_{B}$ with the same
dimension as $\mathcal{H}_{A}$, and with an orthonormal basis $\{|v_{j}%
\rangle\}_{j}$, and understand the mixed state $\hat{\rho}_{A}$ as a reduction
of the pure entangled state $|\psi\rangle_{AB}=\sum_{j}\sqrt{\lambda_{j}%
}|r_{j}\rangle\otimes|v_{j}\rangle$. Hence, a classical mixture of states can
always be transformed into a pure entangled state in a doubled Hilbert space.

\chapter{Quantum operations}
\label{QuantumOperations}
\section{Quantum operations as evolution of a reduced system}

Consider a system $Q$ with associated Hilbert space $\mathcal{H}_{Q}$ subject
to the actions of an experimentalist named Quinn; \textit{quantum operations}
appear as an answer to the following question: are unitary transformations and
projective measurements the only type of operations that Quinn is allowed to
perform onto the system $Q$?

The answer to this question is `no': Quinn can always append an auxiliary
system $L$ with associated Hilbert space $\mathcal{H}_{L}$, apply unitaries
and projective measurements on the joint system, and finally dismiss (trace
out) the appended system $L$. Quantum operations correspond to these reduced
unitaries and projective measurements as felt by the system $Q$ alone.

As we are going to prove in the reminding of the section \cite{Nielsen00book}%
,\ any quantum operation can be represented by a map of the type%
\begin{equation}
\mathcal{E}[\hat{\rho}]=\sum_{k=1}^{K}\hat{E}_{k}\hat{\rho}\hat{E}%
_{k}^{\dagger},\label{OperatorSum}%
\end{equation}
where $K$ can be selected at will by Quinn, and where the only restriction on
the operators $\{\hat{E}_{k}\}_{k=1,2,...,K}$ (which act onto $\mathcal{H}%
_{Q}$) is that they must satisfy%
\begin{equation}
\mathrm{tr}\left\{  \left[  \sum_{k=1}^{K}\hat{E}_{k}^{\dagger}\hat{E}%
_{k}\right]  \hat{\rho}\right\}  \leq1\text{ }\forall\hat{\rho},
\end{equation}
what is denoted symbolically by%
\begin{equation}
\sum_{k=1}^{K}\hat{E}_{k}^{\dagger}\hat{E}_{k}\leq\hat{I};
\end{equation}
in other words, the map can be \textit{trace preserving} or \textit{trace
decreasing}. If the state of the system $Q$ was $\hat{\rho}_{Q}$ prior to the
quantum operation, it becomes%
\begin{equation}
\hat{\rho}_{Q}^{\prime}=\frac{\mathcal{E}[\hat{\rho}_{Q}]}{\mathrm{tr}%
\{\mathcal{E}[\hat{\rho}_{Q}]\}},
\end{equation}
after it. The operators $\{\hat{E}_{k}\}_{k=1,2,...,K}$ are known as
\textit{Kraus operators}, and the expression (\ref{OperatorSum}) as an
\textit{operator-sum representation} of the quantum operation.

\bigskip

Let's see how by applying unitaries and measurements onto the joint Hilbert
space $\mathcal{H}_{Q}\otimes\mathcal{H}_{L}$, the reduced dynamics of system
$Q$ is described by a map of the type (\ref{OperatorSum}). Suppose that the
initial state of the joint system is%
\begin{equation}
\hat{\rho}_{QL}=\hat{\rho}_{Q}\otimes|\varphi_{L}\rangle\langle\varphi_{L}|;
\end{equation}
note that assuming that system $L$ is in a pure state is still completely
general, because if it is in a mixed state, this can be purified by
introducing another extra system, and renaming the joint system as $L$.

If Quinn applies a global unitary $\hat{U}$, the state evolves to%
\begin{equation}
\hat{\rho}_{QL}^{\prime}=\hat{U}(\hat{\rho}_{Q}\otimes|\varphi_{L}%
\rangle\langle\varphi_{L}|)\hat{U}^{\dagger}.
\end{equation}
Introducing now an orthonormal basis in $\mathcal{H}_{L}$ given by
$\{|l_{k}\rangle\}_{k=1,2,...,d_{L}}$, the reduced state of the system $Q$ is
written as%
\begin{equation}
\hat{\rho}_{Q}^{\prime}=\mathrm{tr}_{L}\{\hat{\rho}_{QL}^{\prime}\}=\sum
_{k=1}^{d_{L}}\langle l_{k}|\hat{U}(\hat{\rho}_{Q}\otimes|\varphi_{L}%
\rangle\langle\varphi_{L}|)\hat{U}^{\dagger}|l_{k}\rangle=\sum_{k=1}^{d_{L}%
}\hat{E}_{k}\hat{\rho}_{Q}\hat{E}_{k}^{\dagger},
\end{equation}
where the operators $\hat{E}_{k}=\langle l_{k}|\hat{U}|\varphi_{L}\rangle$ act
onto $\mathcal{H}_{Q}$. Note that $\mathrm{tr}_{Q}\{\hat{\rho}_{Q}^{\prime
}\}=1$, and hence, the reduced unitary corresponds to a trace-preserving map
like (\ref{OperatorSum}) with a number of Kraus operators $K$ given by the
dimension of the Hilbert space of the appended system $L$.

If, on the other hand, Quinn applies a non-selective measurement of a joint
observable with diagonal representation
\begin{equation}
\hat{Y}=\sum_{j=1}^{d_{Q}\cdot d_{L}}y_{j}\hat{P}_{j},
\end{equation}
where $\{\hat{P}_{j}=|y_{j}\rangle\langle y_{j}|\}_{j=1,2,...,d_{Q}\cdot
d_{L}}$ are the projectors onto its eigenvectors $\{|y_{j}\rangle
\in\mathcal{H}_{Q}\otimes\mathcal{H}_{L}\}_{j=1,2,...,d_{Q}\cdot d_{L}}$, the
state evolves to%
\begin{equation}
\hat{\rho}_{QL}^{\prime}=\sum_{j=1}^{d_{Q}\cdot d_{L}}\hat{P}_{j}(\hat{\rho
}_{Q}\otimes|\varphi_{L}\rangle\langle\varphi_{L}|)\hat{P}_{j}.
\end{equation}
The reduced state of system $Q$ can be written then as%
\begin{equation}
\hat{\rho}_{Q}^{\prime}=\mathrm{tr}_{L}\{\hat{\rho}_{QL}^{\prime}\}=\sum
_{j=1}^{d_{Q}\cdot d_{L}}\sum_{k=1}^{d_{L}}\langle l_{k}|\hat{P}_{j}(\hat
{\rho}_{Q}\otimes|\varphi_{L}\rangle\langle\varphi_{L}|)\hat{P}_{j}%
|l_{k}\rangle=\sum_{j=1}^{d_{Q}\cdot d_{L}}\sum_{k=1}^{d_{L}}\hat{E}_{k}%
^{(j)}\hat{\rho}_{Q}\hat{E}_{k}^{(j)\dagger},
\end{equation}
where $\hat{E}_{k}^{(j)}=\langle l_{k}|\hat{P}_{j}|\varphi_{L}\rangle$. Again,
it is immediate to check that $\mathrm{tr}_{Q}\{\hat{\rho}_{Q}^{\prime}\}=1$,
and hence, the reduced non-selective measurement is a trace-preserving quantum
operation with a number of Kraus operators given by $d_{Q}\cdot d_{L}^{2}$.

The reduced dynamics of a selective measurement of the joint observable
$\hat{Y}$ is a little more subtle. Assume that after the measurement Quinn
obtains the outcome $y_{j}$, so that, accordingly, the state $\hat{\rho}_{QL}$
collapses to%
\begin{equation}
\hat{\rho}_{QL}^{(j)}=p_{j}^{-1}\hat{P}_{j}(\hat{\rho}_{Q}\otimes|\varphi
_{L}\rangle\langle\varphi_{L}|)\hat{P}_{j},
\end{equation}
where%
\begin{equation}
p_{j}=\mathrm{tr}\{\hat{P}_{j}\hat{\rho}_{QL}\},
\end{equation}
is the probability for the outcome $y_{j}$ to appear after the measurement. In
this case it is easy to rewrite the reduced state of system $Q$ as%
\begin{equation}
\hat{\rho}_{Q}^{(j)}=\mathrm{tr}_{L}\{\hat{\rho}_{QL}^{(j)}\}=p_{j}^{-1}%
\sum_{k=1}^{d_{L}}\hat{E}_{k}^{(j)}\hat{\rho}_{Q}\hat{E}_{k}^{(j)\dagger
}\text{,}%
\end{equation}
and the corresponding probability as%
\begin{equation}
p_{j}=\mathrm{tr}_{Q}\left\{  \left[  \sum_{k=1}^{d_{L}}\hat{E}_{k}%
^{(j)\dagger}\hat{E}_{k}^{(j)}\right]  \hat{\rho}_{Q}\right\}  \leq1,
\end{equation}
where $\hat{E}_{k}^{(j)}=\langle l_{k}|\hat{P}_{j}|\varphi_{L}\rangle$. Hence,
similarly to the previous cases, the reduced selective measurement is
described by a map of the type (\ref{OperatorSum}), but the map is trace
decreasing in general.

\bigskip

Indeed, an important theorem states that any trace preserving quantum
operation can be written as the reduced evolution of system $Q$ after the
application of a joint unitary transformation acting onto the Hilbert space
$\mathcal{H}_{Q}\otimes\mathcal{H}_{L}$. On the other hand, trace decreasing
operations require an extra measurement of a joint observable, and the desired
quantum operation is accomplished only when a particular outcome appears in
the measurement, what happens with a probability equal to the trace of the map
(it requires then post-selection). Hence, while trace preserving operations
can be implemented deterministically---that is, at will by `pushing a
button'---trace decreasing operations can only be applied
probabilistically---meaning that we need to wait for the appareance of a
particular outcome in some measurement---. We will sometimes refer to system
$L$ as the \textit{environment}, and a representation of the quantum operation
in terms of reduced unitaries and measurements acting onto the system and the
environment is known as a \textit{Stinespring dilation}.

Finally, it is possible to show \cite{Nielsen00book} that two sets of Kraus
operators $\{\hat{E}_{k}\}_{k=1,2,...,K}$ and $\{\hat{F}_{l}\}_{l=1,2,...,L}$
lead to the same quantum operation if there exists a left-unitary matrix
$\mathcal{U}$ for which%
\begin{equation}
\hat{E}_{k}=\sum_{l}\mathcal{U}_{kl}\hat{F}_{l}\text{,}%
\end{equation}
where if $K\neq L$, then $|K-L|$ zeros must be included in the set with less
Kraus operators, so that $\mathcal{U}$ is a square matrix.

\section{Generalized measurements and POVMs}
\label{GenMesPOVM}

From the previous discussion it should be clear that the most general
measurement that one can perform on a system is completely described by a
collection of trace-decreasing quantum operations $\{\mathcal{E}%
_{j}\}_{j=1,2,...,J>1}$, which forms a complete set, that is,%
\begin{equation}
\sum_{j=1}^{J}\mathrm{tr}\{\mathcal{E}_{j}[\hat{\rho}]\}=1\text{ }\forall
\hat{\rho}\text{.}%
\end{equation}

The generalized measurements whose associated set of quantum operations
$\{\mathcal{E}_{j}\}_{j=1,2,...,J>1}$ are described by a single Kraus
operator, that is,%
\begin{equation}
\mathcal{E}_{j}[\hat{\rho}]=\hat{E}_{j}\hat{\rho}\hat{E}_{j}^{\dagger}%
\end{equation}
with%
\begin{equation}
\mathrm{tr}\{\hat{E}_{j}^{\dagger}\hat{E}_{j}\hat{\rho}\}\leq1\text{ }%
\forall\hat{\rho}\text{ \ \ \ \ and \ \ \ \ }\sum_{j=1}^{J}\hat{E}%
_{j}^{\dagger}\hat{E}_{j}=\hat{I}\text{,}%
\end{equation}
are very special, because it can be shown that its simplest Stinespring
dilation does not require a joint unitary, just a joint projective
measurement. The set $\{\hat{\Pi}_{j}=\hat{E}_{j}^{\dagger}\hat{E}%
_{j}\}_{j=1,2,...,J}$ is known as a \textit{positive operator-valued measure}
(POVM), while the operators $\{\hat{E}_{j}\}_{j=1,2,...,J}$ are called the
\textit{measurement operators}.

This generalized measurements are the closest ones to projective measurements;
the POVM $\{\hat{\Pi}_{j}\}_{j=1,2,...,J}$ plays the role of the spectral
decomposition of the measured observable, while the measurement operators
$\{\hat{E}_{j}\}_{j=1,2,...,J}$ (uniquely defined from the POVM up to a
left-unitary transformation) play the role of the projectors. The probability
of observing the outcome `$j$' is%
\begin{equation}
p_{j}=\mathrm{tr}\{\hat{\Pi}_{j}\hat{\rho}\},
\end{equation}
if the system was in the state $\hat{\rho}$ prior to the measurement, after
which it collapses to the state
\begin{equation}
\hat{\rho}_{j}=p_{j}^{-1}\hat{E}_{j}\hat{\rho}\hat{E}_{j}^{\dagger},
\end{equation}
if the measurement is selective, or the state%
\begin{equation}
\hat{\rho}^{\prime}=\sum_{j=1}^{J}\hat{E}_{j}\hat{\rho}\hat{E}_{j}^{\dagger},
\end{equation}
if it is non-selective.

As a simple application of POVM-based measurements, consider the following
problem. Suppose that someone picks one state out of the set $\{|\varphi
_{1}\rangle,|\varphi_{2}\rangle\}$ and asks us to find with a single
measurement which one did he/she picked. If the states are orthogonal, this is
trivial: we make a projective measurement defined by the projectors $\{\hat
{P}_{1}=|\varphi_{1}\rangle\langle\varphi_{1}|,\hat{P}_{2}=|\varphi_{2}%
\rangle\langle\varphi_{2}|\}$, and check which outcome appeared. The problem
is that it is simple to prove that when the states are not orthogonal, there
is no strategy based on projective measurements allowing us to determine which
state was given to us. However, we can design a strategy based on POVMs which
will allow us to perform the needed task, although it does not work all the time.

Consider the POVM $\{\hat{\Pi}_{1}=k_{1}(\hat{I}-|\varphi_{1}\rangle
\langle\varphi_{1}|),\hat{\Pi}_{2}=k_{2}(\hat{I}-|\varphi_{2}\rangle
\langle\varphi_{2}|),\hat{\Pi}_{3}=\hat{I}-\hat{\Pi}_{1}-\hat{\Pi}_{2}\}$.
Suppose that we get the outcome `1'; then, we know for sure that we got the
state $|\varphi_{2}\rangle$ as the probability of observing `1' when the state
is $|\varphi_{1}\rangle$ is zero, that is, $\langle\varphi_{1}|\hat{\Pi}%
_{1}|\varphi_{1}\rangle=0$. The opposite happens when we get the outcome `2',
we know for sure that $|\varphi_{1}\rangle$ was given to us, because
$\langle\varphi_{2}|\hat{\Pi}_{2}|\varphi_{2}\rangle=0$. Finally, when we get
the outcome `3' we don't know which state we had, but at least we never make a
misidentification of the state.

\section{Local operations and classical communication protocols}

Finally in this chapter on quantum operations, we consider a very important
class of operations performed onto a bipartite system as the one considered in
Chapter 2.

Suppose that Alice and Bob are in distant locations, so that one does not have
access to the part of the system belonging to the other. In this scenario, it
is natural to think that the most general class of operations that can be
performed on the joint system are local operations (arbitrary operations
acting only on $A$ or $B$ independently) in a correlated fashion (Alice and
Bob can communicate by phone to decide together what to do). Quantum
operations of this kind are known as \textit{local operations and classical
communication} (LOCC) \textit{protocols}, and play a central role in many
problems of quantum information (we already saw one, the characterization of
entangled states).

As an example, consider the following prototypical protocol: Bob performs a
generalized measurement described by the set $\{\mathcal{E}_{B}^{(j)}%
\}_{j=1,2,...,J}$ of trace decreasing quantum operations, and communicates the
outcome, say `$j$', to Alice, who applies a trace-preserving quantum operation
$\mathcal{E}_{A}^{(j)}$ from a pre-agreed set $\{\mathcal{E}_{A}%
^{(j)}\}_{j=1,2,...,J}$ in one-to-one correspondence with the possible
outcomes of Bob's measurements. Let us denote by $\{\hat{B}_{k}^{(j)}%
\}_{k=1,2,...,K_{j}}$ the Kraus operators associated to $\mathcal{E}_{B}%
^{(j)}$ (note that when $K_{j}=1$ $\forall j$, Bob's measurement is a
POVM-based measurement), and by $\{\hat{A}_{m}^{(j)}\}_{m=1,2,...,M_{j}}$ the
ones associated to $\mathcal{E}_{A}^{(j)}$. The resulting possible maps will
be given by%
\begin{equation}
\mathcal{E}_{\mathrm{LOCC}}^{(j)}=\mathcal{E}_{A}^{(j)}[\mathcal{E}_{B}%
^{(j)}[\hat{\rho}]]=\sum_{k=1}^{K_{j}}\sum_{m=1}^{M_{j}}(\hat{A}_{m}%
^{(j)}\otimes\hat{I})(\hat{I}\otimes\hat{B}_{k}^{(j)})\hat{\rho}(\hat{I}\otimes\hat{B}%
_{k}^{(j)\dagger})(\hat{A}_{m}^{(j)\dagger}\otimes\hat{I}).
\end{equation}
We will find this type of LOCC protocols many times along the notes, which we
will call \textit{one-way-direct} LOCC protocols.

Note that trace preserving LOCC protocols can only decrease the entanglement
of the state shared by Alice and Bob, as intuition says. What is a little more
surprising is that trace-decreasing LOCC protocols can enhance the
entanglement (we shall find one example of this when studying photon addition
and subtraction), what is a further example of how much counter-intuitive
quantum mechanics can be. Of course, a complete set of trace decreasing LOCC
maps can only decrease the entanglement on average, showing that any local
operation able to enhance the entanglement is intrinsically probabilistic.

\chapter{Majorization in quantum mechanics}
\label{Majorization}
In this section we introduce a relation between probability distributions
called \textit{majorization} \cite{Arnold87book}, and connect it to a couple
of important questions in quantum mechanics \cite{Nielsen00book}.

\section{The concept of majorization}

Majorization appeared as a way to order vectors in terms of their disorder, in
an effort to understand when one probability distribution can be built from
another by randomizing the latter.

Take two probability distributions $\mathbf{p}=\operatorname{col}(p_{1}%
,p_{2},...,p_{d})$ and $\mathbf{q}=\operatorname{col}(q_{1},q_{2},...,q_{d})$,
where $d$ can be infinite. We say that $\mathbf{p}$ majorizes $\mathbf{q}$,
and denote it by $\mathbf{p}\succ\mathbf{q}$, if and only if%
\begin{equation}
\sum_{n=1}^{m}p_{n}^{\downarrow}\geq\sum_{n=1}^{m}q_{n}^{\downarrow}\text{
\ \ \ \ }\forall m<d\text{,}%
\end{equation}
where $\mathbf{p}^{\downarrow}$ and $\mathbf{q}^{\downarrow}$ are the original
vectors with their components rearranged in decreasing order.

This characterization of the majorization relation is interesting from an
operational point of view, since it is easy to check numerically if two
vectors satisfy this condition. Nevertheless, it can be proven that
$\mathbf{p}\succ\mathbf{q}$ is strictly equivalent to two other operational relations:

\begin{itemize}
\item[(m1)] For every concave function $h(x)$, it is satisfied $\sum_{n=1}%
^{d}h(p_{n})\leq\sum_{n=1}^{d}h(q_{n})$.

\item[(m2)] $\mathbf{q}$ can be obtained from $\mathbf{p}$ as $\mathbf{q}%
=D\mathbf{p}$, where $D$ is a column-stochastic matrix\footnote{A square
matrix is column-stochastic if its elements are real and positive, its columns
sum to one, and its rows sum to less than one. Most of the literature on the
connection between majorization and quantum information studies
finite-dimensional systems, in which case it can be shown that
column-stochastic matrices are also doubly-stochastic (columns and rows both
sum to one). One needs the slightly more general definition of
column-stochastic to cope with infinite dimensional spaces, as we will do in
the next chapter.}.
\end{itemize}

These relations a very interesting from an interpretational point of view.
First, note that the entropy is a concave function, and hence, the first
relation says that the entropy of $\mathbf{q}$ is larger than the entropy of
$\mathbf{p}$; now, as we have seen more entropy means more information loss,
which can be somehow interpreted as more disorder. Second, note that any
column-stochastic matrix can be written as a convex sum of permutations;
hence, the second relation says that $\mathbf{q}$ can be obtained from
$\mathbf{p}$ by applying a random mixture of permutations on the latter.
Hence, both (m1) and (m2) seem to state that $\mathbf{q}$ is more disordered
than $\mathbf{p}$.

\section{Majorization and ensemble decompositions of a state}

As a first simple application of majorization to quantum mechanics, we give an
answer to the following question: under which conditions can we find an
ensemble decomposition based on a probability distribution $\mathbf{w}$, say
$\{w_{k},|\varphi_{k}\rangle\}_{k=1,2,...,N}$, of a density operator
$\hat{\rho}$ with diagonal representation $\hat{\rho}=\sum_{n=1}^{d}%
\lambda_{n}|r_{n}\rangle\langle r_{n}|$?

The answer is rather simple: only when $\boldsymbol{\lambda}\succ\mathbf{w}$.
Of course, if $N\neq d$, zeros are added in the vector with smaller
dimensionality to match the dimensions of $\boldsymbol{\lambda}$ and
$\mathbf{w}$.

This result is indeed a simple consequence of what we learned in Chapter
\ref{Isolated} concerning the freedom to represent a given density operator by
different ensemble decompositions. In particular, if $\hat{\rho}$ can be
represented by the ensembles $\{w_{k},|\varphi_{k}\rangle\}_{k}$ and
$\{\lambda_{n},|r_{n}\rangle\}_{n}$, there must exist a left-unitary matrix
$\mathcal{U}$ for which%
\begin{equation}
\sqrt{w_{l}}|\varphi_{l}\rangle=\sum_{n}\mathcal{U}_{kn}\sqrt{\lambda_{n}%
}|r_{n}\rangle;
\end{equation}
now, taking the inner product of this expression with itself, and using the
orthonormality of the $\{|r_{n}\rangle\}_{n}$ set, we get%
\begin{equation}
w_{k}=\sum_{n}\left\vert \mathcal{U}_{kn}\right\vert ^{2}\lambda_{n}.
\end{equation}
Finally, as $\mathcal{U}$ is left-unitary, the matrix with elements
$\left\vert \mathcal{U}_{kn}\right\vert ^{2}$ is column-stochastic.

\section{Majorization and the transformation of entangled states}

The next application of majorization theory to quantum mechanics appears when
answering another question of paramount importance in information theory:
given a bipartite state $|\psi\rangle_{AB}\in\mathcal{H}_{A}\otimes
\mathcal{H}_{B}$, under which conditions can it be transformed
\textit{deterministically} into another state $|\varphi\rangle_{AB}%
\in\mathcal{H}_{A}\otimes\mathcal{H}_{B}$ if Alice and Bob are allowed to use
only LOCC protocols?

By `deterministically' we mean that given a complete set of LOCC protocols
$\{\mathcal{E}_{\mathrm{LOCC}}^{(j)}\}_{j}$, the transformation $|\psi
\rangle_{AB}\rightarrow|\varphi\rangle_{AB}$ succeeds for all of them, that
is, $\mathcal{E}_{\mathrm{LOCC}}^{(j)}[|\psi\rangle_{AB}\langle\psi
|]\propto|\varphi\rangle_{AB}\langle\varphi|$ $\forall j$. If, for example,
the transformation works only for $\mathcal{E}_{\mathrm{LOCC}}^{(j=0)}$, but
not for the rest,\ then Alice and Bob will be able to transform $|\psi
\rangle_{AB}$ into $|\varphi\rangle_{AB}$ only some of the time, in particular
with probability $p_{0}=\mathrm{tr}\{\mathcal{E}_{\mathrm{LOCC}}^{(0)}%
[|\psi\rangle_{AB}\langle\psi|]\}$; hence the transformation fails with
probability $(1-p_{0})$, that is, Alice and Bob's strategy works only probabilistically.

To answer this question, consider the diagonal representations of the reduced
states $\hat{\rho}_{A}^{\psi}=\mathrm{tr}_{B}\{|\psi\rangle_{AB}\langle
\psi|\}=\sum_{n=1}^{d}\lambda_{n}^{\psi}|r_{n}\rangle_{\psi}\langle r_{n}|$
and $\hat{\rho}_{A}^{\varphi}=\mathrm{tr}_{B}\{|\varphi\rangle_{AB}%
\langle\varphi|\}=\sum_{n=1}^{d}\lambda_{n}^{\varphi}|r_{n}\rangle_{\varphi
}\langle r_{n}|$. Then, Alice and Bob can transform $|\psi\rangle_{AB}$ into
$|\varphi\rangle_{AB}$ via an LOCC strategy, if and only if
$\boldsymbol{\lambda}^{\psi}\prec\boldsymbol{\lambda}^{\varphi}$, in which
case we use the symbolic notation $|\psi\rangle_{AB}\prec|\varphi\rangle_{AB}$.

Note that since the entanglement entropy is a concave function of the
eigenvalues of the reduced density operator, this majorization relation
implies that $|\psi\rangle_{AB}$ can only be transformed deterministically via
an LOCC protocol into states of lower entanglement, that is, $E[|\psi
\rangle_{AB}]>E[|\varphi\rangle_{AB}]$.

It is also possible to prove that if $|\psi\rangle_{AB}$ can be transformed
into $|\varphi\rangle_{AB}$ deterministically via an LOCC protocol, it can
always be done with a one-way-direct LOCC protocol of the following form: Bob
performs a measurement described by some POVM $\{\hat{\Pi}_{j}\}_{j=1,2,...,J}%
$ and communicates the outcome, say `$j_{0}$', to Alice, who applies a unitary
quantum operation $\hat{A}_{j_{0}}$ chosen from a pre-agreed set of unitaries
$\{\hat{A}_{j}\}_{j=1,2,...,J}$ in one to one correspondence with the possible
outcomes of Bob's measurement. Hence, if $\{\hat{B}_{j}\}_{j=1,2,...,J}$ are
the measurement operators associated to Bob's POVM, the transformation is
accomplished as%
\begin{equation}
|\varphi\rangle_{AB}\propto(\hat{A}_{j}\otimes\hat{I})(\hat{I}\otimes\hat
{B}_{j})|\psi\rangle_{AB}\text{ \ \ \ \ }\forall j\text{.}%
\end{equation}

\chapter{Quantum information with continuous variables}
\label{ContinuousVariables}
\section{Infinite dimensional Hilbert spaces: the Harmonic Oscillator}

The \textit{harmonic oscillator} is the prototypical system which is described
quantum mechanically by an infinite dimensional Hilbert space. To see this,
let's find the eigenstates of its Hamiltonian, which is given by%
\begin{equation}
\hat{H}=\frac{\hat{p}^{2}}{2m}+\frac{m\omega^{2}}{2}\hat{q}^{2},
\end{equation}
being $m$ the mass of the oscillator and $\omega$ its oscillation frequency.
According to the postulates of quantum mechanics, the \textit{position}
$\hat{q}$ and \textit{momentum} $\hat{p}$ satisfy the commutation relation%
\begin{equation}
\lbrack\hat{q},\hat{p}]=\mathrm{i}\hbar.
\end{equation}
We will always work with dimensionless versions of them, the so-called X- and
P-\textit{quadratures} (although we may keep using the names `position' and
`momentum' most of the time)%
\begin{equation}
\hat{X}=\sqrt{\frac{2\omega m}{\hbar}}\hat{q},\text{ \ \ and \ \ }\hat
{P}=\sqrt{\frac{2}{\hbar\omega m}}\hat{p},
\end{equation}
which satisfy the commutator%
\begin{equation}
\lbrack\hat{X},\hat{P}]=2\mathrm{i}\text{,} \label{XPcommutator}%
\end{equation}
and in terms of which the Hamiltonian reads%
\begin{equation}
\hat{H}=\frac{\hbar\omega}{4}\left(  \hat{X}^{2}+\hat{P}^{2}\right)  \text{.}%
\end{equation}

In order to find the eigensystem of this operator, we decompose the
quadratures as%
\begin{equation}
\hat{X}=\hat{a}^{\dagger}+\hat{a},\text{ \ \ and \ \ }\hat{P}=\mathrm{i}%
(\hat{a}^{\dagger}-\hat{a}),
\end{equation}
where the operators $\hat{a}$ and $\hat{a}^{\dagger}$, known as the
\textit{annihilation} and \textit{creation} \textit{operators}, satisfy the
commutation relation%
\begin{equation}
\lbrack\hat{a},\hat{a}^{\dagger}]=1\text{;}%
\end{equation}
in terms of these operators, the Hamiltonian is rewritten as%
\begin{equation}
\hat{H}=\hbar\omega(\hat{a}^{\dagger}\hat{a}+1/2),
\end{equation}
and hence the problem has been reduced to finding the eigensystem of the
so-called \textit{number operator} $\hat{N}=\hat{a}^{\dagger}\hat{a}$.

Let us call $n$ to a generic real number contained in the spectrum of $\hat
{N}$, whose corresponding eigenvector we denote by $|n\rangle$, so that,
$\hat{N}|n\rangle=n|n\rangle$. We normalize the vectors to one by definition,
that is, $\langle n|n\rangle=1$ $\forall n$. The eigensystem of $\hat{N}$ is
readily found from the following two properties:

\begin{itemize}
\item $\hat{N}$ is a positive operator, as for any vector $|\psi\rangle$ it is
satisfied $\langle\psi|\hat{N}|\psi\rangle=\left(  \hat{a}|\psi\rangle,\hat
{a}|\psi\rangle\right)  \geq0$. When applied to its eigenvectors, this
property forbids the existence of negative eigenvalues, that is, $n\geq0$.

\item Using the commutation relation\footnote{This is straightforward to find
by using the property $[\hat{A}\hat{B},\hat{C}]=\hat{A}[\hat{B},\hat{C}%
]+[\hat{A},\hat{C}]\hat{B}$, valid for any three operators $\hat{A}$, $\hat
{B}$, and $\hat{C}$.} $[\hat{N},\hat{a}]=-\hat{a}$, it is trivial to show that
the vector $\hat{a}|n\rangle$ is also an eigenvector of $\hat{N}$ with
eigenvalue $n-1$. Similarly, from the commutation relation $[\hat{N},\hat
{a}^{\dagger}]=\hat{a}^{\dagger}$ it is found that the vector $\hat
{a}^{\dagger}|n\rangle$ is an eigenvector of $\hat{N}$ with eigenvalue $n+1$.
\end{itemize}

These two properties imply that the spectrum of $\hat{N}$ is the set of
natural numbers $n\in\left\{  0,1,2,...\right\}  \equiv%
\mathbb{N}
$, and that the eigenvector $|0\rangle$ corresponding to $n=0$ must satisfy
$\hat{a}|0\rangle=0$; otherwise it would be possible to find negative
eigenvalues, hence contradicting the positivity of $\hat{N}$.\ Thus, the set
of eigenvectors $\left\{  |n\rangle\right\}  _{n\in%
\mathbb{N}
}$ is an infinite, countable set. Moreover, using the property $\hat
{a}|0\rangle=0$ and the commutation relations, it is easy to prove that the
eigenvectors corresponding to different eigenvalues are orthogonal, that is,
$\langle n|m\rangle=\delta_{nm}$. Finally, according to the axioms of quantum
mechanics only the vectors normalized to one are physically relevant. Hence,
we conclude that the Hilbert space generated by the eigenvectors of $\hat{N}$
is isomorphic to $l^{2}\left(  \infty\right)  $---see Section
\ref{InfiniteHilbert}---, and hence it is an infinite-dimensional Hilbert space.

Summarizing, we have been able to prove that the Hilbert space associated to
the one-dimensional harmonic oscillator is infinite-dimensional. In the
process, we have explicitly built an orthonormal basis of this space by using
the eigenvectors $\left\{  |n\rangle\right\}  _{n\in%
\mathbb{N}
}$ of the number operator $\hat{N}$, with the annihilation and creation
operators $\{\hat{a},\hat{a}^{\dagger}\}$ allowing us to move through this set
as
\begin{equation}
\hat{a}|n\rangle=\sqrt{n}|n-1\rangle,\text{ \ \ \ \ and \ \ \ \ }\hat
{a}^{\dagger}|n\rangle=\sqrt{n+1}|n+1\rangle, \label{DownUp}%
\end{equation}
the factors in the square roots being easily found from normalization requirements.

In contrast to the number operator, which has a discrete spectrum, the
quadrature operators possess a pure continuous spectrum; let us focus on the
$\hat{X}$ operator, whose eigenvectors we denote by $\{|x\rangle\}_{x\in%
\mathbb{R}
}$ with corresponding eigenvalues $\{x\}_{x\in%
\mathbb{R}
}$, that is,%
\begin{equation}
\hat{X}|x\rangle=x|x\rangle.
\end{equation}
In order to prove that $\hat{X}$ has a pure continuous spectrum, just note
that, from the relation%
\begin{equation}
\exp\left(  \frac{\mathrm{i}}{2}y\hat{P}\right)  \hat{X}\exp\left(
-\frac{\mathrm{i}}{2}y\hat{P}\right)  =\hat{X}+y,
\end{equation}
which is easily found via the Baker-Campbell-Haussdorf lemma\footnote{This
lemma reads%
\begin{equation}
e^{\hat{B}}\hat{A}e^{-\hat{B}}=\sum_{n=0}^{\infty}\frac{1}{n!}%
\underset{n}{\underbrace{[\hat{B},[\hat{B},...[\hat{B},}}\hat{A}%
\underset{n}{\underbrace{]...]]}}, \label{BCHlemma}%
\end{equation}
and is valid for two general operators $\hat{A}$ and $\hat{B}$.}, it follows
that if $|x\rangle$ is an eigenvector of $\hat{X}$ with $x$ eigenvalue, then
the vector $\exp(-\mathrm{i}y\hat{P}/2)|x\rangle$ is also an eigenvector of
$\hat{X}$ with eigenvalue $x+y$. Now, as this holds for any real $y$, we
conclude that the spectrum of $\hat{X}$ is the whole real line. Moreover,
being a self-adjoint operator, one can use the eigensystem $\hat{X}$ as a
continuous basis of the Hilbert space of the oscillator by using the Dirac
normalization $\langle x|y\rangle=\delta(x-y)$. The same results can be
obtained for the $\hat{P}$ operator, whose eigenvectors we denote by
$\{|p\rangle\}_{p\in%
\mathbb{R}
}$ with corresponding eigenvalues $\{p\}_{p\in%
\mathbb{R}
}$, that is,%
\begin{equation}
\hat{P}|p\rangle=p|p\rangle.
\end{equation}

It is not difficult to prove that there exists a Fourier transform relation
between the position and momentum bases, that is,%
\begin{equation}
|p\rangle=\int_{-\infty}^{+\infty}\frac{dx}{\sqrt{4\pi}}\exp\left(
\frac{\mathrm{i}}{2}px\right)  |x\rangle\text{ \ }\Longleftrightarrow\text{
\ }|x\rangle=\int_{-\infty}^{+\infty}\frac{dp}{\sqrt{4\pi}}\exp\left(
-\frac{\mathrm{i}}{2}px\right)  |p\rangle.
\end{equation}
To this aim we now prove that%
\begin{equation}
\langle x|p\rangle=\frac{1}{\sqrt{4\pi}}\exp(\mathrm{i}xp/2).
\label{xy-scalar}%
\end{equation}
First note that the commutator $[\hat{X},\hat{P}]=2\mathrm{i}$ implies that%
\begin{equation}
\langle x|\hat{P}|x^{\prime}\rangle=\frac{2\mathrm{i}\delta\left(
x-x^{\prime}\right)  }{x-x^{\prime}},
\end{equation}
and hence%
\begin{equation}
\langle x|\hat{P}|\psi\rangle=\int_{%
\mathbb{R}
}dx^{\prime}\frac{2\mathrm{i}\delta\left(  x-x^{\prime}\right)  }{x-x^{\prime
}}\langle x^{\prime}|\psi\rangle=\int_{%
\mathbb{R}
}dx^{\prime}\frac{2\mathrm{i}\delta\left(  x-x^{\prime}\right)  }{x-x^{\prime
}}\left[  \langle x|\psi\rangle+(x^{\prime}-x)\frac{d\langle x|\psi\rangle
}{dx}+\sum_{n=2}^{\infty}\frac{(x^{\prime}-x)^{n}}{n!}\frac{d^{n}\langle
x|\psi\rangle}{dx^{n}}\right]  ;
\end{equation}
the order zero of the Taylor expansion is zero because the Kernel is
antisymmetric around $x$, while the terms of order two or above give zero as
well after integrating them. This means that%
\begin{equation}
\langle x|\hat{P}|\psi\rangle=-2\mathrm{i}\frac{d\langle x|\psi\rangle}{dx},
\label{MomentumDerivative}%
\end{equation}
which applied to $|\psi\rangle=|p\rangle$ yields the differential equation%
\begin{equation}
p\langle x|p\rangle=-2\mathrm{i}\frac{d\langle x|p\rangle}{dx},
\end{equation}
which has (\ref{xy-scalar}) as its solution, the factor $1/\sqrt{4\pi}$ coming
from the Dirac normalization of the $|p\rangle$ vectors.

As an example of the use of these continuous representations, we now find the
position representation of the number states, which we write as%
\begin{equation}
|n\rangle=\int_{%
\mathbb{R}
}dx\psi_{n}(x)|x\rangle.
\end{equation}
As a first step we find the projection of vacuum onto a position eigenstate,
the so-called \textit{ground state wave function} $\psi_{0}(x)=\langle
x|0\rangle$, from%
\begin{equation}
0=\langle x|\hat{a}|0\rangle=\frac{1}{2}\langle x|(\hat{X}+\mathrm{i}\hat
{P})|0\rangle=\frac{1}{2}\left(  x+2\frac{d}{dx}\right)  \psi_{0}(x),
\end{equation}
where we have used (\ref{MomentumDerivative}), which is a differential
equation for $\psi_{0}(x)$ having%
\begin{equation}
\psi_{0}(x)=\frac{1}{(2\pi)^{1/4}}\exp(-x^{2}/4),
\end{equation}
as its solution; the factor $(2\pi)^{-1/4}$ is found by imposing the
normalization%
\begin{equation}
\langle0|0\rangle=\int_{%
\mathbb{R}
}dx\psi_{0}^{2}(x)=1\text{.}%
\end{equation}
Now, the projection of any number state $|n\rangle$ onto a position eigenstate
(the $n^{\mathrm{th}}$ \textit{excited wave function}) is found from the
ground state wave function as%
\begin{equation}
\psi_{n}(x)=\langle x|n\rangle=\frac{1}{\sqrt{n!}}\langle x|\hat{a}^{\dagger
n}|0\rangle=\frac{1}{\sqrt{n!}2^{n}}\langle x|\left(  \hat{X}-\mathrm{i}%
\hat{P}\right)  ^{n}|0\rangle=\frac{1}{\sqrt{n!}2^{n}}\left(  x-2\frac{d}%
{dx}\right)  ^{n}\psi_{0}(x),
\end{equation}
which, reminding the Rodrigues formula for the Hermite polynomials%
\begin{equation}
H_{n}(x/\sqrt{2})=2^{-n/2}\exp(x^{2}/4)\left(  x-2\frac{d}{dx}\right)
^{n}\exp(-x^{2}/4),
\end{equation}
leads to the simple expression%
\begin{equation}
\psi_{n}(x)=\frac{1}{\sqrt{2^{n+1/2}\pi^{1/2}n!}}H_{n}(x/\sqrt{2})\exp
(-x^{2}/4)\text{.}%
\end{equation}

Note that not being vectors contained in the Hilbert space of the oscillator
(they cannot be properly normalized), the position and momentum eigenvectors
cannot correspond to physical states; nevertheless, we will see that they can
be understood as a (unphysical) limit of some physical states (the squeezed states).

Finally, let us stress that even though all that we are going to discuss in
what follows applies to a general bosonic system, that is, a system described
by a collection of harmonic oscillators, we will always have in mind the
electromagnetic field (light, in particular), which can be described as a set
of \textit{modes} with well defined polarization, frequency, and spatial
profile, each of which behaves as the mechanical harmonic oscillator that we
have introduced.

\section{The quantum harmonic oscillator in phase space I: The Wigner
function}

As the position and momentum do not have common eigenstates, and moreover,
their eigenstates cannot correspond to physical states of the oscillator, one
concludes that these observables cannot take definite values in quantum
mechanics; given the state $\hat{\rho}$, the best one can offer is the\textit{
probability density function} which will dictate the statistics of a
measurement of these observables, that is, $P(x)\equiv\langle x|\hat{\rho
}|x\rangle$ and $P(p)\equiv\langle p|\hat{\rho}|p\rangle$. In other words,
quantum mechanically there are not well defined trajectories in phase space,
the position and momentum of the oscillator are always affected by some
(\textit{quantum}) \textit{noise}.

It follows naturally the following question: is it then possible to describe
quantum mechanics as a probability distribution defined in phase space which
blurs the classical trajectories? As we are about to see, the answer is only
partially positive, as quantum noise is a lot much subtle than common
classical noise.

A logical way of building such a phase space distribution, say $W_{\hat{\rho}%
}(x,p)$, is as the one having the probability density functions $P(x)$ and
$P(p)$ as its marginals, that is,%
\begin{equation}
P(x)=\int_{%
\mathbb{R}
}dpW_{\hat{\rho}}(x,p),\text{ \ \ and \ \ }P(p)=\int_{%
\mathbb{R}
}dxW_{\hat{\rho}}(x,p)\text{.}%
\end{equation}
It is possible to show that this distribution is uniquely defined by
\cite{Schleich01book}%
\begin{equation}
W_{\hat{\rho}}(x,p)=\frac{1}{4\pi}\int_{%
\mathbb{R}
}dy\exp\left(  -\frac{\mathrm{i}}{2}py\right)  \langle x+y/2|\hat{\rho
}|x-y/2\rangle, \label{WignerMidPoint}%
\end{equation}
which is known as the \textit{Wigner function}. It is immediate to check that
this distribution has the proper marginals, and that it is normalized, i.e.,
$\int_{%
\mathbb{R}
^{2}}dxdpW_{\hat{\rho}}(x,p)=1$. The proof of its uniqueness is not that
simple though.

As we will see in a moment, given the state $\hat{\rho}$ of the oscillator,
the quantum expectation value of an observable $\hat{O}(\hat{X},\hat{P})$ can
be evaluated via the Wigner function as%
\begin{equation}
\langle\hat{O}(\hat{X},\hat{P})\rangle=\mathrm{tr}\{\hat{\rho}\hat{O}\}=\int_{%
\mathbb{R}
^{2}}dxdpW_{\hat{\rho}}(x,p)O^{(s)}(x,p), \label{SymExpectWigner}%
\end{equation}
where $O^{(s)}(x,p)$ is obtained by writing $\hat{O}(\hat{X},\hat{P})$ as a
symmetric function of $\hat{X}$ and $\hat{P}$ with the help of the commutation
relation (\ref{XPcommutator}), and changing these operators by the real
variables $x$ and $p$, respectively.

Taking into account that the prescription to find the quantum operator
associated to a classical observable $O(x,p)$ consists precisely in symmetrize
it with respect to $x$ and $p$, and then change the position and momentum by the
corresponding self-adjoint operators (what guarantees the self-adjointness of
the remaining operator), this result reinforces the interpretation of
$W_{\hat{\rho}}(x,p)$ as a probability distribution in phase space, and hence
of quantum mechanics as noise acting onto the classical trajectories. However,
we will find out later that this distribution can be negative, and hence it is
not a true 2D probability density function. Note that, actually, if
$W_{\hat{\rho}}(x,p)$ was positive all over phase space for any $\hat{\rho}$,
quantum mechanics could be simulated with classical noise, while quantum
mechanics has been proven itself to go well beyond classical mechanics in many experiments.

Given a state $\hat{\rho}$ of the oscillator, expression (\ref{WignerMidPoint}%
) allows us to compute the corresponding Wigner function. However, there is a
much more convenient way of writing the Wigner distribution, which is based in
the so-called \textit{displacement operator}%
\begin{equation}
\hat{D}(\mathbf{r})=\exp\left[  \frac{\mathrm{i}}{2}\mathbf{\hat{R}}^{T}%
\Omega\mathbf{r}\right]  , \label{DisOpReal}%
\end{equation}
where $\mathbf{r}=\operatorname{col}(x,p)$ is the coordinate vector in phase
space (which is usually denoted by the \textit{displacement} of the operator),
$\mathbf{\hat{R}}=\operatorname{col}(\hat{X},\hat{P})$ is the corresponding
vector operator, and $\Omega=%
\begin{bmatrix}
0 & 1\\
-1 & 0
\end{bmatrix}
$ is known as the \textit{symplectic form}. Note that with this matrix
notation the position-momentum commutators read $[\hat{R}_{j},\hat{R}%
_{l}]=2\mathrm{i}\Omega_{jl}$. We will learn more physical things about the
displacement operator in the following sections, but for now, just take it as
a useful mathematical object.

Note that using the formula
\begin{equation}
\exp(\hat{A}+\hat{B})=\exp(-[\hat{A},\hat{B}]/2)\exp(\hat{A})\exp(\hat{B}),
\label{DisentanglingLemma}%
\end{equation}
valid for operators $\hat{A}$ and $\hat{B}$ which commute with their
commutator, this operator can be written as a concatenation of two individual
momentum and position translations plus some phase:%
\begin{equation}
\hat{D}(\mathbf{r})=\exp\left[  -\frac{\mathrm{i}}{4}px\right]  \exp\left[
\frac{\mathrm{i}}{2}p\hat{X}\right]  \exp\left[  -\frac{\mathrm{i}}{2}x\hat
{P}\right]  ,
\end{equation}
which allows us to write%
\begin{align}
\mathrm{tr}\{\hat{D}(\mathbf{r)}\}  &  =e^{-\mathrm{i}px/4}\int_{%
\mathbb{R}
^{2}}dx^{\prime}dp^{\prime}\langle x^{\prime}|e^{\mathrm{i}p\hat{X}%
/2}e^{-\mathrm{i}x\hat{P}/2}|p^{\prime}\rangle\langle p^{\prime}|x^{\prime
}\rangle\\
&  =4\pi e^{-\mathrm{i}px/4}\underset{\delta(p)}{\underbrace{\int_{%
\mathbb{R}
}\frac{dx^{\prime}}{4\pi}e^{\mathrm{i}px^{\prime}/2}}}\underset{\delta
(x)}{\underbrace{\int_{%
\mathbb{R}
}\frac{dp^{\prime}}{4\pi}e^{\mathrm{i}xp^{\prime}/2}}}\text{,}%
\end{align}
arriving to the identity%
\begin{equation}
\mathrm{tr}\{\hat{D}(\mathbf{r)}\}=4\pi\delta^{(2)}(\mathbf{r})\text{,}%
\end{equation}
which will be useful in many situations. Another important property, trivially
proved from (\ref{DisentanglingLemma}), is%
\begin{equation}
\hat{D}(\mathbf{r})\hat{D}(\mathbf{r}^{\prime})=\exp\left(  -\frac{\mathrm{i}%
}{4}\mathbf{r}^{T}\Omega\mathbf{r}^{\prime}\right)  \hat{D}(\mathbf{r}%
+\mathbf{r}^{\prime}), \label{DispConcatenation}%
\end{equation}
and therefore, except for a phase, the concatenation of two displacement
operators is equivalent to a single displacement operator with the sum of the
displacements. Note that the phase is zero only when the condition
$xp^{\prime}=px^{\prime}$ is met, although it plays no physical role when
applied to a state of the system.

The first step in order to find the Wigner function of a given state
$\hat{\rho}$ is to define the \textit{characteristic function}%
\begin{equation}
\chi_{\hat{\rho}}(\mathbf{r)}=\mathrm{tr}\{\hat{D}(\mathbf{r})\hat{\rho
}\}\text{ \ \ \ \ }\iff\text{ \ \ \ \ }\hat{\rho}=\int_{%
\mathbb{R}
^{2}}\frac{d^{2}\mathbf{r}}{4\pi}\chi_{\hat{\rho}}(\mathbf{r)}\hat{D}%
^{\dagger}(\mathbf{r})\text{,} \label{CharacteristicDef}%
\end{equation}
and then the Wigner function is obtained as its Fourier transform%
\begin{equation}
W_{\hat{\rho}}(\mathbf{r})=\int_{%
\mathbb{R}
^{2}}\frac{d^{2}\mathbf{s}}{(4\pi)^{2}}\chi_{\hat{\rho}}(\mathbf{s}%
)e^{\frac{\mathrm{i}}{2}\mathbf{s}^{T}\Omega\mathbf{r}}\text{ \ \ \ \ }%
\iff\text{ \ \ \ \ }\chi_{\hat{\rho}}(\mathbf{s})=\int_{%
\mathbb{R}
^{2}}d^{2}\mathbf{r}W_{\hat{\rho}}(\mathbf{r})e^{-\frac{\mathrm{i}}%
{2}\mathbf{s}^{T}\Omega\mathbf{r}}\text{;} \label{WignerDef}%
\end{equation}
it is not difficult to show that this alternative definition of the Wigner
function leads to the original one given by (\ref{WignerMidPoint}). However,
this definition simplifies a lot many derivations. For example, from
(\ref{CharacteristicDef}) and (\ref{WignerDef}), it is immediate to prove that%
\begin{equation}
\mathrm{tr}\{\hat{\rho}\}=1\text{ \ \ \ \ }\Longrightarrow\text{ \ \ \ \ }%
\chi_{\hat{\rho}}(\mathbf{0)}=1\text{ \ \ \ \ }\Longrightarrow\text{
\ \ \ \ }\int_{%
\mathbb{R}
^{2}}d^{2}\mathbf{r}W_{\hat{\rho}}(\mathbf{r})=1,
\end{equation}
that is, the normalization of the Wigner function. Moreover, evaluating the
trace of (\ref{CharacteristicDef}) in the position eigenbasis, we get%
\begin{equation}
\chi_{\hat{\rho}}(0,p\mathbf{)}=\int dx\langle x|e^{\mathrm{i}p\hat{X}/2}%
\hat{\rho}|x\rangle=\int dxe^{\mathrm{i}px/2}\langle x|\hat{\rho}|x\rangle,
\end{equation}
which using the right hand side of (\ref{WignerDef}) directly implies that%
\begin{equation}
\langle x|\hat{\rho}|x\rangle=\int_{%
\mathbb{R}
}dpW_{\hat{\rho}}(\mathbf{r}),
\end{equation}
that is, the Wigner function has the position probability density function as
one of its marginals. The other marginal is the momentum probability density
function, as is proved from $\chi_{\hat{\rho}}(x,0\mathbf{)}$ in a similar fashion.

Going one step further, using this definitions it is actually quite simple to
prove (\ref{SymExpectWigner}). In particular, it is easy to check that the
expectation value of the symmetrically ordered product $(\hat{X}^{j}\hat
{P}^{l})^{(s)}$ can be obtain from the characteristic function as%
\begin{equation}
\langle(\hat{X}^{j}\hat{P}^{l})^{(s)}\rangle=(-)^{j}(2\mathrm{i})^{j+l}\left.
\frac{\partial^{j+l}}{\partial p^{j}\partial x^{l}}\chi_{\hat{\rho}%
}(\mathbf{r)}\right\vert _{\mathbf{r}=\mathbf{0}},
\end{equation}
leading to%
\begin{equation}
\langle(\hat{X}^{j}\hat{P}^{l})^{(s)}\rangle=\int_{%
\mathbb{R}
^{2}}dxdpW_{\hat{\rho}}(\mathbf{r})x^{j}p^{l},
\end{equation}
after using (\ref{CharacteristicDef}), in agreement with
(\ref{SymExpectWigner}).

In general we will not deal with a single harmonic oscillator (a single mode
of the light field), but with a collection of, say, $N$ harmonic oscillators
($N$ modes of light). Let us define again the coordinate vector in the
complete phase space by $\mathbf{r}=(x_{1},p_{1},x_{2},p_{2},...,x_{N},p_{N}%
)$, and the corresponding vector operator%
\begin{equation}
\mathbf{\hat{R}}=\operatorname{col}(\hat{X}_{1},\hat{P}_{1},\hat{X}_{2}%
,\hat{P}_{2},...,\hat{X}_{N},\hat{P}_{N}),
\end{equation}
in terms of which the commutation relations can be rewritten as%
\begin{equation}
\lbrack\hat{R}_{j},\hat{R}_{l}]=2\mathrm{i}(\Omega_{N})_{jl},
\end{equation}
where%
\begin{equation}
\Omega_{N}=%
{\displaystyle\bigoplus\limits_{m=1}^{N}}
\Omega=%
\begin{bmatrix}
\Omega &  &  & \\
& \Omega &  & \\
&  & \ddots & \\
&  &  & \Omega
\end{bmatrix}
\text{ \ \ \ \ }(=-\Omega^{T}=-\Omega^{-1})\text{,} \label{OmegaDef}%
\end{equation}
is the symplectic form of $N$ modes (in what follows we will suppress the
subindex $N$ except when needed). The state of the system acts now onto the
tensor product of the Hilbert spaces of the modes, and so does the
displacement operator, which is now defined as%
\begin{equation}
\hat{D}(\mathbf{r})=\exp\left[  \frac{\mathrm{i}}{2}\mathbf{\hat{R}}^{T}%
\Omega\mathbf{r}\right]  =\hat{D}(\mathbf{r}_{1})\otimes\hat{D}(\mathbf{r}%
_{2})\otimes...\otimes\hat{D}(\mathbf{r}_{N}),
\end{equation}
and satisfies\footnote{Note that the 2$N$-dimensional Dirac delta function can
be written as%
\begin{equation}
\delta^{(2N)}(\mathbf{s})=\int_{%
\mathbb{R}
^{2N}}\frac{d^{2N}\mathbf{r}}{(4\pi)^{2N}}e^{\frac{\mathrm{i}}{2}%
\mathbf{r}^{T}\Omega\mathbf{s}}.
\end{equation}
} $\mathrm{tr}\{\hat{D}(\mathbf{r)}\}=(4\pi)^{N}\delta^{(2N)}(\mathbf{r})$, as
well as (\ref{DispConcatenation}); the characteristic function is defined as
before%
\begin{equation}
\chi_{\hat{\rho}}(\mathbf{r)}=\mathrm{tr}\{\hat{D}(\mathbf{r})\hat{\rho
}\}\text{ \ \ \ \ }\iff\text{ \ \ \ \ }\hat{\rho}=\int_{%
\mathbb{R}
^{2N}}\frac{d^{2N}\mathbf{r}}{(4\pi)^{N}}\chi_{\hat{\rho}}(\mathbf{r)}\hat
{D}^{\dagger}(\mathbf{r})\text{,} \label{CharacteristicDefN}%
\end{equation}
and the Wigner function as its $2N$-dimensional Fourier transform%
\begin{equation}
W_{\hat{\rho}}(\mathbf{r})=\int_{%
\mathbb{R}
^{2N}}\frac{d^{2N}\mathbf{s}}{(4\pi)^{2N}}\chi_{\hat{\rho}}(\mathbf{s}%
)e^{\frac{\mathrm{i}}{2}\mathbf{s}^{T}\Omega\mathbf{r}}\text{ \ \ \ \ }%
\iff\text{ \ \ \ \ }\chi_{\hat{\rho}}(\mathbf{s})=\int_{%
\mathbb{R}
^{2N}}d^{2N}\mathbf{r}W_{\hat{\rho}}(\mathbf{r})e^{-\frac{\mathrm{i}}%
{2}\mathbf{s}^{T}\Omega\mathbf{r}}\text{.} \label{WignerDefN}%
\end{equation}

We saw in the previous chapters that there is an operation that plays an
important role when dealing with composite Hilbert spaces: the partial trace.
For example, in the case of the $N$ harmonic oscillators being in a state
$\hat{\rho}$, imagine that we want to trace out the last one, obtaining the
reduced state of the remaining $N-1$ oscillators, $\hat{\rho}_{R}%
=\mathrm{tr}_{N}\{\hat{\rho}\}$; let's see what this means in phase space.
From (\ref{CharacteristicDef}) we see that the characteristic function of the
reduced state $\hat{\rho}_{R}$ is just the original one with the phase space
coordinates of the traced oscillator set to zero, that is,%
\begin{equation}
\chi_{R}(\mathbf{r}_{\{N-1\}}\mathbf{)}=\mathrm{tr}\{\hat{D}(\mathbf{r}%
_{\{N-1\}})\hat{\rho}_{R}\}=\chi_{\hat{\rho}}(\mathbf{r}_{\{N-1\}}%
,\mathbf{0),}%
\end{equation}
where we use the notation $\mathbf{r}_{\{N-1\}}=\operatorname{col}%
(\mathbf{r}_{1},\mathbf{r}_{2},...,\mathbf{r}_{N-1})$, and therefore, the
Wigner function associated to the reduced state $\hat{\rho}_{R}$ can be found
by integrating out the phase-space variables of the corresponding oscillator,
that is,%
\begin{equation}
W_{R}(\mathbf{r}^{(n-1)})=\int_{%
\mathbb{R}
^{2(N-1)}}\frac{d^{2(N-1)}\mathbf{s}_{\{N-1\}}}{(4\pi)^{2(N-1)}}\chi
_{R}(\mathbf{s}_{\{N-1\}})e^{\frac{\mathrm{i}}{2}\mathbf{s}_{\{N-1\}}%
^{T}\Omega_{N-1}\mathbf{r}_{\{N-1\}}}=\int d^{2}\mathbf{r}_{N}W_{\hat{\rho}%
}(\mathbf{r}_{\{N-1\}},\mathbf{r}_{N}).
\end{equation}

Let us end this section with an interesting example of a Wigner function: that
of a number state $\hat{\rho}=|n\rangle\langle n|$. Although not
straightforwardly, it possible to show that the corresponding Wigner function
is given by%
\begin{equation}
W_{|n\rangle}(x,p)=\frac{(-1)^{n}}{2\pi}L_{n}(x^{2}+p^{2})\exp\left(
-\frac{x^{2}+p^{2}}{2}\right)  ,\label{NumberWigner}%
\end{equation}
where $L_{n}(z)$ is the Laguerre polynomial of order $n$, which can be found
from the Rodrigues formula%
\begin{equation}
L_{n}(z)=\frac{\exp(z)}{n!}\frac{d^{n}}{dz^{n}}\left[  z^{n}\exp(-z)\right]  .
\end{equation}
Note that, as commented above, this function has negative regions (except for $n=0$, corresponding to the vacuum state), and therefore, it cannot be simulated with any source of classical noise. For example, for odd $n$ it is always negative at the origin of phase space, since $L_n(0)=1 \;\forall n$.

\section{Gaussian continuous variables systems}

\subsection{Gaussian states}

\subsubsection{General definition}

A particularly important class of quantum mechanical states of the harmonic
oscillator are the so-called \textit{Gaussian states}, that is, states which
have a Gaussian Wigner function; as we will see, these are the type of light
states which are naturally generated in the laboratory, although we will also
show how to design experimental schemes whose purpose is the generation of
non-Gaussian states.

The Wigner function of an arbitrary Gaussian state has then the
form\footnote{When dealing with Gaussian states, the following integral is
quite useful:%
\begin{equation}
\int_{%
\mathbb{R}
^{2N}}d^{2N}\mathbf{r}\exp\left(  -\frac{1}{2}\mathbf{r}^{T}A\mathbf{r+x}%
^{T}\mathbf{r}\right)  =\frac{(2\pi)^{N}}{\sqrt{\det A}}\exp\left(  \frac
{1}{2}\mathbf{x}^{T}A^{-1}\mathbf{x}\right)  , \label{GaussianIntegral}%
\end{equation}
where $\mathbf{r\in}%
\mathbb{R}
^{2N}$ and $A$ is a nonsingular $2N\times2N$ matrix.}%
\begin{equation}
W(\mathbf{r})=\frac{1}{2\pi\sqrt{\det V}}\exp\left[  -\frac{1}{2}\left(
\mathbf{r}-\mathbf{\bar{r}}\right)  ^{T}V^{-1}\left(  \mathbf{r}%
-\mathbf{\bar{r}}\right)  \right]  , \label{SingleWigner}%
\end{equation}
where we have defined the \textit{mean vector}%
\begin{equation}
\mathbf{\bar{r}}=\langle\mathbf{\hat{R}}\rangle=\operatorname{col}\left(
\langle\hat{X}\rangle,\langle\hat{P}\rangle\right)  \mathbf{,}%
\end{equation}
and the \textit{covariance matrix}%
\begin{equation}
V=%
\begin{bmatrix}
\langle\delta\hat{X}^{2}\rangle & \frac{1}{2}\langle\{\delta\hat{X},\delta
\hat{P}\}\rangle\\
\frac{1}{2}\langle\{\delta\hat{X},\delta\hat{P}\}\rangle & \langle\delta
\hat{P}^{2}\rangle
\end{bmatrix}
,
\end{equation}
whose elements are given by%
\begin{equation}
V_{jl}=\frac{1}{2}\langle\{\delta\hat{R}_{j},\delta\hat{R}_{l}\}\rangle;
\end{equation}
in this expression we have used the notation $\delta\hat{A}=\hat{A}%
-\langle\hat{A}\rangle$, and denoted the anticommutator by curly brackets.
Note that number states do not belong to these class of states (save vacuum,
as we show below), but we will show that many other several interesting states
do. Note also that Gaussian states are completely defined by their first and
second moments, which is why we will sometimes denote by $\hat{\rho
}_{\mathrm{G}}(\mathbf{\bar{r}},V)$ a given Gaussian state. Finally, we would
like to remark that the mean photon number of the state is given by%
\begin{equation}
\langle\hat{N}\rangle=\frac{1}{4}\mathrm{tr}V+\frac{1}{4}\mathbf{\bar{r}}%
^{2}-\frac{1}{2}. \label{SingleMeanNumber}%
\end{equation}

Classically, any covariance matrix is allowed, as long as it is real,
symmetric, and positive definite. A quantum mechanical harmonic oscillator has
the added constrain $\det V\geq1$, what comes from the uncertainty principle
of position and momentum; indeed, the following proof is quite reminiscent of
the proof of the uncertainty principle:
\begin{align}
\det V  &  =\langle\delta\hat{X}^{2}\rangle\langle\delta\hat{P}^{2}%
\rangle-\frac{1}{4}\langle\{\delta\hat{X},\delta\hat{P}\}\rangle
^{2}\underset{\mathrm{Schwartz}}{\geq}|\langle\delta\hat{X}\delta\hat
{P}\rangle|^{2}-\frac{1}{4}\langle\{\delta\hat{X},\delta\hat{P}\}\rangle^{2}\\
&  =\frac{1}{4}|\underset{\mathrm{imaginary}}{\underbrace{\langle\lbrack
\delta\hat{X},\delta\hat{P}]\rangle}}+\underset{\mathrm{\operatorname{real}%
}}{\underbrace{\langle\{\delta\hat{X},\delta\hat{P}\}\rangle}}|^{2}-\frac
{1}{4}\langle\{\delta\hat{X},\delta\hat{P}\}\rangle^{2}=1.
\end{align}
A Gaussian state corresponding to any real, symmetric covariance matrix
satisfying this condition is physically achievable. Note that this condition
directly implies the positivity of the covariance matrix.

In the case of dealing with a Gaussian state of $N$ modes, their Wigner
function takes the following form in the complete phase space:%
\begin{equation}
W(\mathbf{r})=\frac{1}{(2\pi)^{N}\sqrt{\det V}}\exp\left[  -\frac{1}{2}\left(
\mathbf{r}-\mathbf{\bar{r}}\right)  ^{T}V^{-1}\left(  \mathbf{r}%
-\mathbf{\bar{r}}\right)  \right]  , \label{MultiWigner}%
\end{equation}
where now $\mathbf{\bar{r}}$\ is a vector with $2N$ components $\bar{r}%
_{j}=\langle\hat{R}_{j}\rangle$, and $V$ is a $2N\times2N$ matrix with
elements $V_{jl}=\langle\{\delta\hat{R}_{j}\delta\hat{R}_{l}\}\rangle/2$. Note
that the mean photon number is given in this multi-mode case by%
\begin{equation}
\sum_{j=1}^{N}\langle\hat{N}_{j}\rangle=\frac{1}{4}\mathrm{tr}V+\frac{1}%
{4}\mathbf{\bar{r}}^{2}-\frac{N}{2}. \label{TotalMeanNumber}%
\end{equation}

In this case, the necessary and sufficient condition for a real, symmetric
$2N\times2N$ matrix $V$ to correspond to a physical quantum state of $N$
oscillators is that \cite{Simon94}%
\begin{equation}
V+\mathrm{i}\Omega\geq0, \label{UncertaintyN}%
\end{equation}
where the inequality must be understood as \textquotedblleft positive
semidefinite\textquotedblright, what is again linked to the uncertainty
principle, and implies the positivity of $V$. It is possible to prove that any
Gaussian state for which the eigenvalues of $V+\mathrm{i}\Omega$ are zero
(equivalently, which saturates the uncertainty principle) is pure, and there
exists no other non-Gaussian state with the same covariance matrix.

In many situations it is interesting to understand the state of the $N$
oscillators as a bipartite state of $M$ oscillators plus another $M^{\prime}$
oscillators ($M+M^{\prime}=N$), in which case we talk about an $M\times
M^{\prime}$ continuous variable system. Consider a Gaussian state $\hat{\rho
}_{\mathrm{G}}(\mathbf{\bar{r}},V)$ of the $N$ oscillators, whose mean vector
and covariance matrix we write as%
\begin{equation}
\mathbf{\bar{r}}=\operatorname{col}(\mathbf{\bar{x}},\mathbf{\bar{x}}^{\prime
}),\text{ \ \ \ \ and \ \ \ \ }V=%
\begin{bmatrix}
W & C\\
C^{T} & W^{\prime}%
\end{bmatrix}
, \label{BipartiteGaussian}%
\end{equation}
where $\mathbf{\bar{x}}\in%
\mathbb{R}
^{2M}$, $\mathbf{\bar{x}}^{\prime}\in%
\mathbb{R}
^{2M^{\prime}}$, $W$ and $W^{\prime}$ real, symmetric matrices of dimensions
$2M\times2M$ and $2M^{\prime}\times2M^{\prime}$, respectively, and $C$ a real
matrix of dimensions $2M\times2M^{\prime}$; then, it is easy to prove that the
state of the first $M$ modes after tracing out the remaining $M^{\prime}$
modes, is the Gaussian state%
\begin{equation}
\hat{\rho}_{\mathrm{G}}(\mathbf{\bar{x}},W)=\mathrm{tr}%
_{M+1,M+2,...,M+M^{\prime}}\{\hat{\rho}_{\mathrm{G}}(\mathbf{\bar{r}},V)\},
\label{PartialGaussianTrace}%
\end{equation}
that is, tracing out a mode in a Gaussian state is equivalent to remove its
corresponding entries in the mean vector, as well as its rows and columns in
the covariance matrix. In order to prove this we make use of the
characteristic function, which for the general Gaussian Wigner function
(\ref{MultiWigner}) takes the form%
\begin{equation}
\chi(\mathbf{s})=\exp\left[  -\frac{1}{8}\mathbf{s}^{T}\Omega V\Omega
^{T}\mathbf{s}+\frac{\mathrm{i}}{4}\mathbf{\bar{r}}^{T}\Omega\mathbf{s}%
\right]  ;
\end{equation}
then, by substituting (\ref{BipartiteGaussian}) in this expression, and
remembering that tracing out any mode is equivalent to setting to zero the
corresponding phase space variables in the characteristic function, we can
write the characteristic function of the reduced state of the first $M$ modes
as%
\begin{equation}
\chi_{R}(\mathbf{s}_{\{M\}})=\exp\left[  -\frac{1}{8}\mathbf{s}_{\{M\}}%
^{T}\Omega_{M}W\Omega_{M}\mathbf{s}_{\{M\}}+\frac{\mathrm{i}}{4}%
\mathbf{\bar{x}}^{T}\Omega_{M}\mathbf{s}_{\{M\}}\right]  ,
\end{equation}
what proves (\ref{PartialGaussianTrace}).

Note finally that when the states of the partitions are uncorrelated, that is,%
\begin{equation}
\hat{\rho}_{\mathrm{G}}(\mathbf{\bar{r}},V)=\hat{\rho}_{\mathrm{G}%
}(\mathbf{\bar{x}},W)\otimes\hat{\rho}_{\mathrm{G}}(\mathbf{\bar{x}}^{\prime
},W^{\prime}),
\end{equation}
the covariance matrix of the Wigner function (\ref{MultiWigner}) can be
written as a direct sum%
\begin{equation}
V=W\oplus W^{\prime}.
\end{equation}

\subsubsection{Examples of Gaussian states}

\textbf{The vacuum state.} As commented above, number states are not Gaussian;
nevertheless, there is one exception: the \textit{vacuum state} $|0\rangle$.
To see this, just note that its Wigner function (\ref{NumberWigner}) can be
written as%
\begin{equation}
W_{|0\rangle}(x,p)=\frac{1}{2\pi}\exp\left(  -\frac{x^{2}+p^{2}}{2}\right)  ,
\end{equation}
that is, a Gaussian distribution like (\ref{SingleWigner}) with%
\begin{equation}
\mathbf{\bar{r}}=\mathbf{0},\text{ \ \ \ \ and \ \ \ \ }V=%
\begin{bmatrix}
1 & 0\\
0 & 1
\end{bmatrix}
=\mathcal{I}_{2\times2}.
\end{equation}
Hence, in Gaussian notation $|0\rangle\langle0|=\hat{\rho}_{\mathrm{G}%
}(\mathbf{0},\mathcal{I}_{2\times2})$.

\textbf{Thermal states.} As explained in Chapter \ref{Isolated}, the mixedness
of the state of a system can be related to the amount of information which has
been lost to another system inaccessible to us, that is, to the amount of
correlations shared with this second system. Given a system whose associated
Hilbert space has dimension $d$, and is spanned by some orthonormal basis
$\{|j\rangle\}_{j=1,2,...,d}$, we already saw that\ its maximally mixed state
is%
\begin{equation}
\hat{\rho}_{\mathrm{MM}}=\frac{1}{d}\sum_{j=1}^{d}|j\rangle\langle j|=\frac
{1}{d}\hat{I}_{d},
\end{equation}
where $\hat{I}_{d}$ is the identity of the $d$-dimensional Hilbert space.
Being proportional to the identity, this state is invariant under changes of
basis, and hence, the eigenvalues of any observable of the system are equally
likely; this is in concordance with what one expects intuitively from a state
which has leaked the maximum amount of information to another system.

For infinite dimensional Hilbert spaces ($d\rightarrow\infty$) this state is
not physical since it has infinite energy, that is, $\mathrm{tr}\{\hat{\rho
}\hat{N}\}\rightarrow\infty$. Hence, the question of which is the maximally
mixed state in infinite dimension makes sense only if one adds the energy
constraint $\mathrm{tr}\{\hat{\rho}\hat{N}\}=\bar{n}$, where $\bar{n}$ is a
positive real. It is possible to show that the state which maximizes the von
Neumann entropy subject to this constraint is%
\begin{equation}
\hat{\rho}_{\mathrm{th}}(\bar{n})=\sum_{n=0}^{\infty}\frac{\bar{n}^{n}%
}{\left(  1+\bar{n}\right)  ^{1+n}}|n\rangle\langle n|\text{,}%
\end{equation}
which is still diagonal in the number state basis, but has not a flat
distribution for the number of photons. Interestingly, it can be appreciated
that this distribution is the one expected for a collection of bosons at
thermal equilibrium with an environment, the \textit{Bose-Einstein
distribution}. This is why this state is known as the \textit{thermal state},
whose von Neumann entropy reads%
\begin{equation}
S[\hat{\rho}_{\mathrm{th}}(\bar{n})]=(\bar{n}+1)\log(\bar{n}+1)-\bar{n}%
\log\bar{n}\doteqdot S_{\mathrm{th}}(\bar{n}).
\end{equation}

It is not difficult to see that this state is Gaussian (later on we will
actually prove it by simple means), and that it is defined by a zero mean
vector, and a covariance matrix%
\begin{equation}
V_{\mathrm{th}}(\bar{n})=(2\bar{n}+1)\mathcal{I}_{2\times2}\text{,}%
\end{equation}
that is, $\hat{\rho}_{\mathrm{th}}(\bar{n})=\hat{\rho}_{\mathrm{G}}%
[\mathbf{0},V_{\mathrm{th}}(\bar{n})]$. Note that the vacuum state can be seen
as a thermal state with zero mean photon number.

In the next section we will learn that any $N$-mode Gaussian state can be
decomposed into $N$ uncorrelated thermal states (Williamson's theorem), and
hence thermal states can be seen as the most fundamental Gaussian states.

\subsection{Gaussian unitaries}

\subsubsection{General definition}

Consider a unitary transformation $\hat{U}=\exp(\hat{H}/\mathrm{i}\hbar)$
acting on the state of the light field. We say that such a unitary is
Gaussian, if it maps Gaussian states into Gaussian states.

Let us define the vector operators%
\begin{equation}
\mathbf{\hat{a}}=\operatorname{col}(\hat{a}_{1},\hat{a}_{2},...,\hat{a}%
_{N}),\text{ \ \ and \ \ }\mathbf{\hat{a}}^{+}=\operatorname{col}(\hat{a}%
_{1}^{\dagger},\hat{a}_{2}^{\dagger},...,\hat{a}_{N}^{\dagger})\text{;}%
\end{equation}
it is quite intuitive that any Gaussian unitary will come from a Hamiltonian
having only linear or bilinear terms, that is,
\begin{equation}
\hat{H}=\mathrm{i}\hbar\left(  \boldsymbol{\alpha}\cdot\mathbf{\hat{a}}%
^{+}+\mathbf{\hat{a}}^{+T}\mathcal{F}\mathbf{\hat{a}}+\mathbf{\hat{a}}%
^{+T}\mathcal{G}\mathbf{\hat{a}}^{+}\right)  +\mathrm{H.c.},
\end{equation}
for some vector $\boldsymbol{\alpha}\in%
\mathbb{C}
^{N}$, and some symmetric, complex $N\times N$ matrices $\mathcal{F}$ and
$\mathcal{G}$. Let's try to understand why this is so by analyzing the
physical meaning of each term. The first term corresponds to the injection of
photons in the modes; the second term comprises all the energy shifts $\hat
{a}_{j}^{\dagger}\hat{a}_{j}$, as well as the creation of a photon in one mode
via the annihilation of a photon of another mode, $\hat{a}_{j}^{\dagger}%
\hat{a}_{l}$; the last term takes into account the possibility of generating
two photons simultaneously, $\hat{a}_{j}^{\dagger}\hat{a}_{l}^{\dagger}$. In
other words, all the possible one-body and two-body interactions are taken
into account in this Hamiltonian. It is obvious that if we want to transform
Gaussian states into Gaussian states, the Hamiltonian cannot contain many-body
interactions beyond these, because otherwise one would create correlations
which go beyond the ones captured by the covariance matrix, and hence the
final state would not be completely characterized by its first and second moments.

In the Heisenberg picture, a Gaussian unitary would then induce a so-called
\textit{Bogoliubov transformation}%
\begin{equation}
\mathbf{\hat{a}}\longrightarrow\hat{U}^{\dagger}\mathbf{\hat{a}}\hat
{U}=\mathcal{A}\mathbf{\hat{a}}+\mathcal{B}\mathbf{\hat{a}}^{\dagger
}+\boldsymbol{\alpha},
\end{equation}
where the form of the complex $N\times N$ matrices $\mathcal{A}$ and
$\mathcal{B}$\ in terms of $(\boldsymbol{\alpha},\mathcal{F},\mathcal{G})$ is
unimportant for our purposes; the only restriction on these is that they have
to satisfy $\mathcal{AB}^{T}=\mathcal{BA}^{T}$ and $\mathcal{AA}^{\dagger
}=\mathcal{BB}^{\dagger}+\mathcal{I}_{N\times N}$, in order to preserve the
commutation relations of the creation and annihilation operators
($\mathcal{I}_{N\times N}$ is the identity matrix of dimension $N$).

Instead of writing this linear transformation in terms of the boson operators,
one can write it in terms of the position and momenta, or more compactly, in
terms of the $\mathbf{\hat{R}}$ vector operator:%
\begin{equation}
\mathbf{\hat{R}}\longrightarrow\hat{U}^{\dagger}\mathbf{\hat{R}}\hat
{U}=\mathcal{S}\mathbf{\hat{R}}+\mathbf{d}, \label{PosMomTrans}%
\end{equation}
where, once again, the dependence of $\mathbf{d}\in%
\mathbb{R}
^{2N}$ and the real $2N\times2N$ matrix $\mathcal{S}$ in the previous
transformation parameters is unimportant for our purposes; the only relevant
thing is that, in order to preserve the commutation relations of the
quadratures, $\mathcal{S}$ has to satisfy%
\begin{equation}
\mathcal{S}\Omega\mathcal{S}^{T}=\Omega, \label{SymplecticDef}%
\end{equation}
that is, $\mathcal{S}$ must be a \textit{symplectic matrix}. In the following
we will denote Gaussian unitaries by $\hat{U}_{\mathrm{G}}(\mathbf{d}%
,\mathcal{S})$ to stress the fact that they are completely characterized by
$\mathbf{d}$ and $\mathcal{S}$.

The transformation induced onto the system by the Gaussian unitary $\hat
{U}_{\mathrm{G}}(\mathbf{d},\mathcal{S})$ is easily described in the
Schr\"{o}dinger picture as well if the states are represented by the Wigner
function. To see this, let us first note that the displacement operator is
transformed by the action of this unitary as%
\begin{equation}
\hat{U}_{\mathrm{G}}^{\dagger}\hat{D}(\mathbf{r)}\hat{U}_{\mathrm{G}}%
=\exp\left[  \frac{\mathrm{i}}{2}(\mathcal{S}\mathbf{\hat{R}}+\mathbf{d)}%
^{T}\Omega\mathbf{r}\right]  \underset{\mathcal{S}^{T}\Omega=\Omega
\mathcal{S}^{-1}}{\underbrace{=}}\exp\left[  \frac{\mathrm{i}}{2}%
(\mathbf{\hat{R}}^{T}\Omega\mathcal{S}^{-1}+\mathbf{d}^{T}\Omega
)\mathbf{r}\right]  =\hat{D}(\mathcal{S}^{-1}\mathbf{r})\exp\left[
\frac{\mathrm{i}}{2}\mathbf{d}^{T}\Omega\mathbf{r}\right]  ,
\end{equation}
where the identity $\mathcal{S}^{T}\Omega=\Omega\mathcal{S}^{-1}$ follows from
(\ref{SymplecticDef}) and (\ref{OmegaDef}). Therefore, given the initial state
$\hat{\rho}$ of the system, the corresponding characteristic function is
transformed as%
\begin{equation}
\chi_{\hat{U}_{\mathrm{G}}\hat{\rho}\hat{U}_{\mathrm{G}}^{\dagger}%
}(\mathbf{r)}=\mathrm{tr}\{\hat{D}(\mathbf{r})\hat{U}_{\mathrm{G}}\hat{\rho
}\hat{U}_{\mathrm{G}}^{\dagger}\}=e^{\frac{\mathrm{i}}{2}\mathbf{d}^{T}%
\Omega\mathbf{r}}\mathrm{tr}\{\hat{D}(\mathcal{S}^{-1}\mathbf{r})\hat{\rho
}\}=e^{-\frac{\mathrm{i}}{2}\mathbf{r}^{T}\Omega\mathbf{d}}\chi_{\hat{\rho}%
}(\mathcal{S}^{-1}\mathbf{r)}, \label{CharacEvoUni}%
\end{equation}
and the Wigner function as%
\begin{equation}
W_{\hat{U}_{\mathrm{G}}\hat{\rho}\hat{U}_{\mathrm{G}}^{\dagger}}%
(\mathbf{r})=\int_{%
\mathbb{R}
^{2N}}\frac{d^{2N}\mathbf{s}}{(4\pi)^{2N}}\chi_{\hat{\rho}}(\mathcal{S}%
^{-1}\mathbf{s)}e^{\frac{\mathrm{i}}{2}\mathbf{s}^{T}\Omega(\mathbf{r}%
-\mathbf{d})}\underset{\mathbf{z}=\mathcal{S}^{-1}\mathbf{s}}{\underbrace{=}%
}\int_{%
\mathbb{R}
^{2N}}\frac{d^{2N}\mathbf{z}}{(4\pi)^{2N}}\chi_{\hat{\rho}}(\mathbf{z)}%
e^{\frac{\mathrm{i}}{2}\mathbf{z}^{T}\mathcal{S}^{T}\Omega(\mathbf{r}%
-\mathbf{d})}\underset{\mathcal{S}^{T}\Omega=\Omega\mathcal{S}^{-1}%
}{\underbrace{=}}W_{\hat{\rho}}[\mathcal{S}^{-1}(\mathbf{r}-\mathbf{d})].
\label{WignerEvoUni}%
\end{equation}

In the case of Gaussian states the situation is even more simple: one only
needs to find how the transformation affects the first and second moments of
the state. Using (\ref{PosMomTrans}), it is straightforward to show that the
transformation induced by the Gaussian unitary $\hat{U}_{\mathrm{G}%
}(\mathbf{d},\mathcal{S})$ on the mean vector $\mathbf{\bar{r}}$ and the
covariance matrix $V$ of any state is%
\begin{equation}
\mathbf{\bar{r}}\longrightarrow\mathcal{S}\mathbf{\bar{r}}+\mathbf{d},\text{
\ \ \ \ and \ \ \ \ }V\longrightarrow\mathcal{S}V\mathcal{S}^{T}\text{.}
\label{UnitaryTransSym}%
\end{equation}

Taking again the $M\times M^{\prime}$ partition of the multi-mode system, note
that when the unitary transformation acts independently on each partition,
that is,%
\begin{equation}
\hat{U}_{\mathrm{G}}(\mathbf{d},\mathcal{S})=\hat{U}_{\mathrm{G}}%
(\mathbf{l},\mathcal{L})\otimes\hat{U}_{\mathrm{G}}(\mathbf{l}^{\prime
},\mathcal{L}^{\prime}),
\end{equation}
where $\mathbf{l}\in%
\mathbb{R}
^{2M}$, $\mathbf{l}^{\prime}\in%
\mathbb{R}
^{2M^{\prime}}$, and $\mathcal{L}$ and $\mathcal{L}^{\prime}$ are symplectic
matrices of dimensions $2M\times2M$ and $2M^{\prime}\times2M^{\prime}$,
respectively, we can write%
\begin{equation}
\mathbf{d}=(\mathbf{l},\mathbf{l}^{\prime}),\text{ \ \ \ \ and \ \ \ \ }%
\mathcal{S}=\mathcal{L}\oplus\mathcal{L}^{\prime}.
\end{equation}

Finally, we would like to remark that a Gaussian unitary transformation is
called \textit{passive} when it conserves the number of mean photons
$\sum_{j=1}^{N}\langle\hat{a}_{j}^{\dagger}\hat{a}_{j}\rangle$, and
\textit{active} if it changes it. Now, given the transformation
(\ref{UnitaryTransSym}), and reminding that the mean number of total photons
is proportional to the square modulus of the mean vector, $\mathbf{\bar{r}%
}^{2}$, and the trace of the covariance matrix, $\mathrm{tr}V$, see
(\ref{TotalMeanNumber}), a Gaussian unitary will be passive if and only if%
\begin{equation}
\mathbf{d}=0\text{, \ \ \ \ and \ \ \ \ }\mathcal{S}^{T}\mathcal{S}%
=\mathcal{I}_{2N\times2N}\text{,}%
\end{equation}
the second condition meaning that its associated symplectic transformation
must be \textit{orthogonal}, that is, $\mathcal{S}^{T}=\mathcal{S}^{-1}$.

\subsubsection{Examples of Gaussian unitaries and more Gaussian states}

\textbf{The displacement operator and coherent states.} Consider the unitary
operator%
\begin{equation}
\hat{D}(\alpha)=\exp(\alpha\hat{a}^{\dagger}-\alpha^{\ast}\hat{a}),
\label{DisOp}%
\end{equation}
which, using the formula (\ref{DisentanglingLemma}) can be written in the
following two equivalent ways%
\begin{equation}
\hat{D}^{\left(  \mathrm{n}\right)  }\left(  \alpha\right)  =\exp\left(
-|\alpha|^{2}/2\right)  \exp\left(  \alpha\hat{a}^{\dagger}\right)
\exp\left(  -\alpha^{\ast}\hat{a}\right)  ,\text{ \ \ \ or \ \ }\hat
{D}^{\left(  \mathrm{a}\right)  }\left(  \alpha\right)  =\exp\left(
|\alpha|^{2}/2\right)  \exp\left(  -\alpha^{\ast}\hat{a}\right)  \exp\left(
\alpha\hat{a}^{\dagger}\right)  \text{,}%
\end{equation}
which we will refer to as its \textit{normal} and \textit{antinormal forms}, respectively.

Using the Baker-Campbell-Haussdorf lemma (\ref{BCHlemma}), it is fairly simple
to prove that this operator transforms the annihilation operator as%
\begin{equation}
\hat{a}\rightarrow\hat{D}^{\dagger}(\alpha)\hat{a}\hat{D}(\alpha)=\hat
{a}+\alpha,
\end{equation}
or, in terms of the quadratures%
\begin{equation}
\hat{X}\rightarrow\hat{D}^{\dagger}(\alpha)\hat{X}\hat{D}(\alpha)=\hat
{X}+x_{\alpha},\text{ \ \ \ \ and \ \ \ \ }\hat{P}\rightarrow\hat{D}^{\dagger
}(\alpha)\hat{P}\hat{D}(\alpha)=\hat{P}+p_{\alpha},
\end{equation}
where $\alpha=(x_{\alpha}+\mathrm{i}p_{\alpha})/2$ with $x_{\alpha}\in%
\mathbb{R}
$ and $p_{\alpha}\in%
\mathbb{R}
$. This unitary operator is then called the \textit{displacement operator}
because it allows us to perform translations in phase space; indeed, it is
exactly the operator that we defined in the previous section, see
(\ref{DisOpReal}), what is easily shown by rewriting (\ref{DisOp}) in terms of
the position and momentum operators.

As a Gaussian unitary we have $\hat{D}(\alpha)=\hat{U}_{\mathrm{G}}%
(\mathbf{d}_{\alpha},\mathcal{I}_{2\times2})$ with%
\begin{equation}
\mathbf{d}_{\alpha}=\operatorname{col}(x_{\alpha},p_{\alpha})\text{.}%
\end{equation}

The states obtained by displacing the vacuum state are known as
\textit{coherent states}. Using the normal form $\hat{D}^{(\mathrm{n})}%
(\alpha)$ of the displacement operator, it is easy to obtain%
\begin{equation}
|\alpha\rangle=\hat{D}(\alpha)|0\rangle=\sum_{n=0}^{\infty}\frac{\exp
(-|\alpha|^{2}/2)\alpha^{n/2}}{n!}|n\rangle.
\end{equation}
They are Gaussian states with the same covariance matrix as vacuum, but with a
non-zero mean vector, that is,%
\begin{equation}
\mathbf{\bar{r}}=\mathbf{d}_{\alpha},\text{ \ \ \ \ and \ \ \ \ }%
V=\mathcal{I}_{2\times2}\text{,}%
\end{equation}
or in Gaussian notation, $|\alpha\rangle\langle\alpha|=\hat{\rho}_{\mathrm{G}%
}(\mathbf{d}_{\alpha},\mathcal{I}_{2\times2})$. Hence, these states have the
same noise properties as vacuum, but describe a bright light beam, that is, a
light beam with non-zero mean field. In fact, they are a fair approximation to
the state describing the beam coming out from a (phase-locked) laser.

From a mathematical point of view, they are the eigenstates of the
annihilation operator, that is, $\hat{a}|\alpha\rangle=\alpha|\alpha\rangle$.
Later on we will learn that, even though the annihilation operator is not
self-adjoint, we can build a POVM-based measurement which has its eigenvalues
as the possible outcomes (\textit{heterodyne detection}).

\textbf{The squeezing operator and squeezed states}. Consider now the
\textit{squeezing operator}%
\begin{equation}
\hat{S}(r)=\exp(\frac{r}{2}\hat{a}^{2}-\frac{r}{2}\hat{a}^{\dagger2}),
\label{SqOp}%
\end{equation}
where $r\in\lbrack0,\infty\lbrack$. This operator is implemented
experimentally for an optical mode of frequency $\omega_{0}$ by pumping with a
strong laser beam of twice that frequency a crystal with second order
nonlinearity; pairs photons of frequency $\omega_{0}$ are generated via the
so-called \textit{spontaneous parametric down-conversion process}.

Using the Baker-Campbell-Haussdorf lemma (\ref{BCHlemma}), it is again simple
to prove that this operator transforms the annihilation operator as%
\begin{equation}
\hat{a}\rightarrow\hat{S}^{\dagger}(r)\hat{a}\hat{S}(r)=\hat{a}\cosh r-\hat
{a}^{\dagger}\sinh r,
\end{equation}
or, in terms of the quadratures%
\begin{equation}
\hat{X}\rightarrow\hat{S}^{\dagger}(r)\hat{X}\hat{S}(r)=e^{-r}\hat{X},\text{
\ \ \ \ and \ \ \ \ }\hat{P}\rightarrow\hat{S}^{\dagger}(r)\hat{P}\hat
{S}(r)=e^{r}\hat{P},
\end{equation}
so that it is characterized as a Gaussian unitary by $\hat{S}(r)=\hat
{U}_{\mathrm{G}}[\mathbf{0},\mathcal{Q}(r)]$ with%
\begin{equation}
\mathcal{Q}(r)=%
\begin{bmatrix}
e^{-r} & 0\\
0 & e^{r}%
\end{bmatrix}
.
\end{equation}

Applying the squeezing operator to a vacuum state, one obtains a so-called
\textit{squeezed vacuum state}. In the number state basis, this state is
characterized by containing only an even number states, what comes from the
fact that the squeezing operator generates photons in pairs; its explicit
representation in this basis is found to be \cite{Gerry05book}%
\begin{equation}
|r\rangle=\hat{S}(r)|0\rangle=\sum_{n=0}^{\infty}\frac{1}{2^{n}n!}\sqrt
{\frac{(2n)!}{\cosh r}}\tanh^{n}r|2n\rangle. \label{SqueezedState}%
\end{equation}
This Gaussian state has zero mean, and covariance matrix%
\begin{equation}
V_{\mathrm{sq}}(r)=\mathcal{Q}(r)\mathcal{Q}^{T}(r)=%
\begin{bmatrix}
e^{-2r} & 0\\
0 & e^{2r}%
\end{bmatrix}
,
\end{equation}
that is, $|r\rangle\langle r|=\hat{\rho}_{\mathrm{G}}[\mathbf{0}%
,V_{\mathrm{sq}}(r)]$. Note that in the limit $r\rightarrow\infty$ the
variance of the position goes to zero, while the variance of the momentum goes
to infinity, and hence in the limit of infinite squeezing the state
(\ref{SqueezedState}) is an eigenstate of the position operator. Note,
however, that this limit is unphysical, as the number of photons $\langle
\hat{N}\rangle=\sinh^{2}r$ diverges, and hence, an infinite amount of energy
is needed to generate a position eigenstate.

\textbf{The phase shift operator.} The free evolution of an oscillator
(corresponding to the free propagation of an optical mode through a linear
medium) induces the unitary transformation%
\begin{equation}
\hat{R}(\theta)=\exp(-\mathrm{i}\theta\hat{a}^{\dagger}\hat{a}),
\end{equation}
known as the \textit{phase shift operator}, which transforms the annihilation
operator as%
\begin{equation}
\hat{a}\rightarrow\hat{R}^{\dagger}(\theta)\hat{a}\hat{R}(\theta
)=\exp(\mathrm{i}\theta)\hat{a},
\end{equation}
or in terms of the position and momentum%
\begin{equation}
\hat{X}\rightarrow\hat{R}^{\dagger}(\theta)\hat{X}\hat{R}(\theta)=\hat{X}%
\cos\theta+\hat{P}\sin\theta,\text{ \ \ \ \ and \ \ \ \ }\hat{P}%
\rightarrow\hat{R}^{\dagger}(\theta)\hat{P}\hat{R}(\theta)=\hat{P}\cos
\theta-\hat{X}\sin\theta\text{.}%
\end{equation}
Hence, as a Gaussian unitary this transformation is characterized by $\hat
{R}(\theta)=\hat{U}_{\mathrm{G}}[\mathbf{0},\mathcal{R}(\theta)]$, where%
\begin{equation}
\mathcal{R}(\theta)=%
\begin{bmatrix}
\cos\theta & \sin\theta\\
-\sin\theta & \cos\theta
\end{bmatrix}
,
\end{equation}
what shows that a phase shift is equivalent to a \textit{proper rotation} in
phase space.

Note that number states are invariant under this transformation---they are
eigenstates of $\hat{R}(\theta)$---, and hence, thermal states are invariant
under rotations in phase space. This is not the case for coherent or squeezed states.

\textbf{The two-mode squeezing operator and two-mode squeezed states.} All the
unitaries considered so far act on a single mode of the electromagnetic field,
and hence, they cannot be used to induce entanglement between several modes.
In this example we consider the \textit{two-mode squeezing operator}%
\begin{equation}
\hat{S}_{12}(r)=\exp(r\hat{a}_{1}\hat{a}_{2}-r\hat{a}_{1}^{\dagger}\hat{a}%
_{2}^{\dagger}), \label{TMSop}%
\end{equation}
which can be implemented experimentally via a nonlinear crystal as the
squeezing operator (\ref{SqOp}), but now in a regime in which the
down-converted photons are distinguishable either in frequency, and/or
polarization, and/or spatial mode.

Under the action of this operator, the annihilation operators are transformed
as%
\begin{align}
\hat{a}_{1}  &  \rightarrow\hat{S}_{12}^{\dagger}(r)\hat{a}_{1}\hat{S}%
_{12}(r)=\hat{a}_{1}\cosh r-\hat{a}_{2}^{\dagger}\sinh r,\\
\hat{a}_{2}  &  \rightarrow\hat{S}_{12}^{\dagger}(r)\hat{a}_{2}\hat{S}%
_{12}(r)=\hat{a}_{2}\cosh r-\hat{a}_{1}^{\dagger}\sinh r,
\end{align}
or in terms of the quadratures%
\begin{align}
\hat{X}_{1}  &  \rightarrow\hat{S}_{12}^{\dagger}(r)\hat{X}_{1}\hat{S}%
_{12}(r)=\hat{X}_{1}\cosh r-\hat{X}_{2}\sinh r,\\
\text{ \ \ \ }\hat{P}_{1}  &  \rightarrow\hat{S}_{12}^{\dagger}(r)\hat{P}%
_{1}\hat{S}_{12}(r)=\hat{P}_{1}\cosh r+\hat{P}_{2}\sinh r,\\
\hat{X}_{2}  &  \rightarrow\hat{S}_{12}^{\dagger}(r)\hat{X}_{2}\hat{S}%
_{12}(r)=\hat{X}_{2}\cosh r-\hat{X}_{1}\sinh r,\\
\hat{P}_{2}  &  \rightarrow\hat{S}_{12}^{\dagger}(r)\hat{P}_{2}\hat{S}%
_{12}(r)=\hat{P}_{2}\cosh r+\hat{P}_{1}\sinh r.
\end{align}
Hence, as a Gaussian unitary this transformation is characterized by $\hat
{S}_{12}(r)=\hat{U}_{\mathrm{G}}[\mathbf{0},\mathcal{Q}_{12}(r)]$ with%
\begin{equation}
\mathcal{Q}_{12}(r)=%
\begin{bmatrix}
\mathcal{I}_{2\times2}\cosh r & -\mathcal{Z}\sinh r\\
-\mathcal{Z}\sinh r & \mathcal{I}_{2\times2}\cosh r
\end{bmatrix}
,
\end{equation}
where $\mathcal{Z}=\mathrm{diag}(1,-1)$.

Applying the two-mode squeezing operator to a vacuum state, one obtains a
so-called \textit{two-mode squeezed vacuum state}. In the number state basis,
this state is characterized by a perfectly correlated statistics of the number
of quanta in the modes, what again comes from the fact that the two-mode
squeezing operator generates photons in pairs; its explicit representation in
this basis is \cite{Gerry05book}%
\begin{equation}
|r\rangle_{12}=\hat{S}_{12}(r)|0,0\rangle=\frac{1}{\cosh r}\sum_{n=0}^{\infty
}\tanh^{n}r|n,n\rangle, \label{TMSVstate}%
\end{equation}
where we have used the notation $|n\rangle\otimes|m\rangle\equiv|n,m\rangle$.
This Gaussian state has zero mean, and covariance matrix%
\begin{equation}
V_{2\mathrm{sq}}(r)=\mathcal{Q}_{12}(r)\mathcal{Q}_{12}^{T}(r)=%
\begin{bmatrix}
\mathcal{I}_{2\times2}\cosh2r & -\mathcal{Z}\sinh2r\\
-\mathcal{Z}\sinh2r & \mathcal{I}_{2\times2}\cosh2r
\end{bmatrix}
, \label{TMSVcov}%
\end{equation}
that is, $|r\rangle_{12}\langle r|=\hat{\rho}[\mathbf{0},V_{2\mathrm{sq}}(r)]$.

Note that by taking the partial trace onto any of its two modes, the two-mode
squeezed vacuum state becomes a thermal state with mean photon number $\bar
{n}=\sinh^{2}r$, that is $\mathrm{tr}_{2}\{|r\rangle_{12}\langle
r|\}=\mathrm{tr}_{1}\{|r\rangle_{12}\langle r|\}=\hat{\rho}_{\mathrm{th}%
}(\sinh^{2}r)$, and hence the two-mode squeezed vacuum state can be seen as
the purification of a thermal state, what in addition shows that it is the
maximally entangled state in infinite dimension for a fixed energy. We will
come back to the entanglement properties of the two-mode squeezed vacuum state
in Section \ref{GaussianEntanglement}.

\textbf{The beam splitter operator.} We are going to analyze only one more
type of two-mode unitary transformations, the one induced by the so-called
\textit{beam splitter operator}%
\begin{subequations}
\begin{equation}
\hat{B}_{12}\left(  \beta\right)  =\exp\left(  \beta\hat{a}_{1}\hat{a}%
_{2}^{\dagger}-\beta\hat{a}_{1}^{\dagger}\hat{a}_{2}\right)  , \label{BSop}%
\end{equation}
which can be implemented experimentally by, for example, mixing two optical
beams in a beam splitter of \textit{transmissivity} $T=\cos^{2}\beta$.

Under the action of this operator, the annihilation operators are transformed
as%
\end{subequations}
\begin{align}
\hat{a}_{1}  &  \rightarrow\hat{B}^{\dagger}\left(  \beta\right)  \hat{a}%
_{1}\hat{B}\left(  \beta\right)  =\hat{a}_{1}\cos\beta+\hat{a}_{2}\sin\beta\\
\hat{a}_{2}  &  \rightarrow\hat{B}^{\dagger}\left(  \beta\right)  \hat{a}%
_{2}\hat{B}\left(  \beta\right)  =\hat{a}_{2}\cos\beta-\hat{a}_{1}\sin
\beta\text{.}%
\end{align}
or in terms of the quadratures%
\begin{align}
\hat{X}_{1}  &  \rightarrow\hat{B}_{12}^{\dagger}\left(  \beta\right)  \hat
{X}_{1}\hat{B}_{12}\left(  \beta\right)  =\hat{X}_{1}\cos\beta-\hat{X}_{2}%
\sin\beta,\\
\text{ \ \ \ }\hat{P}_{1}  &  \rightarrow\hat{B}_{12}^{\dagger}\left(
\beta\right)  \hat{P}_{1}\hat{B}_{12}\left(  \beta\right)  =\hat{P}_{1}%
\cos\beta-\hat{P}_{2}\sin\beta,\\
\hat{X}_{2}  &  \rightarrow\hat{B}_{12}^{\dagger}\left(  \beta\right)  \hat
{X}_{2}\hat{B}_{12}\left(  \beta\right)  =\hat{X}_{2}\cos\beta+\hat{X}_{1}%
\sin\beta,\\
\hat{P}_{2}  &  \rightarrow\hat{B}_{12}^{\dagger}\left(  \beta\right)  \hat
{P}_{2}\hat{B}_{12}\left(  \beta\right)  =\hat{P}_{2}\cos\beta+\hat{P}_{1}%
\sin\beta.
\end{align}
Hence, as a Gaussian unitary this transformation is characterized by $\hat
{B}_{12}(\beta)=\hat{U}_{\mathrm{G}}[\mathbf{0},\mathcal{B}_{12}(\beta)]$ with%
\begin{equation}
\mathcal{B}_{12}(\beta)=%
\begin{bmatrix}
\mathcal{I}_{2\times2}\cos\beta & -\mathcal{I}_{2\times2}\sin\beta\\
\mathcal{I}_{2\times2}\sin\beta & \mathcal{I}_{2\times2}\cosh\beta
\end{bmatrix}
.
\end{equation}

It is interesting to note when the states of both modes are coherent, they
keep being coherent after the action of the beam splitter transformation, as
$\mathcal{B}_{12}(\beta)\mathcal{B}_{12}^{T}(\beta)=\mathcal{I}_{4\times4}$.
As an example consider the state $|\alpha\rangle\otimes|0\rangle$ (one mode in
an arbitrary coherent state, and the other in vacuum), which has the Gaussian
representation $\hat{\rho}_{\mathrm{G}}(\mathbf{d},\mathcal{I}_{4\times4})$
with%
\begin{equation}
\mathbf{d}=\operatorname{col}(2\operatorname{Re}\{\alpha\},2\operatorname{Im}%
\{\alpha\},0,0);
\end{equation}
after the action of the beam splitter operator, it becomes $\hat{\rho
}_{\mathrm{G}}^{\prime}(\mathbf{d}^{\prime},\mathcal{I}_{4\times4})$ with%
\begin{equation}
\mathbf{d}^{\prime}=\operatorname{col}(2\operatorname{Re}\{\alpha\}\cos
\beta,2\operatorname{Im}\{\alpha\}\cos\beta,2\operatorname{Re}\{\alpha
\}\sin\beta,2\operatorname{Im}\{\alpha\}\sin\beta),
\end{equation}
which is the tensor product of two coherent states, in particular,
$|\alpha\cos\beta\rangle\otimes|\alpha\sin\beta\rangle$. This is exactly what
one expects when a laser field is sent through a beam splitter: part of the
laser is transmitted, and part is reflected.

\subsection{General Gaussian unitaries and states}

In this section we will use symplectic analysis (or better, `symplectic
tricks'), to find interesting facts about general Gaussian unitary
transformations and Gaussian states.

It is well known in symplectic analysis that any $2N\times2N$ symplectic
matrix $\mathcal{S}$ can be decomposed as%
\begin{equation}
\mathcal{S}=\mathcal{K}\left[
{\displaystyle\bigoplus\limits_{j=1}^{N}}
\mathcal{Q}(r_{j})\right]  \mathcal{L},
\end{equation}
where $\mathcal{K}$ and $\mathcal{L}$ are orthogonal, symplectic matrices
(this is known as the \textit{Euler decomposition} of a symplectic matrix, or
as its \textit{Bloch-Messiah reduction}). Physically, this means that a
general $N$-mode unitary transformation can be seen as the concatenation of
three operations: an $N$-port interferometer mixing all the modes\footnote{In
optics, an interferometer is just a collection of beam splitters which mix
optical beams entering through its input ports. They correspond to the most
general pasive Gaussian unitary, and are described by a concatenation of
single-mode phase shifts and two-mode beam splitters.}, $N$ single-mode
squeezers acting independently on each mode, and a second $N$-port interferometer.

As an important example involving two modes, note that the two-mode squeezing
transformation can be written as%
\begin{equation}
\mathcal{Q}_{12}(r)=\mathcal{B}_{12}(-\frac{\pi}{4})\left[  \mathcal{Q}%
(-r)\oplus\mathcal{Q}(r)\right]  \mathcal{B}_{12}(\frac{\pi}{4}),
\end{equation}
what in the Hilbert space means that two-mode squeezed vacuum state can be
obtained by mixing a position squeezed state with a momentum squeezed state in
a 50/50 beam splitter, that is,%
\begin{equation}
|r\rangle_{12}=\hat{B}_{12}(-\pi/4)[|-r\rangle\otimes|r\rangle]\text{;}
\label{TwoMode-OneMode}%
\end{equation}
note that the first beam splitter disappears because the two-mode vacuum state
$|0\rangle\otimes|0\rangle$ is invariant under passive transformations.
Squeezed beams being easily generated in the laboratory, the relation
(\ref{TwoMode-OneMode}) has allowed us to achieve two-mode squeezed beams with
high degree of entanglement.

As a second example, note that for one mode the only passive transformations
are the rotations in phase space, what means that an arbitrary single-mode
Gaussian unitary can be written as the concatenation of a phase shift, a
squeezing operation, a second phase shift, and a final displacement, that is,%
\begin{equation}
\hat{U}_{\mathrm{G}}(\alpha,\theta,r,\phi)=\hat{D}(\alpha)\hat{R}(\theta
)\hat{S}(r)\hat{R}(\phi).
\end{equation}
Now, it is quite intuitive (although it is not trivial to prove) that any
Gaussian state $\hat{\rho}_{\mathrm{G}}$ having von Neumann entropy $S_{0}$
can be obtained by applying a unitary transformation onto the thermal state
$\hat{\rho}_{\mathrm{th}}(\bar{n}_{0})$ with that same
entropy---$S_{\mathrm{th}}(\bar{n}_{0})=S_{0}$---, that is,%
\begin{equation}
\hat{\rho}_{\mathrm{G}}(\alpha,\theta,r,\bar{n}_{0})=\hat{D}(\alpha)\hat
{R}(\theta)\hat{S}(r)\hat{\rho}_{\mathrm{th}}(\bar{n}_{0})\hat{S}^{\dagger
}(r)\hat{R}^{\dagger}(\theta)\hat{D}^{\dagger}(\alpha),
\end{equation}
what means that the covariance matrix of any single-mode Gaussian state can
always be decomposed as%
\begin{equation}
V(\theta,r,\bar{n}_{0})=(2\bar{n}_{0}+1)\mathcal{R}(\theta)\mathcal{Q}%
(2r)\mathcal{R}^{T}(\theta)=(2\bar{n}_{0}+1)\left[
\begin{array}
[c]{cc}%
\cosh2r-\cos2\theta\sinh2r & \sin2\theta\sinh2r\\
\sin2\theta\sinh2r & \cosh2r+\cos2\theta\sinh2r
\end{array}
\right]  ;
\end{equation}
note that the first phase shift has disappeared because thermal states are
invariant under such transformations.

A second interesting \textit{theorem} is that of \textit{Williamson}'s, which
states that any positive $2N\times2N$ symmetric matrix $V$ can be brought to
its diagonal form $V^{\oplus}$ by a symplectic transformation $\mathcal{W}$,
that is,%
\begin{equation}
V=\mathcal{W}V^{\oplus}\mathcal{W}^{T}\text{, \ \ \ \ with \ \ \ \ }V^{\oplus
}=%
{\displaystyle\bigoplus\limits_{j=1}^{N}}
\nu_{j}\mathcal{I}_{2\times2}\text{.}%
\end{equation}
This theorem has a huge application in the world of Gaussian states. Note
that, physically, $V^{\oplus}$ can be seen as the covariance matrix of $N$
independent modes in a thermal state with mean photon numbers $\{\bar{n}%
_{j}=(\nu_{j}-1)/2\}_{j=1,2,...,N}$, while the symplectic transformation
$\mathcal{W}$ corresponds to a Gaussian unitary transformation. Williamson's
theorem is then completely equivalent to state that any $N$-mode Gaussian
state $\hat{\rho}_{\mathrm{G}}(\mathbf{\bar{r}},V)$ can be obtained as%
\begin{equation}
\hat{\rho}_{\mathrm{G}}(\mathbf{\bar{r}},V)=\hat{U}_{\mathrm{G}}%
(\mathbf{\bar{r}},\mathcal{W})\left\{
{\displaystyle\bigotimes\limits_{j=1}^{N}}
\hat{\rho}_{\mathrm{th}}[(\nu_{j}-1)/2]\right\}  \hat{U}_{\mathrm{G}}%
^{\dagger}(\mathbf{\bar{r}},\mathcal{W}).
\end{equation}
The set $\{\nu_{j}\}_{j=1,2,...,N}$ is called the \textit{symplectic spectrum}
of $V$, so that each $\nu_{j}$ is a \textit{symplectic eigenvalue}; it is
possible to show that the symplectic spectrum of $V$ can be computed as the
absolute values of the eigenvalues of the Hermitian matrix \textrm{i}$\Omega
V$.

This decomposition is very important, since it allows us to write many
properties of Gaussian states and covariance matrices in an easy form. For
example, the condition (\ref{UncertaintyN}) which ensures that $V$ is the
covariance matrix of a physical Gaussian state can be rewritten as%
\begin{equation}
V>0\text{, \ \ \ \ and \ \ \ \ }\nu_{j}\geq1\text{ }\forall j\text{.}
\label{SymplecticPhysical}%
\end{equation}
As a second important example, note that as unitary transformations do not
change the von Neumann entropy, the entropy of $\hat{\rho}_{\mathrm{G}%
}(\mathbf{\bar{r}},V)$ can be directly computed as the sum of the entropies of
the corresponding thermal states, that is,%
\begin{equation}
S[\hat{\rho}_{\mathrm{G}}(\mathbf{\bar{r}},V)]=\sum_{j=1}^{N}S_{\mathrm{th}%
}[(\nu_{j}-1)/2]=\sum_{j=1}^{N}g(\nu_{j}), \label{SymplecticEntropy}%
\end{equation}
where we have defined the function%
\begin{equation}
g(x)=\left(  \frac{x+1}{2}\right)  \log\left(  \frac{x+1}{2}\right)  -\left(
\frac{x-1}{2}\right)  \log\left(  \frac{x-1}{2}\right)  , \label{g}%
\end{equation}
which is positive and monotonically increasing for $x\geq1$.

It is particularly simple to evaluate the symplectic eigenvalues in the case
of dealing with one or two modes. In the case of one mode, the trick is to
realize that the determinant of the covariance matrix $V$ is invariant under
symplectic transformations, and hence the sole symplectic eigenvalue reads in
this case%
\begin{equation}
\nu=\sqrt{\det V}\text{.} \label{SingleSymplecticEigen}%
\end{equation}
For two modes, let us write the covariance matrix in the block form%
\begin{equation}
V=%
\begin{bmatrix}
A & C\\
C^{T} & B
\end{bmatrix}
,
\end{equation}
where $A=A^{T}$, $B=B^{T}$, and $C$ are $2\times2$ real matrices. In this
case, there is an extra symplectic invariant, namely $\Delta(V)\doteqdot\det
A+\det B+2\det C$, and hence the symplectic eigenvalues of a general two-mode
Gaussian state can be obtained from%
\begin{equation}
\det V=\nu_{+}^{2}\nu_{-}^{2}\text{, \ \ \ \ and \ \ \ \ }\Delta(V)=\nu
_{+}^{2}+\nu_{-}^{2}\text{,}%
\end{equation}
leading to%
\begin{equation}
\nu_{\pm}^{2}=\frac{\Delta(V)\pm\sqrt{\Delta^{2}(V)-4\det V}}{2}.
\end{equation}
In terms of the two-mode symplectic invariants, the second condition in
(\ref{SymplecticPhysical}) is rewritten as%
\begin{equation}
\det V\geq1\text{, \ \ \ \ and \ \ \ \ }\Delta\leq1+\det V.
\end{equation}
It is particularly relevant the case in which the covariance matrix of the
two-mode Gaussian state is in the so-called \textit{standard form}%
\begin{equation}
V=%
\begin{bmatrix}
a & 0 & c_{1} & 0\\
0 & a & 0 & c_{2}\\
c_{1} & 0 & b & 0\\
0 & c_{2} & 0 & b
\end{bmatrix}
; \label{StandardForm}%
\end{equation}
indeed, it is possible to show that the covariance matrix of any bipartite
Gaussian state can be brought to this standard form via a local Gaussian
unitary transformation $\hat{U}_{\mathrm{G}}=\hat{U}_{\mathrm{G}}%
(\mathbf{0},\mathcal{S}_{1})\otimes\hat{U}_{\mathrm{G}}(\mathbf{0}%
,\mathcal{S}_{2})$. In the particular case $c_{1}=-c_{2}=c>0$, the symplectic
eigenvalues read%
\begin{equation}
\nu_{\pm}=\frac{\sqrt{(a+b)^{2}-4c^{2}}\pm(b-a)}{2},
\end{equation}
and the symplectic matrix $\mathcal{W}$ satisfying $V=\mathcal{W}V^{\oplus
}\mathcal{W}^{T}$ can be explicitly found as%
\begin{equation}
\mathcal{W}=%
\begin{bmatrix}
\omega_{+}\mathcal{I}_{2\times2} & \omega_{-}\mathcal{Z}\\
\omega_{-}\mathcal{Z} & \omega_{+}\mathcal{I}_{2\times2}%
\end{bmatrix}
,
\end{equation}
with%
\begin{equation}
\omega_{\pm}^{2}=\frac{1}{2}\left[  \frac{a+b}{\sqrt{(a+b)^{2}-4c^{2}}}%
\pm1\right]  .
\end{equation}

\subsection{Gaussian bipartite states and Gaussian
entanglement\label{GaussianEntanglement}}

In Chapter \ref{Entanglement} we introduced the concept of entanglement as
correlations between two systems $A$ and $B$ which go beyond the ones allowed
classically. In this section we particularize those ideas to the case of
Gaussian continuous variable states. In the following we consider only two
modes, that is, a $1\times1$ continuous variable system, although we will talk
at the end a little about general $N\times M$ systems.

As explained in Chapter \ref{Entanglement}, the Peres-Horodecki criterion,
that is, the positivity of the partial transpose of the state, is a necessary
condition for a state to be separable. It turns out that it is also a
sufficient criterion for $1\times1$ Gaussian states. It is possible to show
that for continuous variables, transposition is equivalent to a change of sign
of the momenta; hence, the partial transposition operation corresponds to a
change of sign in the corresponding momenta. In the case of a $1\times1$
Gaussian state $\hat{\rho}_{\mathrm{G}}(\mathbf{\bar{r}},V)$, this means that
partial transposition of the second mode is equivalent to the transformation%
\begin{equation}
\mathbf{\tilde{r}}=(\mathcal{I}_{2\times2}\oplus\mathcal{Z})\mathbf{\bar{r}%
}\text{, \ \ \ \ }\tilde{V}=(\mathcal{I}_{2\times2}\oplus\mathcal{Z}%
)V(\mathcal{I}_{2\times2}\oplus\mathcal{Z}). \label{PartialTransMeanCov}%
\end{equation}
The Peres-Horodecki criterion is then reduced to check whether $\tilde{V}$ is
a physical covariance matrix. It is not difficult to prove that $\tilde{V}$ is
positive definite, and hence, the only condition left to analyze is $\tilde
{V}+\mathrm{i}\Omega\geq0$, or, equivalently, $\tilde{\nu}_{\pm}\geq1$ in
terms of the symplectic eigenvalues of $\tilde{V}$ (note that being symmetric
and real, $\tilde{V}$ satisfies Williamson's theorem as well).

Let us consider the example of the two-mode squeezed vacuum state
$|r\rangle_{12}$, for which%
\begin{equation}
\tilde{V}=%
\begin{bmatrix}
\mathcal{I}_{2\times2}\cosh2r & -\mathcal{I}_{2\times2}\sinh2r\\
-\mathcal{I}_{2\times2}\sinh2r & \mathcal{I}_{2\times2}\cosh2r
\end{bmatrix}
,
\end{equation}
which has%
\begin{equation}
\det\tilde{V}=1,\text{ \ \ \ \ and \ \ \ \ }\Delta(\tilde{V})=2(\cosh
^{2}2r+\sinh^{2}2r)=2(1+2\sinh^{2}2r),
\end{equation}
and therefore symplectic eigenvalues%
\begin{equation}
\nu_{\pm}=(1+2\sinh^{2}2r)\pm\sqrt{(1+2\sinh^{2}2r)^{2}-1}.
\end{equation}
For any $r>0$ we have $\nu_{-}<1$, what is a signature of $|r\rangle_{12}$
being an entangled state.

From an experimental point of view, the Peres-Horodecki criterion is quite
demanding, as it requires the full reconstruction of the covariance matrix of
the observed beams. However, there is a simpler separability criterion which
requires only the analysis of the variance of a suitable pair of joint
quadratures (what can be check experimentally via two homodyne measurements,
as we will see later). This criterion, which was introduced simultaneously by
Duan, Giedke, Cirac, and Zoller \cite{Duan00} and by Simon \cite{Simon00},
states that a Gaussian state is separable if for every $\varphi$%
\begin{equation}
W^{\varphi}\doteqdot V[(\hat{X}_{1}^{\varphi}-\hat{X}_{2}^{\varphi})/\sqrt
{2}]+V[(\hat{P}_{1}^{\varphi}+\hat{P}_{2}^{\varphi})/\sqrt{2}]\geq2,
\end{equation}
where%
\begin{align}
\hat{X}^{\varphi}  &  \doteqdot e^{\mathrm{i}\varphi}\hat{a}^{\dagger
}+e^{-\mathrm{i}\varphi}\hat{a}=\hat{X}\cos\varphi+\hat{P}\sin\varphi,\\
\hat{P}^{\varphi}  &  \doteqdot\mathrm{i}(e^{\mathrm{i}\varphi}\hat
{a}^{\dagger}-e^{-\mathrm{i}\varphi}\hat{a})=\hat{P}\cos\varphi-\hat{X}%
\sin\varphi,
\end{align}
and $V(\hat{A})=\langle\delta\hat{A}^{2}\rangle$. It is possible to show that
this is also a sufficient condition for separability in the case of $1\times
N$ Gaussian states. Note that for covariance matrices written in the standard
form (\ref{StandardForm}), the so-called \textit{witness} $W^{\varphi}$
reduces to%
\begin{equation}
W^{\varphi}=a+b+(c_{2}-c_{1})\cos2\varphi.
\end{equation}

Let's come back to the example of the two-mode squeezed vacuum state
$|r\rangle_{12}$; its covariance matrix (\ref{TMSVcov}) being already in
standard form, we get%
\begin{equation}
W^{\varphi}=2(\cosh2r+\sinh2r\cos2\varphi);
\end{equation}
for $\varphi=\pi/2$, the witness reads%
\begin{equation}
W^{\pi/2}=2\exp(-2r),
\end{equation}
which is clearly below 2 for every $r>0$, hence showing once again that
$|r\rangle_{12}$ is indeed an entangled state. In particular, $|r\rangle_{12}$
presents quantum anti-correlations between the position of the oscillators,
and correlations between their momenta, that is,%
\begin{equation}
V[(\hat{P}_{1}-\hat{P}_{2})/\sqrt{2}]=V[(\hat{X}_{1}+\hat{X}_{2})/\sqrt
{2}]=\exp(-2r);
\end{equation}
in the $r\rightarrow\infty$ limit, the state $|r\rangle_{12}$ is therefore an
eigenstate of the $\hat{P}_{1}-\hat{P}_{2}$ and $\hat{X}_{1}+\hat{X}_{2}$
operators given by%
\begin{equation}
|\mathrm{EPR}\rangle=\int_{%
\mathbb{R}
}dp|p,p\rangle=\int_{%
\mathbb{R}
}dx|x,-x\rangle,
\end{equation}
which is exactly the type of state that Einstein, Podolsky, and Rosen
considered in his attempt at proving that quantum mechanics was inconsistent
\cite{Einstein35}.

Let us finally stress that a necessary and sufficient criterion for
separability has been found for the Gaussian states of a general $N\times M$
bipartite continuous variable system. This criterion states that the Gaussian
state $\hat{\rho}_{\mathrm{G}}(\mathbf{\bar{r}},V)$ is separable if and only
if there exists a pair of matrices $V_{A}$ and $V_{B}$ with dimensions
$2N\times2N$ and $2M\times2M$, respectively, for which%
\begin{equation}
V\geq V_{A}\oplus V_{B}\text{.}%
\end{equation}
Of course, this criterion is quite difficult to handle in practice, but
fortunately an equivalent, operationally friendly criterion was introduced by
Giedke, Kraus, Lewenstein, and Cirac, based on the concept of
\textit{nonlinear maps}; we will however not explain this criterion which can
be consulted in \cite{ParisNotes}, or in the original reference
\cite{Giedke01}.

Let's move now to the quantification of the entanglement present in a Gaussian
state. As we commented in Chapter \ref{Entanglement}, this problem has been
only solved for pure states, for which the entanglement entropy is the unique
measure of quantum correlations. For mixed states, however, we have not found
a completely satisfactory measure (the distillable entanglement and the
entanglement of formation are reasonable measures, but cannot be computed for
most states, while the logarithmic negativity is easy to compute but does not
satisfy all the conditions needed for a proper entanglement measure), not even
for the reduced class of continuous variable Gaussian states.

In the following we explain how to compute the entanglement entropy and the
logarithmic negativity for $1\times1$ Gaussian states $\hat{\rho}_{\mathrm{G}%
}(\mathbf{\bar{r}},V)$, whose covariance matrix we write in the same block
form as before%
\begin{equation}
V=%
\begin{bmatrix}
A & C\\
C^{T} & B
\end{bmatrix}
, \label{GeneralTwoModeCov}%
\end{equation}
where $A=A^{T}$, $B=B^{T}$, and $C$ are $2\times2$ real matrices (the
generalization to $N\times M$ Gaussian states is straightforward).

As commented in (\ref{PartialGaussianTrace}), tracing out one mode of a
Gaussian state is equivalent to retaining in the covariance matrix the minor
corresponding to the remaining modes. Hence, to evaluate the entanglement
entropy of the Gaussian state\ $\hat{\rho}_{\mathrm{G}}(\mathbf{\bar{r}},V)$
having covariance matrix (\ref{GeneralTwoModeCov}), one just needs to evaluate
the entropy of the single-mode covariance matrix $A$. This matrix has
$\nu=\sqrt{\det A}$ as its sole symplectic eigenvalue, and therefore, based on
(\ref{SymplecticEntropy}), its entropy---and hence the entanglement entropy of
the corresponding two-mode Gaussian state---is given by%
\begin{equation}
E[\hat{\rho}_{\mathrm{G}}(\mathbf{\bar{r}},V)]=g(\sqrt{\det A}),
\end{equation}
where the function $g(x)$ was defined in (\ref{g}).

For mixed states the entanglement entropy is not even an entanglement
monotone, and hence, it cannot be considered a proper entanglement measure for
such states. One has then to consider other measures, and here we focus on the
logarithmic negativity $E_{N}[\hat{\rho}]$. It is possible to show that for an
arbitrary Gaussian state $\hat{\rho}_{\mathrm{G}}(\mathbf{\bar{r}},V)$, this
entanglement measure can be computed as%
\begin{equation}
E_{N}[\hat{\rho}_{\mathrm{G}}(\mathbf{\bar{r}},V)]=\sum_{j}F(\tilde{\nu}_{j}),
\end{equation}
where%
\begin{equation}
F(x)=\left\{
\begin{array}
[c]{cc}%
-\log x & x<1\\
0 & x\geq1
\end{array}
\right.  ,
\end{equation}
and $\{\tilde{\nu}_{j}\}_{j}$ is the symplectic spectrum of the covariance
matrix $\tilde{V}$ corresponding to\ the partial transposition of $\hat{\rho
}_{\mathrm{G}}(\mathbf{\bar{r}},V)$, which is defined in
(\ref{PartialTransMeanCov}) for a $1\times1$ system.

Let us start by quantifying the entanglement of the two-mode squeezed vacuum
state $|r\rangle_{12}$; being a pure bipartite state, its entanglement is
measured by its entanglement entropy%
\begin{equation}
E[|r\rangle_{12}]=g(\cosh2r),
\end{equation}
which is nothing but the entropy of the reduced thermal state. It is not
difficult to check that, starting at zero for $r=0$, this is a monotonically
increasing function of $r$, as expected.

In the case of the two-mode squeezed vacuum state we can even prove a stronger
result, namely that $|r^{\prime}\rangle_{12}\succ|r\rangle_{12}$ for
$r^{\prime}<r$. Let us write the state as%
\begin{equation}
|\lambda\rangle_{12}=\sum_{n=0}^{\infty}\sqrt{p_{n}(\lambda)}|n,n\rangle,
\end{equation}
where%
\begin{equation}
p_{n}(\lambda)=(1-\lambda^{2})\lambda^{2n},\text{ \ \ \ \ being \ \ \ \ }%
\lambda=\tanh r\text{.}%
\end{equation}
It is fairly simple to check that the triangular matrix%
\begin{equation}
D(\lambda,\lambda^{\prime})=%
\begin{bmatrix}
d_{0} & 0 & 0 & 0 & \cdots\\
d_{1} & d_{0} & 0 & 0 & \cdots\\
d_{2} & d_{1} & d_{0} & 0 & \cdots\\
d_{3} & d_{2} & d_{1} & d_{0} & \cdots\\
\vdots & \vdots & \vdots & \vdots & \ddots
\end{bmatrix}
,\text{ \ \ \ \ with \ \ \ \ }d_{n}=\frac{1-\lambda^{2}}{1-\lambda^{\prime2}%
}[\lambda^{2}-H(n-1)\lambda^{\prime2}]\lambda^{2(n-1)},
\end{equation}
$H(x)$ being the Heaviside step function defined as $H(x)=1$ for $x\geq0$ and
$H(x)=0$ for $x<0$, is column stochastic, and that it connects the Schmidt
distributions $\mathbf{p}(\lambda)$ and $\mathbf{p}(\lambda^{\prime})$ as%
\begin{equation}
\mathbf{p}(\lambda)=D(\lambda,\lambda^{\prime})\mathbf{p}(\lambda^{\prime
})\text{.}%
\end{equation}
Hence, as $\mathbf{p}(\lambda^{\prime})$ can be transformed into
$\mathbf{p}(\lambda)$ via a column stochastic matrix, we conclude that
$|\lambda^{\prime}\rangle_{12}$ majorizes $|\lambda\rangle_{12}$ for
$\lambda^{\prime}<\lambda$. This implies the previous result that we found
concerning the entanglement entropy, namely that $E[|r\rangle_{12}%
]>E[|r\rangle_{12}]$ for $r>r^{\prime}$, and much more, for example, that
$|r^{\prime}\rangle_{12}$ can be transformed into a two-mode squeezed vacuum
state of lower entanglement via an LOCC protocol.

\subsection{Gaussian channels}

\subsubsection{General definition}

In this section we introduce one of the most important objects in the field of
quantum information with continuous variables: \textit{Gaussian channels}. We
call \textit{channel} to any trace preserving quantum operation acting on a
continuous variables system. The channel is Gaussian when it maps Gaussian
states into Gaussian states. As we will see, they receive their name because
they actually model the most important communication channels used in current
technologies, such as fibers or wires.

As we saw, a way of characterizing an arbitrary trace preserving quantum
operation $\mathcal{E}$ is by giving a complete set of Kraus operators
$\{\hat{E}_{k}\}_{k=1,2,...,K}$ which transform a state $\hat{\rho}$ into the
state%
\begin{equation}
\hat{\rho}^{\prime}=\mathcal{E}[\hat{\rho}]=\sum_{k=1}^{K}\hat{E}_{k}\hat
{\rho}\hat{E}_{k}^{\dagger}.
\end{equation}
Gaussian channels, on the other hand, can be characterized by their action on
the first moments of the state. In particular, similarly to Gaussian
unitaries, and as will be clear from the following discussion, Gaussian
channels acting on $N$ modes of the electromagnetic field are characterized by
a vector $\mathbf{d\in%
\mathbb{R}
}^{2N}$, plus two real $2N\times2N$ matrices $\mathcal{K}$ and $\mathcal{N}$,
which transform the mean vector $\mathbf{\bar{r}}$ and covariance matrix $V$
of the state as%
\begin{equation}
\mathbf{\bar{r}\rightarrow}\mathcal{K}\mathbf{\bar{r}}+\mathbf{d}\text{,
\ \ \ \ and \ \ \ \ }V\rightarrow\mathcal{K}V\mathcal{K}^{T}+\mathcal{N}.
\label{ChannelTrans}%
\end{equation}
The matrices $\mathcal{K}$ and $\mathcal{N}$ must satisfy certain conditions
in order to correspond to a true trace preserving quantum operation. First, as
the covariance matrix is symmetric, so has to be the matrix $\mathcal{N}.$
Secondly, in order to map positive operators (like density matrices) into
positive operators, they have to satisfy the following restriction%
\begin{equation}
\mathcal{N}+\mathrm{i}\Omega-\mathrm{i}\mathcal{K}\Omega\mathcal{K}^{T}%
\geq0\text{.}%
\end{equation}
Note that Gaussian unitaries correspond to a Gaussian channel for which
$\mathcal{N}$ is zero, and $\mathcal{K}$ is symplectic. Note also that for
single-mode channels ($N=1$), this last condition can be rewritten as%
\begin{equation}
\mathcal{N}\geq0\text{, \ \ \ \ }\det\mathcal{N}\geq(\det\mathcal{K}-1)^{2}.
\label{PosCondSingChann}%
\end{equation}
Roughly speaking, $\mathcal{K}$ plays the role of the amplification and
attenuation of the channel (plus a possible rotation), while $\mathcal{N}$
includes any source of quantum or classical noise; we will come back to their
physical meaning in the next section.

Indeed, it is quite simple to understand why Gaussian channels correspond to a
transformation of the type (\ref{ChannelTrans}). To this aim, we just need to
remember that any trace preserving quantum operation can be seen as a unitary
transformation acting on the system, plus some environment in a pure state
which is dismissed after the interaction. It is obvious that in order for the
channel to be Gaussian, both the state of the environment $|\psi_{E}\rangle$
and the joint unitary transformation $\hat{U}_{\mathrm{G}}(\mathbf{s}%
,\mathcal{S})$ must be Gaussian. Moreover, as every pure Gaussian state is
connected to vacuum via some Gaussian unitary transformation which can be
included in the joint unitary $\hat{U}_{\mathrm{G}}(\mathbf{s},\mathcal{S})$,
we can take the initial state of the environment as the multi-mode vacuum
state, that is,%
\begin{equation}
|\psi_{E}\rangle=%
{\displaystyle\bigotimes\limits_{j=1}^{N_{E}}}
|0\rangle\doteqdot|vac\rangle,
\end{equation}
where we have assumed that the environment consists in $N_{E}$ modes. The
state $\hat{\rho}$ of the system is then transformed into%
\begin{equation}
\hat{\rho}^{\prime}=\mathrm{tr}_{E}\{\hat{U}_{\mathrm{G}}(\mathbf{s}%
,\mathcal{S})[\hat{\rho}\otimes|vac\rangle\langle vac|]\hat{U}_{\mathrm{G}%
}^{\dagger}(\mathbf{s},\mathcal{S})\}.
\end{equation}
Let us write the Gaussian parameters associated to the joint unitary as%
\begin{equation}
\mathbf{s}=\operatorname{col}(\mathbf{d},\mathbf{d}_{E})\text{, \ \ \ \ and
\ \ \ \ }\mathcal{S}=%
\begin{bmatrix}
\mathcal{S}_{S} & \mathcal{S}_{SE}\\
\mathcal{S}_{ES} & \mathcal{S}_{E}%
\end{bmatrix}
,
\end{equation}
where $\mathbf{d\in%
\mathbb{R}
}^{2N}$, and the real matrices $\mathcal{S}_{S}$, $\mathcal{S}_{E}$,
$\mathcal{S}_{SE}$, and $\mathcal{S}_{ES}$, have dimensions $2N\times2N$,
$2N_{E}\times2N_{E}$, $2N\times2N_{E}$, and $2N_{E}\times2N,$ respectively.
Let us write also the mean and the covariance matrix of the initially
separable joint state of the system plus the environment as%
\begin{equation}
\mathbf{\bar{r}}_{SE}=\operatorname{col}(\mathbf{\bar{r}},\mathbf{0}),\text{
\ \ \ \ and \ \ \ \ }V_{SE}=V\oplus\mathcal{I}_{2N_{E}\times2N_{E}}\text{;}%
\end{equation}
after the unitary, these are transformed into%
\begin{align}
\mathbf{\bar{r}}_{SE}^{\prime}  &  =\mathcal{S}\mathbf{\bar{r}}_{SE}%
=\operatorname{col}(\mathcal{S}_{S}\mathbf{\bar{r}}+\mathbf{d},\mathbf{d}%
_{E})\text{,}\\
V_{SE}^{\prime}  &  =\mathcal{S}V_{SE}\mathcal{S}^{T}=%
\begin{bmatrix}
\mathcal{S}_{S}V\mathcal{S}_{S}^{T}+\mathcal{S}_{SE}\mathcal{S}_{SE}^{T} &
\mathcal{S}_{S}V\mathcal{S}_{S}+\mathcal{S}_{SE}\mathcal{S}_{E}^{T}\\
\mathcal{S}_{ES}V\mathcal{S}_{S}^{T}+\mathcal{S}_{E}\mathcal{S}_{SE}^{T} &
\mathcal{S}_{ES}V\mathcal{S}_{ES}^{T}+\mathcal{S}_{E}\mathcal{S}_{E}^{T}%
\end{bmatrix}
,
\end{align}
so that by tracing out the environment, the transformation onto the mean
vector and the covariance matrix of the system is%
\begin{equation}
\mathbf{\bar{r}}^{\prime}=\mathcal{S}_{S}\mathbf{\bar{r}}+\mathbf{d},\text{
\ \ \ \ and \ \ \ \ }V^{\prime}=\mathcal{S}_{S}V\mathcal{S}_{S}^{T}%
+\mathcal{S}_{SE}\mathcal{S}_{SE}^{T},
\end{equation}
which is exactly the type of transformation introduced in (\ref{ChannelTrans}%
), where we now make the identifications%
\begin{equation}
\mathcal{K}=\mathcal{S}_{S}\text{, \ \ \ \ and \ \ \ \ }\mathcal{N}%
=\mathcal{S}_{SE}\mathcal{S}_{SE}^{T}.
\end{equation}

We would like to stress that it is possible to show that the Stinespring
dilation of any Gaussian channel can be generated by choosing an environment
with less than twice the number of modes of the system, that is, $N_{E}\leq2N$.

Note that the Gaussianity is a property of the channel, not of the state of
the system, that is, one can consider the action of the Gaussian channel onto
non-Gaussian states, as we shall make later. Indeed, the transformation of a
general state $\hat{\rho}$ after passing through the channel receives a very
simple description in terms of characteristic functions. To see this, let us
write the inverse of the symplectic matrix $\mathcal{S}$\ as
\begin{equation}
\mathcal{S}^{-1}=%
\begin{bmatrix}
\mathcal{T}_{S} & \mathcal{T}_{SE}\\
\mathcal{T}_{ES} & \mathcal{T}_{E}%
\end{bmatrix}
=%
\begin{bmatrix}
(\mathcal{S}/\mathcal{S}_{E})^{-1} & -\mathcal{S}_{S}^{-1}\mathcal{S}%
_{SE}(\mathcal{S}/\mathcal{S}_{E})^{-1}\\
-(\mathcal{S}/\mathcal{S}_{E})^{-1}\mathcal{S}_{ES}\mathcal{S}_{S}^{-1} &
(\mathcal{S}/\mathcal{S}_{S})^{-1}%
\end{bmatrix}
,
\end{equation}
where we have used the general block-inversion formula (see the Wikipedia!),
being%
\begin{equation}
\mathcal{S}/\mathcal{S}_{S}=\mathcal{S}_{E}-\mathcal{S}_{ES}\mathcal{S}%
_{S}^{-1}\mathcal{S}_{SE}\text{ \ \ \ \ and \ \ \ \ }\mathcal{S}%
/\mathcal{S}_{E}=\mathcal{S}_{S}-\mathcal{S}_{SE}\mathcal{S}_{E}%
^{-1}\mathcal{S}_{ES},
\end{equation}
the so-called \textit{Schur complements} of $\mathcal{S}_{S}$ and
$\mathcal{S}_{E}$, respectively. Let us also denoting by $\mathbf{r}$ and
$\mathbf{r}_{E}$ the phase space coordinates of the relevant modes and the
environmental modes, respectively, so that the initial characteristic function
can be written as%
\begin{equation}
\chi_{0}(\mathbf{r},\mathbf{r}_{E}\mathbf{)=}\chi_{\hat{\rho}}(\mathbf{r)}%
\chi_{vac}(\mathbf{r}_{E}\mathbf{)}\text{;}%
\end{equation}
then, recalling the transformation of the characteristic function under
Gaussian unitaries (\ref{CharacEvoUni}), and after tracing out the
environmental modes (that is, setting to zero their phase space variables), we
get the output characteristic function%
\begin{equation}
\chi_{\mathcal{E}[\hat{\rho}]}(\mathbf{r})=e^{-\frac{\mathrm{i}}{2}%
\mathbf{r}^{T}\Omega\mathbf{d}}\chi_{\hat{\rho}}(\mathcal{T}_{S}%
\mathbf{r)}\chi_{vac}(\mathcal{T}_{ES}\mathbf{r)}\text{.}%
\end{equation}

In the following, we will denote by $\mathcal{E}(\mathcal{K},\mathcal{N})$ any
Gaussian channel, obviating the displacement $\mathbf{d}$ which can actually
be generated after the channel via a unitary displacement transformation
(\ref{DisOp}), since this does not change the covariance matrix in any way.

\subsubsection{An example: phase-insensitive Gaussian channels}

There is a particularly simple class of single-mode Gaussian channels which
play an important role in communication technologies: the
\textit{phase-insensitive Gaussian channels}, which are defined by%
\begin{equation}
\mathcal{K}=\sqrt{\tau}\mathcal{I}_{2\times2}\text{, \ \ \ \ and
\ \ \ \ }\mathcal{N}=\mu\mathcal{I}_{2\times2},
\end{equation}
where $\tau\geq0$ and $\mu>0$ satisfy%
\begin{equation}
\mu\geq|\tau-1|,
\end{equation}
by virtue of the positivity condition (\ref{PosCondSingChann}). Note that this
class of channels are called \textquotedblleft
phase-insensitive\textquotedblright\ because they are invariant under
rotations in phase space. We will denote them by $\mathcal{C}(\tau,\mu)$.

After crossing the channel, the covariance matrix $V$ of any state is
transformed into%
\begin{equation}
V^{\prime}=\mathcal{K}V\mathcal{K}^{T}+\mathcal{N}=%
\begin{bmatrix}
\tau V_{11}+\mu & \tau V_{12}\\
\tau V_{21} & \tau V_{11}+\mu
\end{bmatrix}
,
\end{equation}
and hence%
\begin{align}
\mathrm{tr}V^{\prime}  &  =\tau\mathrm{tr}V+2\mu,\\
\det V^{\prime}  &  =\tau^{2}\det V+\mu\lbrack\tau\mathrm{tr}V+\mu].
\end{align}
Taking into account that the mean number of photons is proportional to the
trace of the covariance matrix---see (\ref{SingleMeanNumber})---, and the von
Neumann entropy is a monotonically increasing function of its
determinant---see (\ref{SymplecticEntropy}) and (\ref{SingleSymplecticEigen}%
)---, we conclude that $\tau$ acts as an attenuation (for $0\leq\tau<1$) or
amplification (for $\tau>1$) factor, while $\mu$ adds noise (mixedness) to the
state. Note that this implies that quantum mechanics does not allow to
attenuate or amplify a signal without introducing noise (at least
deterministically, that is, via trace preserving operations), what comes from
the fact that the uncertainty principle (which in turn comes from the position
and momentum commutators themselves) must be satisfied at all times.

There are two interesting limiting cases:

\begin{itemize}
\item For $0\leq\tau<1$ and $\mu=1-\tau$ we talk about \textit{pure-loss
channels}, which are a good approximation of the fibers used in current
optical communication technologies. It is fairly simple to check that the
simplest Stinespring dilation of such channels consists in mixing the the
input mode with a single environmental mode in a beam splitter (\ref{BSop})
with mixing angle $\cos^{2}\beta=\tau$ (see Figure). The parameter
$T\doteqdot\cos^{2}\beta$ is known as the \textit{attenuation factor} or
\textit{transmissivity}, and the channel is usually denoted by $\mathcal{L}%
(T)$.

\item For $\tau>1$ and $\mu=\tau-1$, we talk about \textit{quantum-limited
amplifiers}, which are the less noisy (deterministic) amplifiers that quantum
mechanics allow. Again, it is simple to check that the simplest Stinespring
dilation of these channels consists in mixing the input mode with a single
environmental mode in a two-mode squeezer (\ref{TMSop}) with squeezing
parameter satisfying $\cosh^{2}r=\tau$ (see Figure). The parameter
$G\doteqdot\cosh^{2}r$ is known as the \textit{amplification factor}, and the
channel is usually denoted by $\mathcal{A}(G)$.
\end{itemize}

It is simple to see that any phase-insensitive Gaussian channel can be seen as
the concatenation of a pure-loss channel and a quantum-limited amplifier, that
is, $\mathcal{C}(\tau,\mu)=\mathcal{L}(T)\circ\mathcal{A}(G)$, where%
\begin{equation}
\tau=TG,\text{ \ \ \ \ and \ \ \ \ }\mu=G(1-T)+(G-1)\text{.}%
\end{equation}

\section{Measurements in continuous variables systems\label{MeasurementsCV}}

\subsection{General description of measurements in phase
space\label{MeasurementsPhaseSpace}}

In Section \ref{GenMesPOVM} we learned that the most general measurement that
one can perform in a quantum system can always be described by a complete set
of trace-decreasing operations $\{\mathcal{E}_{j}\}_{j=1,2,...,J>1}$, each
corresponding to one of the possible measurement outcomes. When all the
trace-decreasing operations were described by a single Kraus operator, we
talked about a POVM-based measurement, the simplest generalizations of the
familiar projective measurements. In this section we will learn a convenient
way of describing such generalized measurements for continuous variables systems.

Let's start with some useful definitions. Given the initial state of a system
$\hat{\rho}$, and the POVM $\{\hat{\Pi}_{j}=\hat{E}_{j}^{\dagger}\hat{E}%
_{j}\}_{j=1,2,...,J}$, we will denote by%
\begin{equation}
\tilde{\rho}_{j}\doteqdot\mathcal{E}_{j}[\hat{\rho}]=\hat{E}_{j}\hat{\rho}%
\hat{E}_{j}^{\dagger},
\end{equation}
the unnormalized state obtained after the outcome $j$ appears; such an outcome
appears with probability $p_{j}=\mathrm{tr}\{\tilde{\rho}_{j}\}$, and the
normalized state of the system reads $\hat{\rho}_{j}=p_{j}^{-1}\tilde{\rho
}_{j}$. Similarly, and assuming that the system is described as a collection
of $N$ oscillators, we define the corresponding unnormalized characteristic
and Wigner functions as%
\begin{equation}
\tilde{\chi}_{j}(\mathbf{r)}=\mathrm{tr}\{\hat{D}(\mathbf{r})\tilde{\rho}%
_{j}\}\text{ \ \ \ \ and \ \ \ \ }\tilde{W}_{j}(\mathbf{r})=\int_{%
\mathbb{R}
^{2N}}\frac{d^{2N}\mathbf{s}}{(4\pi)^{2N}}\tilde{\chi}_{j}(\mathbf{s}%
)e^{\frac{\mathrm{i}}{2}\mathbf{s}^{T}\Omega\mathbf{r}}\text{,}
\label{UnnormalizedChiW}%
\end{equation}
from which the probability of the corresponding outcome can be obtained as%
\begin{equation}
p_{j}=\tilde{\chi}_{j}(\mathbf{0})=\int_{%
\mathbb{R}
^{2N}}d^{2N}\mathbf{r}\tilde{W}_{j}(\mathbf{r}), \label{POVMprobability}%
\end{equation}
and the normalized functions as%
\begin{equation}
\chi_{j}(\mathbf{r})=\mathrm{tr}\{\hat{D}(\mathbf{r})\hat{\rho}_{j}%
\}=p_{j}^{-1}\tilde{\chi}_{j}(\mathbf{r)}\text{ \ \ \ \ and \ \ \ \ }%
W_{j}(\mathbf{r})=\int_{%
\mathbb{R}
^{2N}}\frac{d^{2N}\mathbf{s}}{(4\pi)^{2N}}\chi_{j}(\mathbf{s})e^{\frac
{\mathrm{i}}{2}\mathbf{s}^{T}\Omega\mathbf{r}}=p_{j}^{-1}\tilde{W}%
_{j}(\mathbf{r).}%
\end{equation}

There are many situations in which the measurement is not applied to the whole
system, but only to one of the modes that conform it; we talk then about
\textit{partial measurements}. Moreover, as we will see in the next sections,
the measurement performed onto a light beam is usually \textit{destructive},
that is, the mode disappears after the measurement is done, so that one has to
trace it out of the system. Assuming that the system has $N+1$ modes, and that
the measurement is applied to the last mode, this means that the
(unnormalized) state of the remaining $N$ modes after the measurement will
be\footnote{Note that the cyclic property applies also for the partial trace
when the operator acts as the identity on the non-traced subspaces. In
particular, consider a bipartite Hilbert space space $\mathcal{H}_{A}%
\otimes\mathcal{H}_{B}$, and let us denote by $\{|a_{j}\rangle\}_{j}$ and
$\{|b_{l}\rangle\}_{l}$ orthonormal bases in the individual subspaces. Taking
this into account, let's prove the following identity:%
\begin{equation}
\mathrm{tr}\{\hat{\rho}(\hat{I}_{A}\otimes\hat{B})\}=\mathrm{tr}\{(\hat{I}%
_{A}\otimes\hat{B})\hat{\rho}\}, \label{PartialCyclicTrace}%
\end{equation}
where $\hat{\rho}$ is an operator acting on the complete space $\mathcal{H}%
_{A}\otimes\mathcal{H}_{B}$. To this aim, we just use an explicit
representation of the $\hat{\rho}$ operator%
\begin{equation}
\hat{\rho}=\sum_{klmn}\rho_{kl,mn}|a_{k}\rangle\langle a_{l}|\otimes
|b_{m}\rangle\langle b_{n}|\text{ \ \ \ \ }\Longrightarrow\mathrm{tr}%
\{\hat{\rho}(\hat{I}_{A}\otimes\hat{B})\}=\sum_{j}\sum_{klmn}\rho
_{kl,mn}|a_{k}\rangle\langle a_{l}|(\langle b_{j}|b_{m}\rangle\langle
b_{n}|\hat{B}|b_{j}\rangle);
\end{equation}
now, introducing the identity $\hat{I}_{B}=\sum_{r}|b_{r}\rangle\langle
b_{r}|$ right before $\hat{B}$, we get
\begin{equation}
\mathrm{tr}\{\hat{\rho}(\hat{I}_{A}\otimes\hat{B})\}=\sum_{r}\sum_{klmn}%
\rho_{kl,mn}|a_{k}\rangle\langle a_{l}|\left[  \langle b_{r}|\hat{B}\left(
\sum_{j}|b_{j}\rangle\langle b_{j}|\right)  |b_{m}\rangle\langle b_{n}%
|b_{r}\rangle\right]  =\sum_{r}\langle b_{r}|(\hat{I}_{A}\otimes\hat
{B})\left(  \sum_{klmn}\rho_{kl,mn}|a_{k}\rangle\langle a_{l}|\otimes
|b_{m}\rangle\langle b_{n}|\right)  |b_{r}\rangle,
\end{equation}
which is exactly the right had side of (\ref{PartialCyclicTrace}).}%
\begin{equation}
\tilde{\rho}_{j}=\mathrm{tr}_{N+1}\{(\hat{I}_{N}\otimes\hat{E}_{j})\hat{\rho
}(\hat{I}_{N}\otimes\hat{E}_{j}^{\dagger})\}=\mathrm{tr}_{N+1}\{(\hat{I}%
_{N}\otimes\hat{\Pi}_{j})\hat{\rho}\},
\end{equation}
where $\hat{\rho}$ is the initial state of the $N+1$ modes (obviously the
first $N$ modes only `feel' the measurement if they share some correlations
with the measured mode). The first interesting feature of such partial,
destructive measurements is that one only needs the POVM $\{\hat{\Pi}%
_{j}\}_{j=1,2,...,J}$ to evaluate the final state of the non-measured modes;
this is contrast to non-destructive measurements, which require knowledge of
the measurement operators $\{\hat{E}_{j}\}_{j=1,2,...,J}$ in order to
understand the final state.

Partial measurements iare easily described in terms of the characteristic and
Wigner functions. In the case of the characteristic function, the derivation
is simple by using both sides of (\ref{CharacteristicDefN}):%
\begin{align}
\tilde{\chi}_{j}(\mathbf{r}_{\{N\}}\mathbf{)}  &  =\mathrm{tr}_{\{N\}}%
\{\hat{D}(\mathbf{r}_{\{N\}})\tilde{\rho}_{j}\}=\mathrm{tr}\{[\hat
{D}(\mathbf{r}_{\{N\}})\otimes\hat{\Pi}_{j}]\hat{\rho}\}\label{PartialCharac}%
\\
&  =\int_{%
\mathbb{R}
^{2(N+1)}}\frac{d^{2(N+1)}\mathbf{s}}{(4\pi)^{N+1}}\chi_{\hat{\rho}%
}(\mathbf{s)}\underset{(4\pi)^{N}\delta^{(2N)}(\mathbf{r}_{\{N\}}%
-\mathbf{s}_{\{N\}})}{\underbrace{\mathrm{tr}_{\{N\}}\{\hat{D}(\mathbf{r}%
_{\{N\}}-\mathbf{s}_{\{N\}})\}}}\underset{\chi_{\hat{\Pi}_{j}}(-\mathbf{s}%
_{N+1}\mathbf{)}}{\underbrace{\mathrm{tr}_{N+1}\{\hat{\Pi}_{j}\hat{D}%
^{\dagger}(\mathbf{s}_{N+1})\}}}=\int_{%
\mathbb{R}
^{2}}\frac{d^{2}\mathbf{r}_{N+1}}{4\pi}\chi_{\hat{\rho}}(\mathbf{r)}\chi
_{\hat{\Pi}_{j}}(-\mathbf{r}_{N+1}\mathbf{).}\nonumber
\end{align}
Using (\ref{UnnormalizedChiW}) and (\ref{WignerDefN}), we derive now the
transformation rule for the Wigner function:%
\begin{align}
\tilde{W}_{j}(\mathbf{r}_{\{N\}})  &  =\int_{%
\mathbb{R}
^{2N}}\frac{d^{2N}\mathbf{s}_{\{N\}}}{(4\pi)^{2N}}\int_{%
\mathbb{R}
^{2}}\frac{d^{2}\mathbf{s}_{N+1}}{4\pi}\chi_{\hat{\rho}}(\mathbf{s}%
_{\{N\}},\mathbf{s}_{N+1}\mathbf{)}\chi_{\hat{\Pi}_{j}}(-\mathbf{s}%
_{N+1}\mathbf{)}e^{\frac{\mathrm{i}}{2}\mathbf{s}_{\{N\}}^{T}\Omega
_{N}\mathbf{r}_{\{N\}}}\label{PartialWigner}\\
&  =\int_{%
\mathbb{R}
^{2}}d^{2}\mathbf{r}_{N+1}\int_{%
\mathbb{R}
^{2(N+1)}}\frac{d^{2(N+1)}\mathbf{s}}{(4\pi)^{2N+1}}\chi_{\hat{\rho}%
}(\mathbf{s)}W_{\hat{\Pi}_{j}}(\mathbf{r}_{N+1})e^{\frac{\mathrm{i}}%
{2}\mathbf{s}^{T}\Omega\mathbf{r}}=4\pi\int_{%
\mathbb{R}
^{2}}d^{2}\mathbf{r}_{N+1}W_{\hat{\rho}}(\mathbf{r)}W_{\hat{\Pi}_{j}%
}(\mathbf{r}_{N+1}).\nonumber
\end{align}
Hence, both for the characteristic and the Wigner functions, the
transformation is obtained by multiplying the initial function of the $N+1$
modes by the function associated to the POVM element $\hat{\Pi}_{j}$ (with the
suitable sign in the argument), and integrating out the measured mode.

\subsection{Photodetection: measuring the photon number}

The most fundamental measurement technique for light is photodetection. As we
shall see with a couple of examples (homodyne detection and on/off detection),
any other scheme used for measuring different properties of light makes use of
photodetection as a part of it.

This technique is based on the photoelectric effect or variations of it. The
idea is that when the light beam that we want to detect impinges a metallic
surface, it is able to release some of the bound electrons of the metal, which
are then collected by an anode. The same happens if light impinges on a
semiconductor surface, though in this case instead of becoming free, valence
electrons are promoted to the conduction band. The most widely used metallic
photodetectors are known as \textit{photo--multiplier tubes}, while those
based on semiconducting films are the so-called \textit{avalanche photo
diodes}. In both cases, each photon is able to create one single electron,
whose associated current would be equally difficult to measure by electronic
means; for this reason, each photoelectron is accelerated towards a series of
metallic plates at increasing positive voltages, releasing then more electrons
which contribute to generate a measurable electric pulse, the
\textit{photopulse}.

It is customarily said that counting photopulses is equivalent to counting
photons, and hence, photodetection is equivalent to a measurement of the
number of photons of the light field. This is a highly idealized situation,
valid only in some limits which we will try to understand now.%

\begin{figure}[t]
\begin{center}
\includegraphics[width=0.5\textwidth]{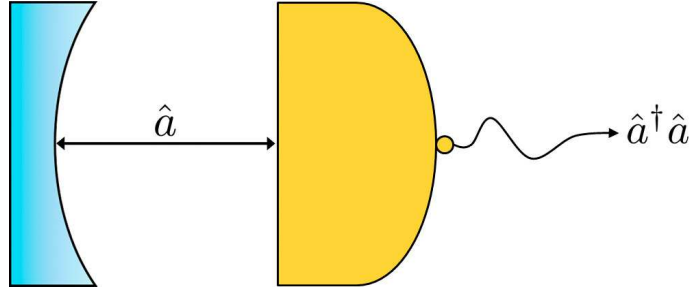}%
\caption{Schematic representation of Mollow's ideal single--mode detection.}%
\label{fDetec1}%
\end{center}
\end{figure}

Consider the following model for a perfectly efficient detection scheme. A
single--mode field with boson operators $\{\hat{a},\hat{a}^{\dagger}\}$
initially in some state $\hat{\rho}$ is kept in continuous interaction with a
photodetector during a time interval $T$. The intuitive picture of such a
scenario is shown in Figure \ref{fDetec1}: A cavity formed by the
photodetector itself and an extra perfectly reflecting mirror contains a
single mode. By developing a microscopic model of the detector and its
interaction with the light mode, Mollow was able to show that the probability
of generating $n$ photoelectrons (equivalently, the probability of observing
$n$ photopulses) during the time interval $T$ is given by \cite{Mollow68}%
\begin{equation}
p_{n}=\left\langle :\frac{\left(  1-e^{-\kappa T}\right)  ^{n}\hat{a}^{\dagger
n}\hat{a}^{n}}{n!}\exp\left[  -(1-e^{-\kappa T})\hat{a}^{\dagger}\hat
{a}\right]  :\right\rangle ,
\end{equation}
where the expectation value has to be evaluated in the initial state
$\hat{\rho}$ of the light mode, and $\kappa$ is some parameter accounting for
the light--detector interaction. Using the operator identity :$\exp\left[
-(1-e^{-\lambda})\hat{a}^{\dagger}\hat{a}\right]  $: $=\exp(-\lambda\hat
{a}^{\dagger}\hat{a})$ \cite{Louisell73book}, and the help of the number state
basis $\{|n\rangle\}_{n\in%
\mathbb{N}
}$, it is straightforward to get%
\begin{equation}
p_{n}=\sum_{m=n}^{\infty}\langle n|\hat{\rho}|n\rangle\frac{m!}{n!(m-n)!}%
\left(  1-e^{-\kappa T}\right)  ^{n}\left(  e^{-\kappa T}\right)
^{m-n}\underset{T\gg\kappa^{-1}}{\longrightarrow}\langle n|\hat{\rho}%
|n\rangle, \label{pn}%
\end{equation}
and hence, for large enough detection times the number of observed pulses
follows the statistics of the number of photons. In other words, this ideal
photodetection scheme is equivalent to measuring the number operator $\hat
{a}^{\dagger}\hat{a}$ as already commented, that is, a projective measurement
with projectors $\{\hat{P}_{n}=|n\rangle\langle n|\}_{n=0,1,2,...}$.

However, in real photodetectors the condition $T\gg\kappa^{-1}$ is hardly met;
one usually defines the \textit{quantum efficiency} $\eta=1-e^{-\kappa T}$,
which in current photodetectors varies from one wavelength to another, and
then photodetection is equivalent to a generalized measurement with POVM
elements $\{\hat{\Pi}_{n}\}_{n=0,1,...}$, being%
\begin{equation}
\hat{\Pi}_{n}=\sum_{m=n}^{\infty}\binom{m}{n}\eta^{n}\left(  1-\eta\right)
^{m-n}|n\rangle\langle n|.
\end{equation}
This POVM-based measurement admits a very simple Stinespring dilation, which
is quite convenient to gain some intuition about optical measurement schemes:
before arriving to a photodetector with unit quantum efficiency, the optical
mode is mixed with an ancillary vacuum mode in a beam splitter of
transmissivity $\cos^{2}\beta=\eta$. In order to prove that this scheme leads
to the same POVM as the detector with finite efficiency, let us compute the
probability of observing $n$ photopulses in the detector. Using the identity%
\begin{equation}
\hat{B}(\beta)=e^{\beta\hat{a}\hat{a}_{E}^{\dagger}-\beta\hat{a}^{\dagger}%
\hat{a}_{E}}=e^{\hat{a}\hat{a}_{E}^{\dagger}\tan\beta}(\cos\beta)^{\hat
{a}^{\dagger}\hat{a}-\hat{a}_{E}^{\dagger}\hat{a}_{E}}e^{-\hat{a}^{\dagger
}\hat{a}_{E}\tan\beta},
\end{equation}
and taking into account that%
\begin{equation}
e^{\zeta\hat{a}\hat{a}_{E}^{\dagger}}|k,0\rangle=\sum_{j=0}^{k}\frac{\zeta
^{j}}{j!}\hat{a}^{j}\hat{a}_{E}^{\dagger j}|k,0\rangle=\sum_{j=0}^{k}\zeta
^{j}\sqrt{\binom{k}{j}}|k-j,j\rangle,
\end{equation}
the state of the system after the beam splitter can be written as%
\begin{equation}
\hat{\rho}_{SE}=\hat{B}(\beta)(\hat{\rho}\otimes|0\rangle\langle0|)\hat
{B}^{\dagger}(\beta)=\sum_{n,m=0}^{\infty}\rho_{nm}\cos^{n+m}\beta\sum
_{j=0}^{n}\sum_{l=0}^{m}\sqrt{\binom{n}{j}\binom{m}{l}}\tan^{j+l}%
\beta|n-j,j\rangle\langle m-l,l|.
\end{equation}
The reduced of the detected mode reads then,%
\begin{equation}
\hat{\rho}_{S}=\mathrm{tr}\{\hat{\rho}_{SE}\}=\sum_{n,m=0}^{\infty}\rho
_{nm}\cos^{n+m}\beta\sum_{k=0}^{\min\{n,m\}}\sqrt{\binom{n}{k}\binom{m}{k}%
}\tan^{2k}\beta|n-k\rangle\langle m-k|,
\end{equation}
and since the detector is taken as ideal, the probability of observing $l$
photopulses is equal to%
\begin{equation}
p_{l}=\langle l|\hat{\rho}_{S}|l\rangle=\sum_{n,m=0}^{\infty}\rho_{nm}%
\cos^{n+m}\beta\sum_{k=0}^{\min\{n,m\}}\sqrt{\binom{n}{k}\binom{m}{k}}%
\tan^{2k}\beta\underset{\delta_{nm}\delta_{k,n-l}}{\underbrace{\delta
_{n-k,l}\delta_{m-k,l}}}=\sum_{n=l}^{\infty}\rho_{nn}\binom{n}{n-l}\cos
^{2n}\beta\tan^{2(n-l)}\beta,
\end{equation}
which coincides with (\ref{pn}) once the identification $\cos^{2}\beta=\eta$
is done.

Apart from the finite quantum efficiency, which accounts for the missed
photons which do not generate photoelectrons in the detector, there is another
source of imperfection in the photodetector: electrons which are pulled out
from the detector without interacting with any photon of the detected mode.
One refers to the corresponding photopulses as \textit{dark counts}, and they
can be modeled within the previous Stinespring dilation in a very simple way:
by assuming that the ancilla mode is not in vacuum but in some other state,
say $\hat{\rho}_{E}$. In this scenario, the POVM elements become%
\begin{equation}
\hat{\Pi}_{n}=\mathrm{tr}\{(\hat{I}\otimes\hat{\rho}_{E})\hat{B}^{\dagger
}(\beta)(|n\rangle\langle n|\otimes\hat{I})\hat{B}(\beta)\}.
\end{equation}

In the following all these imperfections will be ignored, so that we will
assume that photodetection is equivalent to a measurement of the number of
photons of the field impinging the detector. However, it is important to
understand the experimental limitations before proposing any interesting
theoretical protocol, and also to know how to treat them theoretically in case
we find the need of doing a more realistic analysis.%

\begin{figure}[t]
\begin{center}
\includegraphics[width=0.5\linewidth]{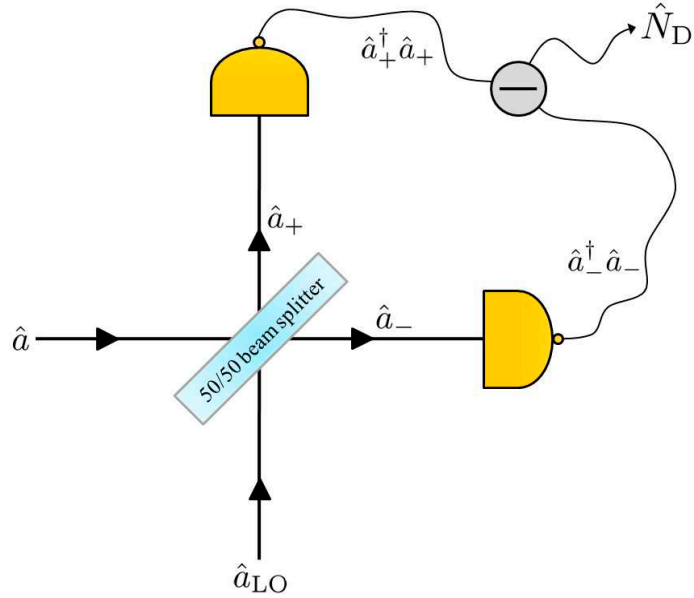}%
\caption{Homodyne detection scheme with ideal photodetectors. When the local
oscillator is in a strong coherent state, this setup gives access to the
quadratures of light.}%
\label{fDetec2}%
\end{center}
\end{figure}

\subsection{Homodyne detection: measuring the quadratures}

Even though the output of the photodetectors can take only integer values
(number of recorded photopulses), they can be arranged to approximately
measure the quadratures of light, which we remind are continuous observables.
This arrangement is called homodyne detection. The basic scheme is shown in
Figure \ref{fDetec2}. The mode we want to measure is mixed in a beam splitter
with another mode, called the \textit{local oscillator}, which is in a
coherent state $|\alpha_{\mathrm{LO}}\rangle$. When the beam splitter is 50/50
the homodyne scheme is said to be \textit{balanced}, and the annihilation
operators of the modes leaving its output ports are given by%
\begin{equation}
\hat{a}_{\pm}=\frac{1}{\sqrt{2}}\left(  \hat{a}\pm\hat{a}_{\mathrm{LO}%
}\right)  \text{,}%
\end{equation}
being $\hat{a}_{\mathrm{LO}}$ the annihilation operator of the local
oscillator mode. These modes are measured with independent photodetectors, and
then the corresponding signals are subtracted. Based on the idealized
photodetection picture of the previous section, this scheme is analogous to a
measurement of the photon number difference%
\begin{equation}
\hat{N}_{\mathrm{D}}=\hat{a}_{+}^{\dagger}\hat{a}_{+}-\hat{a}_{-}^{\dagger
}\hat{a}_{-}=\hat{a}_{\mathrm{LO}}^{\dagger}\hat{a}+\hat{a}_{\mathrm{LO}}%
\hat{a}^{\dagger}\text{.}%
\end{equation}
Taking into account that the local oscillator is in a coherent state with
amplitude $\alpha_{\mathrm{LO}}=|\alpha_{\mathrm{LO}}|\exp(i\varphi)$, and is
not correlated with our measured mode, it is not difficult to show that the
first moments of this operator can be written as%
\begin{subequations}
\begin{align}
\left\langle \hat{N}_{\mathrm{D}}\right\rangle  &  =|\alpha_{\mathrm{LO}%
}|\left\langle \hat{X}^{\varphi}\right\rangle \\
\left\langle \hat{N}_{\mathrm{D}}^{2}\right\rangle  &  =|\alpha_{\mathrm{LO}%
}|^{2}\left[  \left\langle \hat{X}^{\varphi2}\right\rangle +\frac{\left\langle
\hat{a}^{\dagger}\hat{a}\right\rangle }{|\alpha_{\mathrm{LO}}|^{2}}\right]  ,
\end{align}
where%
\end{subequations}
\begin{equation}
\hat{X}^{\varphi}=e^{-\mathrm{i}\varphi}\hat{a}+e^{\mathrm{i}\varphi}\hat
{a}^{\dagger},
\end{equation}
is a generalized quadrature which coincides with the position and momentum for
$\varphi=0$ and $\pi/2$ respectively. Hence, in the \textit{strong local
oscillator limit} $|\alpha_{\mathrm{LO}}|^{2}\gg\max\{1,$ $\left\langle
\hat{a}^{\dagger}\hat{a}\right\rangle \}$, the output signal of the homodyne
scheme has the mean of a quadrature $\hat{X}^{\varphi}$ of the analyzed mode
(the one selected by the phase of the local oscillator), as well as its same
variance. Moreover, it is simple but tedious to check that all the moments of
$\hat{N}_{\mathrm{D}}$ coincide with those of $\hat{X}^{\varphi}$ in the
strong local oscillator limit, and therefore, balanced homodyne detection can
be seen as a measurement of the corresponding quadrature.

It might be difficult to accept that a measurement of $\hat{N}_{\mathrm{D}}$,
which has a discrete spectrum, can be equivalent to a measurement of $\hat
{X}^{\varphi}$, which has a continuous spectrum. The reconciliation between
this two pictures comes from the condition $|\alpha_{\mathrm{LO}}|^{2}\gg1$,
which essentially means that the local oscillator is very intense, and
therefore, there are so many photons impinging the detectors that the
photopulses are generated at a rate much faster than the response time of the
photodetectors, and the output signal is basically felt as a continuous
photocurrent the observer.

Just as we explained in Section \ref{MeasurementsPhaseSpace}, sometimes it is
interesting to apply a measurement onto one mode out of a collection of modes
(say the last mode of a system with $N+1$ modes); this is what we defined as a
partial measurement. Partial homodyne detection receives a very simple
treatment in terms of Wigner functions. For example, in the ideal case
explained above (strong local oscillator limit), homodyne detection of the
position quadrature is described by the continuous set of projectors
$\{\hat{P}(x)=|x\rangle\langle x|\}_{x\in%
\mathbb{R}
}$; assume that the measurement pops out the outcome\footnote{Being a
measurement of a continuous observable, in real experiments the outcome cannot
be a definite value; instead, one can just ensure that the outcome was in
certain interval according to the precission of the measurement device.
Nevertheless, we will use definite-outcome idealization, since nothing
qualitatively different is introduced otherwise.} $x_{0}$, so that, according
to (\ref{PartialWigner}) and (\ref{PartialCharac}), the characteristic and
Wigner functions of the remaining $N$ modes collapse to (unnormalized)%
\begin{equation}
\tilde{\chi}_{x_{0}}(\mathbf{r}_{\{N\}})=\int_{%
\mathbb{R}
^{2}}\frac{d^{2}\mathbf{r}_{N+1}}{4\pi}\chi_{\hat{\rho}}(\mathbf{r)}%
\chi_{|x_{0}\rangle\langle x_{0}|}(-\mathbf{r}_{N+1}\mathbf{)}\text{
\ \ \ \ and \ \ \ \ }\tilde{W}_{x_{0}}(\mathbf{r}_{\{N\}})=4\pi\int_{%
\mathbb{R}
^{2}}d^{2}\mathbf{r}_{N+1}W_{\hat{\rho}}(\mathbf{r)}W_{|x_{0}\rangle\langle
x_{0}|}(\mathbf{r}_{N+1}),
\end{equation}
where $\hat{\rho}$ is the initial state of the $N+1$ modes. The characteristic
function of the projector $|x_{0}\rangle\langle x_{0}|$ is easily found as%
\begin{equation}
\chi_{|x_{0}\rangle\langle x_{0}|}(\mathbf{r})=\mathrm{tr}\{\hat{D}%
(\mathbf{r})|x_{0}\rangle\langle x_{0}|\}=e^{-\mathrm{i}px/4}\langle
x_{0}|e^{\mathrm{i}p\hat{X}/2}e^{-\mathrm{i}x\hat{P}/2}|x_{0}\rangle
=e^{-\mathrm{i}px/4}e^{\mathrm{i}px_{0}/2}\underset{4\pi\delta
(x)}{\underbrace{\int_{%
\mathbb{R}
}dp_{0}e^{-\mathrm{i}xp_{0}/2}}}\underset{1/4\pi}{\underbrace{\langle
x_{0}|p_{0}\rangle\langle p_{0}|x_{0}\rangle}}=\delta(x)e^{\mathrm{i}px_{0}%
/2},\label{PosEigCharacteristic}%
\end{equation}
where we have used (\ref{xy-scalar}), while Fourier transforming this
expression we obtain the corresponding Wigner function%
\begin{equation}
W_{_{|x_{0}\rangle\langle x_{0}|}}(x,p)=\int_{%
\mathbb{R}
^{2}}\frac{dx^{\prime}dp^{\prime}}{(4\pi)^{2}}\chi_{|x_{0}\rangle\langle
x_{0}|}(x^{\prime},p^{\prime})e^{\frac{\mathrm{i}}{2}x^{\prime}p-\frac
{\mathrm{i}}{2}xp^{\prime}}=\frac{\delta(x-x_{0})}{4\pi}\text{.}%
\label{PosEigWigner}%
\end{equation}
These expressions lead us to the following characteristic and Wigner functions
of the remaining modes:%
\begin{equation}
\tilde{\chi}_{x_{0}}(\mathbf{r}_{\{N\}})=\int_{%
\mathbb{R}
^{2}}\frac{dp}{4\pi}\chi_{\hat{\rho}}(\mathbf{r}_{\{N\}},0,p\mathbf{)}%
e^{-\mathrm{i}px_{0}/2}\text{ \ \ \ \ and \ \ \ \ }\tilde{W}_{x_{0}%
}(\mathbf{r}_{\{N\}})=\int_{%
\mathbb{R}
}dpW_{\hat{\rho}}(\mathbf{r}_{\{N\}},x_{0},p\mathbf{)}%
.\label{PartialHomodyneFunctions}%
\end{equation}
Note that, even though the characteristic (\ref{PosEigCharacteristic}) and
Wigner (\ref{PosEigWigner}) functions of the projector are not
normalizable---what makes sense, since the position eigenstate isn't
either---, the characteristic and Wigner functions of the remaining modes can
be normalized, so that the probability density function associated to the
possible outcomes $x_{0}$ is given by%
\begin{equation}
P(x_{0})=\tilde{\chi}_{x_{0}}(\mathbf{0})=\int_{%
\mathbb{R}
^{2N}}d^{2}\mathbf{r}_{\{N\}}\tilde{W}_{x_{0}}(\mathbf{r}_{\{N\}}).
\end{equation}

Note that the projector $|x_{0}\rangle\langle x_{0}|$ is Gaussian, and
therefore, the partial homodyne measurement maps Gaussian states into Gaussian
states. In the following section we evaluate the collapse of a general
Gaussian state associated to a particular Gaussian POVM-element, and we will
give there the transformation rules for such a particular class of states
under a partial homodyne measurement.

\subsection{Partial on/off detection: de-Gaussification by vacuum removal}

Despite the incredible advances in photodetector technologies, practical
photon-counters are still out of reach. A less demanding detection strategy is
the so-called \textit{on/off} detection, in which the detector gives a signal
whenever one or more photons reach it, but this signal is identical no matter
how many photons triggered it. In the next section we will see that on/off
detection allows for the implementation of interesting operations, such as
photon addition and subtraction onto a light field, which are the most
fundamental non-Gaussian and non-unitary operations that one can think of.

On/off detection is then a projective measurement with two possible outcomes
\textit{off} (`no-click') and \textit{on} (`click!'), with corresponding
projectors%
\begin{equation}
\hat{P}_{off}=|0\rangle\langle0|\text{ \ \ \ and \ \ \ \ }\hat{P}_{on}=\hat
{I}-|0\rangle\langle0|=\sum_{n=1}^{\infty}|n\rangle\langle n|.
\end{equation}
These projectors have very simple characteristic functions%
\begin{equation}
\chi_{\hat{P}_{off}}(\mathbf{r)}=\chi_{|0\rangle\langle0|}(\mathbf{r)}\text{
\ \ \ \ and \ \ \ \ }\chi_{\hat{P}_{on}}(\mathbf{r)}=4\pi\delta^{(2)}%
(\mathbf{r})-\chi_{|0\rangle\langle0|}(\mathbf{r),}%
\end{equation}
and very simple Wigner functions as well%
\begin{equation}
W_{\hat{P}_{off}}(\mathbf{r})=W_{|0\rangle\langle0|}(\mathbf{r})\text{
\ \ \ \ and \ \ \ \ }W_{\hat{P}_{on}}(\mathbf{r})=\frac{1}{4\pi}%
-W_{|0\rangle\langle0|}(\mathbf{r).}%
\end{equation}

Applied as a partial measurement (as usual, the measurement is applied onto
the last mode of a system with $N+1$ modes), on/off detection makes the system
evolve from some initial state $\hat{\rho}$ of the $N+1$ oscillators, to a
reduced state of the first $N$ oscillators with (unnormalized) characteristic
and Wigner functions
\begin{equation}
\tilde{\chi}_{off}(\mathbf{r}_{\{N\}}\mathbf{)}=\int_{%
\mathbb{R}
^{2}}\frac{d^{2}\mathbf{r}_{N+1}}{4\pi}\chi_{\hat{\rho}}(\mathbf{r)}%
\chi_{|0\rangle\langle0|}(-\mathbf{r}_{N+1}\mathbf{)}\text{ \ \ \ \ }%
\Longleftrightarrow\text{ \ \ \ \ }\tilde{W}_{off}(\mathbf{r}_{\{N\}}%
)=4\pi\int_{%
\mathbb{R}
^{2}}d^{2}\mathbf{r}_{N+1}W_{\hat{\rho}}(\mathbf{r)}W_{|0\rangle\langle
0|}(\mathbf{r}_{N+1})
\end{equation}
or ($\hat{\rho}_{N}=\mathrm{tr}_{N+1}\{\hat{\rho}\}$ is the reduced initial
state of the $N$ first oscillators)%
\begin{equation}
\tilde{\chi}_{on}(\mathbf{r}_{\{N\}}\mathbf{)}=\chi_{\hat{\rho}_{N}%
}(\mathbf{r}_{\{N\}}\mathbf{)}-\tilde{\chi}_{off}(\mathbf{r}_{\{N\}}%
\mathbf{)}\text{ \ \ \ \ }\Longleftrightarrow\text{ \ \ \ \ }\tilde{W}%
_{on}(\mathbf{r}_{\{N\}})=W_{\hat{\rho}_{N}}(\mathbf{r)-}\tilde{W}%
_{off}(\mathbf{r}_{\{N\}}), \label{OnCharWig}%
\end{equation}
depending on the result of the measurement.

Note that $W_{\hat{P}_{off}}(\mathbf{r})$ is Gaussian, and therefore, the
\textit{off }event projects the state of the non-measured modes into another
Gaussian state. This is not the case for the \textit{on} event, whose
associated Wigner function $W_{\hat{P}_{on}}(\mathbf{r})$ is not Gaussian, and
therefore, it can be used as a \textit{de-Gaussifying} operation. In order to
understand this better, let's analyze the case in which the initial state is a
general Gaussian state $\hat{\rho}_{\mathrm{G}}(\mathbf{d},V)$ of the form
(\ref{BipartiteGaussian}) with $M=N$ and $M^{\prime}=1$, whose mean vector and
covariance matrix we write as%
\begin{equation}
\mathbf{\bar{r}}=(\mathbf{\bar{r}}_{\{N\}},\mathbf{\bar{r}}_{N+1})\text{,
\ \ \ \ and \ \ \ \ }V=\left[
\begin{array}
[c]{cc}%
V_{\{N\}} & C\\
C^{T} & V_{N+1}%
\end{array}
\right]  ,
\end{equation}
where $\mathbf{\bar{r}}_{\{N\}}\in%
\mathbb{R}
^{2N}$, $\mathbf{\bar{r}}_{N+1}\in%
\mathbb{R}
^{2}$, $V_{\{N\}}$ and $V_{N+1}$ are real, symmetric matrices of dimensions
$2N\times2N$ and $2\times2$, respectively, while $C$ is a real $2N\times2$
matrix. Using the Gaussian integral (\ref{GaussianIntegral}), it is
straightforward to prove that the probability of the \textit{off} event is%
\begin{equation}
p_{off}=\frac{2}{\sqrt{\det(V_{N+1}+\mathcal{I}_{2\times2})}}\exp\left[
-\frac{1}{8}\mathbf{\bar{r}}_{N+1}^{T}(V_{N+1}+\mathcal{I}_{2\times
2})\mathbf{\bar{r}}_{N+1}\right]  ,\label{poff}%
\end{equation}
while the corresponding output state is the Gaussian $\hat{\rho}_{off}%
=\hat{\rho}_{\mathrm{G}}(\mathbf{\bar{r}}_{off},V_{off})$ with%
\begin{equation}
\mathbf{\bar{r}}_{off}=\mathbf{\bar{r}}_{\{N\}}-C(V_{N+1}+\mathcal{I}%
_{2\times2})^{-1}\mathbf{\bar{r}}_{N+1}\text{ \ \ \ \ and \ \ \ \ }%
V_{off}=V_{\{N\}}+C(V_{N+1}+\mathcal{I}_{2\times2})^{-1}C^{T}.
\end{equation}
The probability of the \textit{on} event is then $p_{on}=1-p_{off}$, which,
based on (\ref{OnCharWig}), has the associated output Wigner function%
\begin{equation}
W_{on}(\mathbf{r}_{\{N\}})=(1-p_{off})^{-1}[W_{\hat{\rho}_{\mathrm{G}%
}(\mathbf{d}_{\{N\}},V_{\{N\}})}(\mathbf{r}_{\{N\}})-p_{off}W_{\hat{\rho
}_{\mathrm{G}}(\mathbf{d}_{off},V_{off})}(\mathbf{r}_{\{N\}})]\label{Won}%
\end{equation}
Note that even though this Wigner function is not Gaussian, it is a simple
combination of two Gaussians (a \textquotedblleft negative\textquotedblright%
\ mixture in particular), and hence this way of de-Gaussification is very
convenient from the theoretical point of view, since all the tools of Gaussian
states and operations can be used.

Let us consider a simple example: we have two modes in the two-mode squeezed
vacuum state (\ref{TMSVstate}), and we perform an on/off detection onto the
second mode. Given the photon number correlation between the modes, it is
obvious that whenever the outcome is \textit{off}, the first mode gets
projected into the vacuum state, $\hat{\rho}_{off}=|0\rangle\langle0|$; on the
other hand, if the outcome is \textit{on} the state of the first mode will
collapse to the (unnormalized) mixture%
\begin{equation}
\hat{\rho}_{on}=\sum_{n=1}^{\infty}\tanh^{n}r|n\rangle\langle n|\text{.}%
\end{equation}
According to (\ref{poff}), and using the mean vector and covariance matrix of
the two-mode squeezed vacuum state (\ref{TMSVcov}), the probability for the
\textit{off} event is $p_{off}=1/\cosh^{2}r$, and it is simple to check that
(\ref{Won}) leads to $\mathbf{\bar{r}}_{off}=\mathbf{0}$ and\ $V_{off}%
=\mathcal{I}_{2\times2}$, corresponding to the vacuum state. The probability
of the \textit{on} event reads then $p_{on}=\tanh^{2}r$, and the corresponding
Wigner function is%
\begin{equation}
W_{on}(\mathbf{r}_{1})=\sinh^{-2}r[\cosh^{2}rW_{\hat{\rho}_{\mathrm{th}}%
(\sinh^{2}r)}(\mathbf{r}_{1})-W_{|0\rangle\langle0|}(\mathbf{r}_{1})],
\end{equation}
that is, a ``negative mixture'' of a thermal
and a vacuum state. Since the weight of the vacuum state is always larger than that of the thermal state, this Wigner functions always has a negative central region, surrounded by a positive one,
what shows the non-Gaussian character of the state. Moreover, this central negative region has more or less the same size irrespective of the squeezing value; indeed, it is very simple to show that the radius of the central negative region is%
\begin{equation}
R_-=\sqrt{\frac{2\log_{e}(1+\tanh^{2}r)}{\tanh^{2}r}},
\end{equation}
which is a monotonically increasing function of the squeezing, but is lower and upper bounded by $1$ and $\log_{e}^{1/2}4\approx1.18$, and hence varies very little with $r$. In contrast, the positive region gets larger as the squeezing increases, what comes from the thermal component of the state. Finally, note that for small squeezing parameter, the state tends to the
$|1\rangle$ Fock state. In fact, the next step of these notes will consist in showing that if a mode interacts very weakly via the two-mode squeezing or beam splitter interactions with a vacuum mode, the \textit{on} detection of this second mode signals, respectively, the approximate application of the $\hat{a}^{\dagger}$ or $\hat{a}$ operators onto the principal mode.

\appendix

\chapter{The mathematical language of quantum mechanics: Hilbert spaces}
\label{MathematicsQuantum}
Just as classical mechanics is formulated in terms of the mathematical
language of differential calculus and its extensions, quantum mechanics takes
linear algebra (and Hilbert spaces in particular) as its fundamental grammar.
In this section I' introduce the concept of Hilbert space, and discuss the
properties of some operators which will play important roles in the formalism
of quantum mechanics.

\section{Finite--dimensional Hilbert spaces}

In essence, a Hilbert space is a \textit{complex vector space} in which an
\textit{inner product} is defined. Let us define first these terms as are used
in this notes.

A \textit{complex vector space} is a set $\mathcal{V}$, whose elements will be
called \textit{vectors} or \textit{kets} and will be denoted by $\left\{
\left\vert a\right\rangle ,\left\vert b\right\rangle ,\left\vert
c\right\rangle ,...\right\}  $ ($a$, $b$, and $c$ may correspond to any
suitable label), in which the following two operations are defined: the
\textit{vector addition}, which takes two vectors $\left\vert a\right\rangle $
and $\left\vert b\right\rangle $ and creates a new vector inside $\mathcal{V}$
denoted by $\left\vert a\right\rangle +\left\vert b\right\rangle $; and the
\textit{multiplication by a scalar}, which takes a complex number $\alpha\in%
\mathbb{C}
$ (in this section Greek letters will represent complex numbers) and a vector
$\left\vert a\right\rangle $ to generate a new vector in $\mathcal{V}$ denoted
by $\alpha\left\vert a\right\rangle $.

The following additional properties must be satisfied:

\begin{enumerate}
\item The vector addition is commutative and associative, that is, $\left\vert
a\right\rangle +\left\vert b\right\rangle =\left\vert b\right\rangle
+\left\vert a\right\rangle $ and $\left(  \left\vert a\right\rangle
+\left\vert b\right\rangle \right)  +\left\vert c\right\rangle =\left\vert
a\right\rangle +\left(  \left\vert b\right\rangle +\left\vert c\right\rangle
\right)  $.

\item There exists a null vector $\left\vert null\right\rangle $ such that
$\left\vert a\right\rangle +\left\vert null\right\rangle =\left\vert
a\right\rangle $.

\item $\alpha\left(  \left\vert a\right\rangle +\left\vert b\right\rangle
\right)  =\alpha\left\vert a\right\rangle +\alpha\left\vert b\right\rangle .$

\item $\left(  \alpha+\beta\right)  \left\vert a\right\rangle =\alpha
\left\vert a\right\rangle +\beta\left\vert a\right\rangle .$

\item $\left(  \alpha\beta\right)  \left\vert a\right\rangle =\alpha\left(
\beta\left\vert a\right\rangle \right)  .$

\item $1\left\vert a\right\rangle =\left\vert a\right\rangle .$
\end{enumerate}

From these properties it can be proved that the null vector is unique, and can
be built from any vector $\left\vert a\right\rangle $ as $0\left\vert
a\right\rangle $; hence, in the following we denote it simply by $\left\vert
null\right\rangle \equiv0$. It can also be proved that any vector $\left\vert
a\right\rangle $ has a unique \textit{antivector} $\left\vert -a\right\rangle
$ such that $\left\vert a\right\rangle +\left\vert -a\right\rangle =0$, which
is given by $\left(  -1\right)  \left\vert a\right\rangle $ or simply
$-\left\vert a\right\rangle $.

An \textit{inner product} is an additional operation defined in the complex
vector space $\mathcal{V}$, which takes two vectors $\left\vert a\right\rangle
$ and $\left\vert b\right\rangle $ and associates them a complex number. It
will be denoted by $\langle a|b\rangle$ or sometimes also by $\left(
\left\vert a\right\rangle ,\left\vert b\right\rangle \right)  $, and must
satisfy the following properties:

\begin{enumerate}
\item $\langle a|a\rangle>0$ if $\left\vert a\right\rangle \neq0.$

\item $\langle a|b\rangle=\langle b|a\rangle^{\ast}.$

\item $\left(  \left\vert a\right\rangle ,\alpha\left\vert b\right\rangle
\right)  =\alpha\langle a|b\rangle.$

\item $\left(  \left\vert a\right\rangle ,\left\vert b\right\rangle
+\left\vert c\right\rangle \right)  =\langle a|b\rangle+\langle a|c\rangle.$
\end{enumerate}

The following additional properties can be proved from these ones:

\begin{itemize}
\item $\left\langle null\right.  \left\vert null\right\rangle =0.$

\item $\left(  \alpha\left\vert a\right\rangle ,\left\vert b\right\rangle
\right)  =\alpha^{\ast}\langle a|b\rangle.$

\item $\left(  \left\vert a\right\rangle +\left\vert b\right\rangle
,\left\vert c\right\rangle \right)  =\langle a|c\rangle+\langle b|c\rangle.$

\item $\left\vert \langle a|b\rangle\right\vert ^{2}\leq\langle a|a\rangle
\langle b|b\rangle$
\end{itemize}

Note that for any vector $\left\vert a\right\rangle $, one can define the
object $\left\langle a\right\vert \equiv\left(  \left\vert a\right\rangle
,\cdot\right)  $, which will be called a \textit{dual vector} or a
\textit{bra}, and which takes a vector $\left\vert b\right\rangle $ to
generate the complex number $\left(  \left\vert a\right\rangle ,\left\vert
b\right\rangle \right)  \in%
\mathbb{C}
$. It can be proved that the set formed by all the dual vectors corresponding
to the elements in $\mathcal{V}$ is also a vector space, which will be called
the \textit{dual space} and will be denoted by $\mathcal{V}^{+}$. Within this
picture, the inner product can be seen as an operation which takes a bra
$\left\langle a\right\vert $ and a ket $\left\vert b\right\rangle $ to
generate the complex number $\langle a|b\rangle$, a \textit{bracket}. This
whole \textit{bra-c-ket} notation is due to Dirac.

In the following we assume that any time a bra $\left\langle a\right\vert $ is
applied to a ket $\left\vert b\right\rangle $, the complex number $\langle
a|b\rangle$ is formed, so that objects like $|b\rangle\langle a|$ generate
kets when applied to kets from the left, $\left(  |b\rangle\langle a|\right)
|c\rangle=\left(  \langle a|c\rangle\right)  |b\rangle$, and bras when applied
to bras from the right, $\langle c|\left(  |b\rangle\langle a|\right)
=\left(  \langle c|b\rangle\right)  \langle a|$. Technically, $|b\rangle
\langle a|$ is called an \textit{outer product}.

A vector space equipped with an inner product is called an \textit{Euclidean
space}. In the following we give some important definitions and properties
which are needed in order to understand the concept of Hilbert space:

\begin{itemize}
\item The vectors $\left\{  \left\vert a_{1}\right\rangle ,\left\vert
a_{2}\right\rangle ,...,\left\vert a_{m}\right\rangle \right\}  $ are said to
be \textit{linearly independent} if the relation $\alpha_{1}\left\vert
a_{1}\right\rangle +\alpha_{2}\left\vert a_{2}\right\rangle +...+\alpha
_{m}\left\vert a_{m}\right\rangle =0$ is satisfied only for $\alpha_{1}%
=\alpha_{2}=...=\alpha_{m}=0$, as otherwise one of them can be written as a
linear combination of the rest.

\item The \textit{dimension} of the vector space is defined as the maximum
number of linearly independent vectors, and can be finite or infinite.

\item If the dimension of an Euclidean space is $d<\infty$, it is always
possible to build a set of $d$ orthonormal vectors $E=\left\{  \left\vert
e_{j}\right\rangle \right\}  _{j=1,2,..,d}$ satisfying $\langle e_{j}%
|e_{l}\rangle=\delta_{jl}$, such that any other vector $\left\vert
a\right\rangle $ can be written as a linear superposition of them, that is,
$\left\vert a\right\rangle =\sum_{j=1}^{d}a_{j}\left\vert e_{j}\right\rangle
$, being the $a_{j}$'s some complex numbers. This set is called an
\textit{orthonormal basis} of the Euclidean space $\mathcal{V}$, and the
coefficients $a_{j}$ of the expansion can be found as $a_{j}=\langle
e_{j}|a\rangle$. The column formed with the expansion coefficients, which is
denoted by $\operatorname{col}\left(  a_{1},a_{2},...,a_{d}\right)  $, is
called a \textit{representation} of the vector $\left\vert a\right\rangle $ in
the basis $E$.

Note that the set $E^{+}=\left\{  \left\langle e_{j}\right\vert \right\}
_{j=1,2,..,d}$ is an orthonormal basis in the dual space $\mathcal{V}^{+}$, so
that any bra $\left\langle a\right\vert $ can be expanded then as
$\left\langle a\right\vert =\sum_{j=1}^{d}a_{j}^{\ast}\left\langle
e_{j}\right\vert $. The representation of the bra $\left\langle a\right\vert $
in the basis $E$ corresponds to the row formed by its expansion coefficients,
and is denoted by $\left(  a_{1}^{\ast},a_{2}^{\ast},...,a_{n}^{\ast}\right)
$. Note that if the representation of $\left\vert a\right\rangle $ is seen as
a $d\times1$ matrix, the representation of $\left\langle a\right\vert $ can be
obtained as its $1\times d$ conjugate--transpose matrix.

Note finally that the inner product of two vectors $\left\vert a\right\rangle
$ and $\left\vert b\right\rangle $ reads $\langle a|b\rangle=\sum_{j=1}%
^{d}a_{j}^{\ast}b_{j}$ when represented in the same basis, which is the matrix
product of the representations of $\left\langle a\right\vert $ and $\left\vert
b\right\rangle $.
\end{itemize}

For finite dimension, an Euclidean space is a \textit{Hilbert space}. However,
in most applications of quantum mechanics (and certainly in quantum optics),
one has to deal with infinite--dimensional vector spaces. We will treat them
after the following section.

\section{Linear operators in finite--dimensional Hilbert
spaces\label{LinearOperators}}

We now discuss the concept of linear operator, as well as analyze the
properties of some important classes of operators. Only finite--dimensional
Hilbert spaces are considered in this section, we will generalize the
discussion to infinite--dimensional Hilbert spaces in the next section.

We are interested in maps $\hat{L}$ (operators will be denoted with
`\symbol{94}' throughout the notes) which associate to any vector $\left\vert
a\right\rangle $ of a Hilbert space $\mathcal{H}$ another vector denoted by
$\hat{L}\left\vert a\right\rangle $ in the same Hilbert space. If the map
satisfies%
\begin{equation}
\hat{L}\left(  \alpha\left\vert a\right\rangle +\beta\left\vert b\right\rangle
\right)  =\alpha\hat{L}\left\vert a\right\rangle +\beta\hat{L}\left\vert
b\right\rangle ,
\end{equation}
then it is called a \textit{linear operator}. For our purposes this is the
only class of interesting operators, and hence we will simply call them
\textit{operators} in the following.

Before discussing the properties of some important classes of operators, we
need some definitions:

\begin{itemize}
\item Given an orthonormal basis $E=\left\{  \left\vert e_{j}\right\rangle
\right\}  _{j=1,2,..,d}$ in a Hilbert space $\mathcal{H}$ with dimension
$d<\infty$,\ any operator $\hat{L}$ has a representation; while bras and kets
are represented by $d\times1$ and $1\times d$ matrices (rows and columns),
respectively, an operator $\hat{L}$ is represented by a $d\times d$ matrix
with \textit{elements} $L_{jl}=(\left\vert e_{j}\right\rangle ,\hat
{L}\left\vert e_{l}\right\rangle )\equiv\left\langle e_{j}\right\vert \hat
{L}\left\vert e_{l}\right\rangle $. An operator $\hat{L}$ can then be expanded
in terms of the basis $E$ as $\hat{L}=\sum_{j,l=1}^{d}L_{jl}\left\vert
e_{j}\right\rangle \left\langle e_{l}\right\vert $. It follows that the
representation of the vector $\left\vert b\right\rangle =\hat{L}\left\vert
a\right\rangle $ is just the matrix multiplication of the representation of
$\hat{L}$ by the representation of $\left\vert a\right\rangle $, that is,
$b_{j}=\sum_{l=1}^{d}L_{jl}a_{l}$.

\item The\textit{ addition} and \textit{product }of two operators $\hat{L}$
and $\hat{K}$, denoted by $\hat{L}+\hat{K}$ and $\hat{L}\hat{K}$,
respectively, are defined by their action onto any vector $\left\vert
a\right\rangle $: $(\hat{L}+\hat{K})\left\vert a\right\rangle =\hat
{L}\left\vert a\right\rangle +\hat{K}\left\vert a\right\rangle $ and $\hat
{L}\hat{K}\left\vert a\right\rangle =\hat{L}(\hat{K}\left\vert a\right\rangle
)$. It follows that the representation of the addition and the product are,
respectively, the sum and the multiplication of the corresponding matrices,
that is, $(\hat{L}+\hat{K})_{jl}=L_{jl}+K_{jl}$ and $(\hat{L}\hat{K}%
)_{jl}=\sum_{k=1}^{d}L_{jk}K_{kl}$.

\item Note that while the addition is commutative, the product is not in
general. This leads us to the notion of \textit{commutator}, defined for two
operators $\hat{L}$ and $\hat{K}$ as $[\hat{L},\hat{K}]=\hat{L}\hat{K}-\hat
{K}\hat{L}$. When $[\hat{L},\hat{K}]=0$, we say that the operators
\textit{commute}.

\item Given an operator $\hat{L}$, its \textit{trace} is defined as the sum of
the diagonal elements of its matrix representation, that is, $\mathrm{tr}%
\{\hat{L}\}=\sum_{j=1}^{d}L_{jj}$. It may seem that this definition is
basis--dependent, as in general the elements $L_{jj}$ are different in
different bases. However, we will see that the trace is invariant under any
change of basis.

The trace has two important properties. It is \textit{linear} and
\textit{cyclic}, that is, given two operators $\hat{L}$ and $\hat{K}$,
$\mathrm{tr}\{\hat{L}+\hat{K}\}=\mathrm{tr}\{\hat{L}\}+\mathrm{tr}\{\hat{K}\}$
and $\mathrm{tr}\{\hat{L}\hat{K}\}=\mathrm{tr}\{\hat{K}\hat{L}\}$, as is
trivially proved.

\item We say that a vector $\left\vert l\right\rangle $ is an
\textit{eigenvector} of an operator $\hat{L}$ if $\hat{L}\left\vert
l\right\rangle =\lambda\left\vert l\right\rangle $; $\lambda\in%
\mathbb{C}
$ is called its associated \textit{eigenvalue}. The set of all the eigenvalues
of an operator is called its \textit{spectrum}.
\end{itemize}

\bigskip

We can pass now to describe some classes of operators which play important
roles in quantum mechanics.

\bigskip

\textbf{The identity operator.} The \textit{identity operator}, denoted by
$\hat{I}$, is defined as the operator which maps any vector onto itself. Its
representation in any basis is then $I_{jl}=\delta_{jl}$, so that it can
expanded as%
\begin{equation}
\hat{I}=\sum_{j=1}^{d}\left\vert e_{j}\right\rangle \left\langle
e_{j}\right\vert \text{.}%
\end{equation}
This expression is known as the \textit{completeness relation} of the basis
$E$; alternatively, it is said that the set $E$ forms a \textit{resolution of
the identity}.

Note that the expansion of a vector $\left\vert a\right\rangle $ and its dual
$\left\langle a\right\vert $ in the basis $E$ is obtained just by application
of the completeness relation from the left and the right, respectively.
Similarly, the expansion of an operator $\hat{L}$ is obtained by application
of the completeness relation both from the right and the left at the same time.

\bigskip

\textbf{The inverse of an operator.} The \textit{inverse }of an operator
$\hat{L}$, denoted by $\hat{L}^{-1}$, is defined as that satisfying $\hat
{L}^{-1}\hat{L}=\hat{L}\hat{L}^{-1}=\hat{I}$.

\bigskip

\textbf{An operator function}. Consider a real function $f\left(  x\right)  $
which can be expanded in powers of $x$ as $f\left(  x\right)  =\sum
_{m=0}^{\infty}f_{m}x^{m}$; given an operator $\hat{L}$, we define the
\textit{operator} \textit{function} $\hat{f}(\hat{L})=\sum_{m=0}^{\infty}%
f_{m}\hat{L}^{m}$, where $\hat{L}^{m}$ means the product of $\hat{L}$ with
itself $m$ times.

\bigskip

\textbf{The adjoint of an operator.} Given an operator $\hat{L}$, we define
its \textit{adjoint}, and denote it by $\hat{L}^{\dagger}$, as that satisfying
$(\left\vert a\right\rangle ,\hat{L}\left\vert b\right\rangle )=(\hat
{L}^{\dagger}\left\vert a\right\rangle ,\left\vert b\right\rangle )$ for any
two vectors $\left\vert a\right\rangle $ and $\left\vert b\right\rangle $.
Note that the representation of $\hat{L}^{\dagger}$ corresponds to the
conjugate transpose of the matrix representing $\hat{L}$, that is $(\hat
{L}^{\dagger})_{jl}=L_{lj}^{\ast}$. Note also that the adjoint of a product of
two operators $\hat{K}$ and $\hat{L}$ is given by $(\hat{K}\hat{L})^{\dagger
}=\hat{L}^{\dagger}\hat{K}^{\dagger}$.

\bigskip

\textbf{Self--adjoint operators.} We say that $\hat{H}$ is a
\textit{self--adjoint} if it coincides with its adjoint, that is, $\hat
{H}=\hat{H}^{\dagger}$. A property of major importance for the construction of
the laws of quantum mechanics is that the spectrum $\left\{  h_{j}\right\}
_{j=1,2,...,d}$ of a self--adjoint operator is real. Moreover, its associated
eigenvectors\footnote{We will assume that the spectrum of any operator is
non-degenerate, that is, only one eigenvector corresponds to a given
eigenvalue, as all the operators that appear in this thesis have this
property.} $\left\{  \left\vert h_{j}\right\rangle \right\}  _{j=1,2,...,d}$
form an orthonormal basis of the Hilbert space.

The representation of any operator function $\hat{f}(\hat{H})$ in the
\textit{eigenbasis} of $\hat{H}$ is then $\left[  \hat{f}(\hat{H})\right]
_{jl}=f\left(  h_{j}\right)  \delta_{jl}$, from which follows%
\begin{equation}
\hat{f}(\hat{H})=\sum_{j=1}^{d}f\left(  h_{j}\right)  \left\vert
h_{j}\right\rangle \left\langle h_{j}\right\vert .
\end{equation}
This result is known as the \textit{spectral theorem}.

\bigskip

\textbf{Unitary operators.} We say that $\hat{U}$ is a \textit{unitary
operator} if $\hat{U}^{\dagger}=\hat{U}^{-1}$. The interest of this class of
operators is that they preserve inner products, that is, for any two vectors
$\left\vert a\right\rangle $ and $\left\vert b\right\rangle $ the inner
product $(\hat{U}\left\vert a\right\rangle ,\hat{U}\left\vert b\right\rangle
)$ coincides with $\langle a|b\rangle$. Moreover, it is possible to show that
given two orthonormal bases $E=\left\{  \left\vert e_{j}\right\rangle
\right\}  _{j=1,2,..,d}$ and $E^{\prime}=\{|e_{j}^{\prime}\rangle
\}_{j=1,2,..,d}$, there exists a unique unitary matrix $\hat{U}$ which
connects them as $\{|e_{j}^{\prime}\rangle=\hat{U}\left\vert e_{j}%
\right\rangle \}_{j=1,2,..,d}$, and then any basis of the Hilbert space is
unique up to a unitary transformation.

We can now prove that the trace of an operator is basis--independent. Let us
denote by $\mathrm{tr}\{\hat{L}\}_{E}$ the trace of an operator $\hat{L}$ in
the basis $E$; the trace of this operator in the transformed basis can be
written then as $\mathrm{tr}\{\hat{L}\}_{E^{\prime}}=\mathrm{tr}\{\hat
{U}^{\dagger}\hat{L}\hat{U}\}_{E}$, or using the cyclic property of the trace
and the unitarity of $\hat{U}$, $\mathrm{tr}\{\hat{L}\}_{E^{\prime}%
}=\mathrm{tr}\{\hat{U}\hat{U}^{\dagger}\hat{L}\}=\mathrm{tr}\{\hat{L}\}_{E}$,
which proves that the trace is equal in both bases.

Note finally that a unitary operator $\hat{U}$ can always be written as the
exponential of $\mathrm{i}$--times a self--adjoint operator $\hat{H}$, that
is, $\hat{U}=\exp(\mathrm{i}\hat{H})$.

\bigskip

\textbf{Projection operators.} In general, any self--adjoint operator $\hat
{P}$ satisfying $\hat{P}^{2}=\hat{I}$ is called a \textit{projector}. We are
interested only in those projectors which can be written as the outer product
of a vector $\left\vert a\right\rangle $ with itself, that is, $\hat{P}%
_{a}=\left\vert a\right\rangle \left\langle a\right\vert $; when applied to a
vector $\left\vert b\right\rangle $, this gets \textit{projected} along the
`direction' of $\left\vert a\right\rangle $ as $\hat{P}_{a}\left\vert
b\right\rangle =(\langle a|b\rangle)\left\vert a\right\rangle $.

Note that given an orthonormal basis $E$, we can use the projectors $\hat
{P}_{j}=\left\vert e_{j}\right\rangle \left\langle e_{j}\right\vert $ to
extract the components of a vector $\left\vert c\right\rangle $ as $\hat
{P}_{j}\left\vert c\right\rangle =c_{j}\left\vert e_{j}\right\rangle $. Note
also that the completeness and orthonormality of the basis $E$ implies that
$\sum_{j=1}^{d}\hat{P}_{j}=\hat{I}$ and $\hat{P}_{j}\hat{P}_{l}=\delta
_{jl}\hat{I}$, respectively.

\bigskip

\textbf{Density operators.} A self--adjoint operator $\hat{\rho}$ is called a
\textit{density operator} if it is \textit{positive semidefinite}, that is
$\left\langle a\right\vert \hat{\rho}\left\vert a\right\rangle \geq0$ for any
vector $\left\vert a\right\rangle $, and has unit trace.

The interesting property of density operators is that they `hide' probability
distributions in the diagonal of its representation. To see this just note
that given an orthonormal basis $E$, the self--adjointness and positivity of
$\hat{\rho}$ ensure that all its diagonal elements $\left\{  \rho
_{jj}\right\}  _{j=1,2,...,d}\ $are either positive or zero, that is,
$\rho_{jj}\geq0$ $\forall j$, while the unit trace makes them satisfy
$\sum_{j=1}^{d}\rho_{jj}=1$. Hence, the diagonal elements of a density
operator have all the properties required by a \textit{probability
distribution}.

It is possible to show that a density operator can always be expressed as a
\textit{statistical} or \textit{convex mixture} of projection operators, that
is, $\hat{\rho}=\sum_{k=1}^{M}w_{k}\left\vert a_{k}\right\rangle \left\langle
a_{k}\right\vert $, where $\sum_{k=1}^{M}w_{k}=1$ and the vectors $\left\{
\left\vert a_{k}\right\rangle \right\}  _{k=1}^{M}$ are normalized to one, but
don't need not to be orthogonal (note that in fact $M$ doesn't need to be
equal to $d$). Hence, another way of specifying a density matrix is by a set
of normalized vectors together with some statistical rule for mixing them.
When only one vector $\left\vert a\right\rangle $ contributes to the mixture,
$\hat{\rho}=|a\rangle\langle a|$ is completely specified by just this single
vector, and we say that the density operator is \textit{pure}; otherwise, we
say that it is \textit{mixed}.

\section{Generalization to infinite dimensions\label{InfiniteHilbert}}

As explain in Chapter \ref{ContinuousVariables}, the natural Euclidean space
for quantum optical systems is infinite--dimensional, as each mode of the
electromagnetic field behaves as a harmonic oscillator. Unfortunately, not all
the previous concepts and objects that we have introduced for the
finite--dimensional case are trivially generalized to infinite dimensions; in
this section we discuss this generalization.

The first problem that we meet when dealing with infinite--dimensional
Euclidean spaces is that the existence of a basis $\left\{  \left\vert
e_{j}\right\rangle \right\}  _{j=1,2,...}$ in which any other vector can be
represented as $\left\vert a\right\rangle =\sum_{j=1}^{\infty}a_{j}\left\vert
e_{j}\right\rangle $ is not granted. The class of infinite--dimensional
Euclidean spaces in which these infinite but countable bases exist are called
\textit{Hilbert spaces}, and are the ones that will be appearing in quantum mechanics.

The conditions which ensure that an infinite--dimensional Euclidean space is
indeed a Hilbert space can be found in, for example, reference
\cite{Prugovecky71book}. Here we just want to stress that, quite intuitively,
any infinite--dimensional Hilbert space\footnote{An example of
infinite--dimensional complex Hilbert space consists in the vector space
formed by the complex functions of real variable, say $|f\rangle=f\left(
x\right)  $ with $x\in%
\mathbb{R}
$, with integrable square, that is%
\begin{equation}
\int_{%
\mathbb{R}
}dx|f\left(  x\right)  |^{2}<\infty,
\end{equation}
with the inner product%
\begin{equation}
\langle g|f\rangle=\int_{%
\mathbb{R}
}dxg^{\ast}\left(  x\right)  f\left(  x\right)  \text{.}%
\end{equation}
This Hilbert space is known as the $\mathrm{L}^{2}\left(  x\right)  $ space.}
is \textit{isomorphic} to the space called $l^{2}\left(  \infty\right)  $,
which is formed by the column vectors $\left\vert a\right\rangle
\equiv\operatorname{col}(a_{1},a_{2},...)$ where the set $\{a_{j}\in%
\mathbb{C}
\}_{j=1,2,...}$ satisfies the restriction $\sum_{j=1}^{\infty}|a_{j}%
|^{2}<\infty$, and has the operations $\left\vert a\right\rangle +\left\vert
b\right\rangle =\operatorname{col}(a_{1}+b_{1},a_{2}+b_{2},...)$,
$\alpha\left\vert a\right\rangle =\operatorname{col}(\alpha a_{1},\alpha
a_{2},...)$, and $\langle a|b\rangle=\sum_{j=1}^{\infty}a_{j}^{\ast}b_{j}$.

Most of the previous definitions are directly generalized to Hilbert spaces by
taking $d\rightarrow\infty$ (dual space, representations, operators,...).
However, there is one crucial property of self--adjoint operators which
doesn't hold in this case: its eigenvectors may not form an orthonormal basis
of the Hilbert space. The remainder of this section is devoted to deal with
this problem.

Just as in finite dimension, given an infinite--dimensional Hilbert space
$\mathcal{H}$, we say that one of its vectors $|d\rangle$ is an eigenvector of
the self--adjoint operator $\hat{H}$ if $\hat{H}|d\rangle=\delta|d\rangle$,
where $\delta\in%
\mathbb{C}
$ is called its associated eigenvalue. Nevertheless, it can happen in
infinite--dimensional spaces that some vector $|c\rangle$ not contained in
$\mathcal{H}$ also satisfies the condition $\hat{H}|c\rangle=\chi|c\rangle$,
in which case we call it a \textit{generalized eigenvector}, being $\chi$ its
\textit{generalized eigenvalue}\footnote{In $\mathrm{L}^{2}\left(  x\right)  $
we have two simple examples of self--adjoint operators with eigenvectors not
contained in $\mathrm{L}^{2}\left(  x\right)  $: the so-called $\hat{X}$ and
$\hat{P}$, which, given an arbitrary vector $|f\rangle=f\left(  x\right)  $,
act as $\hat{X}|f\rangle=xf\left(  x\right)  $ and $\hat{P}|f\rangle
=-\mathrm{i}df/dx$, respectively. This is simple to see, as the equations%
\begin{equation}
xf_{X}\left(  x\right)  =Xf_{X}\left(  x\right)  \text{ and }-\mathrm{i}%
\frac{d}{dx}f_{P}\left(  x\right)  =Pf_{P}\left(  x\right)  ,
\end{equation}
have%
\begin{equation}
f_{X}\left(  x\right)  =\delta\left(  x-X\right)  \text{ and }f_{P}\left(
x\right)  =\exp\left(  \mathrm{i}Px\right)  \text{,}%
\end{equation}
as solutions, which are not square--integrable, and hence do not belong to
$\mathrm{L}^{2}\left(  x\right)  $.}. The set of all the eigenvalues of the
self--adjoint operator is called its \textit{discrete }(or \textit{point}%
)\textit{ spectrum} and is a countable set, while the set of all its
generalized eigenvalues is called its \textit{continuous spectrum} and is
uncountable, that is, forms a continuous set \cite{Prugovecky71book} (see also
\cite{Galindo90book}).

In quantum optics one finds two extreme cases: either the observable, say
$\hat{H}$, has a pure discrete spectrum $\{h_{j}\}_{j=1,2,...}$; or the
observable, say $\hat{X}$, has a pure continuous spectrum $\{x\}_{x\in%
\mathbb{R}
}$. It can be shown that in the first case the eigenvectors of the observable
form an orthonormal basis of the Hilbert space, so that we can build a
resolution of the identity as $\hat{I}=\sum_{j=1}^{\infty}\left\vert
h_{j}\right\rangle \left\langle h_{j}\right\vert $, and proceed along the
lines of the previous sections.

In the second case, the set of generalized eigenvectors cannot form a basis of
the Hilbert space in the strict sense, as they do not form a countable set and
do not even belong to the Hilbert space. Fortunately, there are still ways to
treat the generalized eigenvectors of $\hat{X}$ `as if' they were a basis of
the Hilbert space. The idea was introduced by Dirac \cite{Dirac30book}, who
realized that normalizing the generalized eigenvectors as\footnote{This
$\delta\left(  x\right)  $ function the so-called \textit{Dirac--delta
distribution} which is defined by the conditions%
\begin{equation}
\int_{x_{1}}^{x_{2}}dx\delta\left(  x-y\right)  =\left\{
\begin{array}
[c]{cc}%
1 & \text{if }y\in\left[  x_{1},x_{2}\right]  \\
0 & \text{if }y\notin\left[  x_{1},x_{2}\right]
\end{array}
\right.  .
\end{equation}
} $\langle x|y\rangle=\delta\left(  x-y\right)  $, one can define the
following integral operator%
\begin{equation}
\int_{%
\mathbb{R}
}dx|x\rangle\left\langle x\right\vert =\hat{I}_{\mathrm{c}},
\end{equation}
which acts as the identity onto the generalized eigenvectors, that is,
$\hat{I}_{\mathrm{c}}|x\rangle=|x\rangle$; it is then assumed that $\hat
{I}_{\mathrm{c}}$ coincides with the identity in $\mathcal{H}$, so that any
other vector $|a\rangle$ or operator $\hat{L}$ in the Hilbert space can be
expanded as%
\begin{equation}
|a\rangle=\int_{%
\mathbb{R}
}dxa\left(  x\right)  |x\rangle\text{ \ \ \ \ and \ \ \ }\hat{L}=\int_{%
\mathbb{R}
^{2}}dxdyL\left(  x,y\right)  |x\rangle\left\langle y\right\vert
\end{equation}
where the elements $a\left(  x\right)  =\langle x|a\rangle$ and $L\left(
x,y\right)  =\langle x|\hat{L}|y\rangle$ of these \textit{continuous
representations} form complex functions defined in $%
\mathbb{R}
$ and $%
\mathbb{R}
^{2}$, respectively. From now on, we will call \textit{continuous basis} to
the set $\left\{  |x\rangle\right\}  _{x\in%
\mathbb{R}
}$.

Dirac introduced this continuous representations as a `limit to the continuum'
of the countable case; even though this approach was very intuitive, it lacked
of mathematical rigor. Some decades after Dirac's proposal, Gel'fand showed
how to generalize the concept of Hilbert space to include these generalized
representations in full mathematical rigor \cite{Gelfand64book}. The
generalized spaces are called \textit{rigged Hilbert spaces} (in which the
algebra of Hilbert spaces joins forces with the theory of continuous
probability distributions), and working on them it is possible to show that
given any self--adjoint operator, one can use its eigenvectors and generalized
eigenvectors to expand any vector of the Hilbert space.

Note finally that given two vectors $|a\rangle$ and $|b\rangle$ of the Hilbert
space, and a continuous basis $\left\{  |x\rangle\right\}  _{x\in%
\mathbb{R}
}$, we can use their generalized representations to write their inner product
as%
\begin{equation}
\langle a|b\rangle=\int_{%
\mathbb{R}
}dxa^{\ast}\left(  x\right)  b\left(  x\right)  .
\end{equation}
It is also easily proved that the trace of any operator $\hat{L}$ can be
evaluated from its continuous representation on $\left\{  |x\rangle\right\}
_{x\in%
\mathbb{R}
}$ as%
\begin{equation}
\mathrm{tr}\{\hat{L}\}=\int_{%
\mathbb{R}
}dxL\left(  x,x\right)  .
\end{equation}
This has important consequences for the properties of density operators, say
$\hat{\rho}$ for the discussion which follows. We explained at the end of the
last section that when represented on an orthonormal basis of the Hilbert
space, its diagonal elements (which are real owed to its self--adjointness)
can be seen as a probability distribution, because they satisfy $\sum
_{j=1}^{\infty}\rho_{jj}=1$ and $\rho_{jj}\geq0$ $\forall$ $j$. Similarly,
because of its unit trace and positivity, the diagonal elements of its
continuous representation satisfy $\int_{%
\mathbb{R}
}dx\rho\left(  x,x\right)  =1$ and $\rho\left(  x,x\right)  \geq0$ $\forall$
$x$, and hence, the real function $\rho\left(  x,x\right)  $ can be seen as a
\textit{probability density function}.

\section{Composite Hilbert spaces}

In many moments of these notes, we find the need to associate a Hilbert space
to a composite system, the Hilbert spaces of whose parts we now. In this
section we show how to build a Hilbert space $\mathcal{H}$ starting from a set
of Hilbert spaces $\left\{  \mathcal{H}_{A},\mathcal{H}_{B},\mathcal{H}%
_{C}...\right\}  $.

Let us start with only two Hilbert spaces $\mathcal{H}_{A}$ and $\mathcal{H}%
_{B}$ with dimensions $d_{A}$ and $d_{B}$, respectively (which might be
infinite); the generalization to an arbitrary number of Hilbert spaces is
straightforward. Consider a vector space $\mathcal{V}$ with dimension
$\mathrm{dim}(\mathcal{V})=d_{A}\times d_{B}$. We define a map called the
\textit{tensor product} which associates to any pair of vectors $|a\rangle
\in\mathcal{H}_{A}$ and $|b\rangle\in\mathcal{H}_{B}$ a vector in
$\mathcal{V}$ which we denote by $|a\rangle\otimes|b\rangle\in\mathcal{V}$.
This tensor product must satisfy the following properties:

\begin{enumerate}
\item $\left(  |a\rangle+|b\rangle\right)  \otimes|c\rangle=|a\rangle
\otimes|c\rangle+|b\rangle\otimes|c\rangle.$

\item $|a\rangle\otimes\left(  |b\rangle+|c\rangle\right)  =|a\rangle
\otimes|b\rangle+|a\rangle\otimes|c\rangle.$

\item $\left(  \alpha|a\rangle\right)  \otimes|b\rangle=|a\rangle
\otimes\left(  \alpha|b\rangle\right)  .$
\end{enumerate}

If we endorse the vector space $\mathcal{V}$ with the inner product $\left(
|a\rangle\otimes|b\rangle,|c\rangle\otimes|d\rangle\right)  =\langle
a|c\rangle\langle b|d\rangle$, it is easy to show it becomes a Hilbert space,
which in the following will be denoted by $\mathcal{H}=\mathcal{H}_{A}%
\otimes\mathcal{H}_{B}$. Given the bases $E_{A}=\{|e_{j}^{A}\rangle
\}_{j=1,2,...,d_{A}}$ and $E_{B}=\{|e_{j}^{B}\rangle\}_{j=1,2,...,d_{B}}$ of
the Hilbert spaces $\mathcal{H}_{A}$ and $\mathcal{H}_{B}$, respectively, a
basis of the \textit{tensor product Hilbert space }$\mathcal{H}_{A}%
\otimes\mathcal{H}_{B}$ can be built as $E=E_{A}\otimes E_{B}=\{|e_{j}%
^{A}\rangle\otimes|e_{l}^{B}\rangle\}_{l=1,2,...,d_{B}}^{j=1,2,...,d_{A}}$
(note that the notation in the first equality is symbolical).

We will use a more economic notation for the tensor product, namely
$|a\rangle\otimes|b\rangle=|a,b\rangle$, except when the explicit tensor
product symbol is needed for any reason. With this notation the basis of the
tensor product Hilbert space is written as $E=\{|e_{j}^{A},e_{l}^{B}%
\rangle\}_{l=1,2,...,d_{B}}^{j=1,2,...,d_{A}}$.

The tensor product also maps operators acting on $\mathcal{H}_{A}$ and
$\mathcal{H}_{B}$ to operators acting on $\mathcal{H}$. Given two operators
$\hat{L}_{A}$ and $\hat{L}_{B}$ acting on $\mathcal{H}_{A}$ and $\mathcal{H}%
_{B}$, the \textit{tensor product operator} $\hat{L}=\hat{L}_{A}\otimes\hat
{L}_{B}$ is defined in $\mathcal{H}$ as that satisfying $\hat{L}%
|a,b\rangle=(\hat{L}_{A}|a\rangle)\otimes(\hat{L}_{B}|b\rangle)$ for any pair
of vectors $|a\rangle\in\mathcal{H}_{A}$ and $|b\rangle\in\mathcal{H}_{B}$.
When explicit subindices making reference to the Hilbert space on which
operators act on are used, so that there is no room for confusion, we will use
the shorter notations $\hat{L}_{A}\otimes\hat{L}_{B}=\hat{L}_{A}\hat{L}_{B}$,
$\hat{L}_{A}\otimes\hat{I}=\hat{L}_{A}$, and $\hat{I}\otimes\hat{L}_{B}%
=\hat{L}_{B}$.

Note that the tensor product preserves the properties of the operators; for
example, given two self--adjoint operators $\hat{H}_{A}$ and $\hat{H}_{B}$,
unitary operators $\hat{U}_{A}$ and $\hat{U}_{B}$, or density operators
$\hat{\rho}_{A}$ and $\hat{\rho}_{B}$, the operators $\hat{H}_{A}\otimes
\hat{H}_{B}$, $\hat{U}_{A}\otimes\hat{U}_{B}$, and $\hat{\rho}_{A}\otimes
\hat{\rho}_{B}$ are self--adjoint, unitary, and a density operator in
$\mathcal{H}$, respectively. Note that this doesn't mean that any
self--adjoint, unitary, or density operator acting on $\mathcal{H}$ can be
written in a simple tensor product form $\hat{L}_{A}\otimes\hat{L}_{B}$.

\bibliographystyle{IEEEtran}
\addcontentsline{toc}{chapter}{Bibliography}
\bibliography{Carl}

\begin{thebibliography}{10}
\providecommand{\url}[1]{#1}
\csname url@samestyle\endcsname
\providecommand{\newblock}{\relax}
\providecommand{\bibinfo}[2]{#2}
\providecommand{\BIBentrySTDinterwordspacing}{\spaceskip=0pt\relax}
\providecommand{\BIBentryALTinterwordstretchfactor}{4}
\providecommand{\BIBentryALTinterwordspacing}{\spaceskip=\fontdimen2\font plus
\BIBentryALTinterwordstretchfactor\fontdimen3\font minus
  \fontdimen4\font\relax}
\providecommand{\BIBforeignlanguage}[2]{{%
\expandafter\ifx\csname l@#1\endcsname\relax
\typeout{** WARNING: IEEEtran.bst: No hyphenation pattern has been}%
\typeout{** loaded for the language `#1'. Using the pattern for}%
\typeout{** the default language instead.}%
\else
\language=\csname l@#1\endcsname
\fi
#2}}
\providecommand{\BIBdecl}{\relax}
\BIBdecl

\bibitem{Nielsen00book}
M.~A. Nielsen and I.~L. Chuang, \emph{Quantum information and quantum
  computation}.\hskip 1em plus 0.5em minus 0.4em\relax Cambridge University
  Press, 2000.

\bibitem{HorodeckiReview}
R.~Horodecki, P.~Horodecki, M.~Horodecki, and K.~Horodecki, ``Quantum
  entanglement,'' \emph{Rev. Mod. Phys.}, vol.~81, pp. 865--942, 2009.

\bibitem{EisertUN}
J.~Eisert, ``Entanglement in quantum information theory,'' {PhD} dissertation;
  arXiv:quant-ph/061025.

\bibitem{Braunstein05}
S.~L. Braunstein and P.~van Loock, ``Quantum information with continuous
  variables,'' \emph{Rev. Mod. Phys.}, vol.~77, pp. 513--577, 2005.

\bibitem{Weedbrook12}
C.~Weedbrook, S.~Pirandola, R.~Garc\'ia-Patr\'on, N.~J. Cerf, T.~C. Ralph,
  J.~H. Shapiro, and S.~Lloyd, ``Gaussian quantum information,'' \emph{Rev.
  Mod. Phys.}, vol.~84, pp. 621--669, May 2012.

\bibitem{NavarreteThesis}
C.~Navarrete-Benlloch, \emph{Contributions to the quantum optics of multi-mode
  optical parametric oscillators}.\hskip 1em plus 0.5em minus 0.4em\relax {PhD}
  dissertation, 2011.

\bibitem{Arnold87book}
B.~Arnold, \emph{Majorization and the Lorenz order}.\hskip 1em plus 0.5em minus
  0.4em\relax Springer-Verlag Lecture Notes in Statistics 43, 1987.

\bibitem{Schleich01book}
W.~P. Schleich, \emph{Quantum optics in phase space}.\hskip 1em plus 0.5em
  minus 0.4em\relax Wiley-VCH, 2001.

\bibitem{Simon94}
R.~Simon, N.~Mukunda, and B.~Dutta, ``Quantum-noise matrix for multimode
  systems: U(\textit{n}) invariance, squeezing, and normal forms,'' \emph{Phys.
  Rev. A}, vol.~49, pp. 1567--1583, Mar 1994.

\bibitem{Gerry05book}
C.~C. Gerry and P.~L. Knight, \emph{Introductory quantum optics}.\hskip 1em
  plus 0.5em minus 0.4em\relax Cambridge University Press, 2005.

\bibitem{Duan00}
L.-M. Duan, G.~Giedke, J.~I. Cirac, and P.~Zoller, ``Inseparability criterion
  for continuous variable systems,'' \emph{Phys. Rev. Lett.}, vol.~84, p. 2722,
  2000.

\bibitem{Simon00}
R.~Simon, ``Peres-horodecki separability criterion for continuous variable
  systems,'' \emph{Phys. Rev. Lett.}, vol.~84, p. 2726, 2000.

\bibitem{Einstein35}
A.~Einstein, B.~Podolsky, and N.~Rosen, ``Can quantum-mechanical description of
  physical reality be considered complete?'' \emph{Phys. Rev.}, vol.~47, pp.
  777--780, 1935.

\bibitem{ParisNotes}
A.~Ferraro, S.~Olivares, and M.~G.~A. Paris, ``Gaussian states in continuous
  variable quantum information,'' lecture notes; arXiv:quant-ph/0503237.

\bibitem{Giedke01}
G.~Giedke, B.~Kraus, M.~Lewenstein, and J.~I. Cirac, ``Entanglement criteria
  for all bipartite gaussian states,'' \emph{Phys. Rev. Lett.}, vol.~87, p.
  167904, 2001.

\bibitem{Mollow68}
B.~R. Mollow, ``Quantum theory of field attenuation,'' \emph{Phys. Rev.}, vol.
  168, pp. 1896--1919, 1968.

\bibitem{Louisell73book}
W.~H. Louisell, \emph{Quantum statistical properties of radiation}.\hskip 1em
  plus 0.5em minus 0.4em\relax John Wiley \& Sons, 1973.

\bibitem{Prugovecky71book}
E.~Prugove\v{c}ky, \emph{Quantum mechanics in Hilbert space}.\hskip 1em plus
  0.5em minus 0.4em\relax Academic Press, 1971.

\bibitem{Galindo90book}
A.~Galindo and P.~Pascual, \emph{Quantum Mechanics I}.\hskip 1em plus 0.5em
  minus 0.4em\relax Springer Verlag, 1990.

\bibitem{Dirac30book}
P.~A.~M. Dirac, \emph{The principles of quantum mechanics}.\hskip 1em plus
  0.5em minus 0.4em\relax Oxford university press, 1930.

\bibitem{Gelfand64book}
I.~M. Gelfand and N.~Y. Vilenkin, \emph{Generalized Functions, Vol. IV}.\hskip
  1em plus 0.5em minus 0.4em\relax Academic Press, 1964.

\end{thebibliography}

\end{document}